\documentclass[12pt,a4paper]{uq-thesis}
\pdfoutput=1

\usepackage{graphicx}
\usepackage{amsbsy}
\usepackage{amsmath}
\usepackage{color}	
\usepackage{epsfig}
\usepackage[margin=15pt,font=small,labelfont=bf]{caption}
\usepackage{setspace}
\usepackage{amsfonts}
\usepackage{epsfig}
\usepackage[numbers]{natbib}
\usepackage{listings}
\usepackage{subfig}

\renewcommand{\emph}[1]{{\it{#1}}}
\newcommand{\Tr}{\mathrm{Tr}\,}
\newcommand{\etal}{\textit{et al. }}

\title{Numerical Simulations of the Ising Model on the Union Jack Lattice}
\author{Vincent Anthony Mellor}
\school{School of Mathematics and Physics}
\thesistype{Doctor of Philosophy}
\submitdate{18th November 2010}

\begin{document}
\beforepreface

\originalitystatement

\contributionstatementjointworks{No jointly-authored works.}
 
\contributionstatementthesis{Advice and suggestions throughout the degree from my supervisors Dr. Jon Links and Dr. Katrina Hibberd. }

\otherdegreestatement{None}

\incorporatedworkstatement{None}

\additionalworkstatement{None}

\newpage
\acknowledgment{I would like to thank Dr Jon Links and Dr Katrina Hibberd for their supervision and advice during my research. Without their advice and alternate viewpoints I would not have thought of investigating some of the problems I did. I would also like to thank my parents for funding my research, and their belief that it was possible, even when I felt like I had hit a brick wall. Equally I would like to give mention to Dr Higuchi of the University of York for initially introducing me to the Ising model in two dimensions.}

\abstract{
The Ising model is famous model for magnetic substances in Statistical Physics, and has been greatly studied in many forms. It was solved in one-dimension by Ernst Ising in 1925 and in two-dimensions without an external magnetic field by Lars Onsager in 1944. In this thesis we look at the anisotropic Ising model on the Union Jack lattice. This lattice is one of the few exactly solvable models which exhibits a re-entrant phase transition and so is of great interest.

Initially we cover the history of the Ising model and some possible applications outside the traditional magnetic substances. Background theory will be presented before briefly discussing the calculations for the one-dimensional and two-dimensional models. After this we will focus on the Union Jack lattice and specifically the work of Wu and Lin in their 1987 paper ``Ising model on the Union Jack lattice as a free fermion model.'' \cite{Wu1987}. Next we will develop a mean field prediction for the Union Jack lattice after first discussing mean field theory for other lattices. Finally we will present the results of numerical simulations. These simulations will be performed using a Monte Carlo method, specifically the Metropolis-Hastings algorithm, to simulate a Markov chain. Initially we calibrate our simulation program using the triangular lattice, before going on to run simulations for Ferromagnetic, Antiferromagnetic and Metamagnetic systems on the Union Jack lattice.
}



\afterpreface

\chapter{Introduction}
\label{ch:Intro}
The Ising model is popular in statistical physics as a model for magnetic substances. The model can also be applied to systems outside of magnetic substances, such as diffusion of a gas and alloying in metals. Initially given to Ernst Ising as his PhD project by Wilhem Lenz in 1920, it is still the subject of many studies to this day. While Ising looked at the one-dimensional chain \cite{Ising1925}, it was extended by Lars Onsager to the two-dimensional square lattice model without an external magnetic field in 1944 \cite{Onsager1944}. The non-planar two-dimensional model, that is one that is with an external magnetic field, and models with higher dimensions, still remain unsolved except in rare cases \cite{Lin1994a}.

In this thesis, we will investigate the Ising model on the Union Jack lattice and re-entrant phase transitions from the point of view of average magnetisation. In particular, we will investigate the results of Wu and Lin from their 1987 \cite{Wu1987,Wu1988} and 1989 \cite{Wu1989} papers. For this investigation we use three methods in the thesis. First we will perform a theoretical analysis of their results. Next we will develop a mean field simulation of the lattice and compare the results to the original exact solution. Finally we will perform numerical simulations on the various systems and compare these to the theoretical results.

\section{Summary of the thesis}
\label{sec:Insum}
In Chapter \ref{ch:Theory} we will briefly revise some background information that will be required for the rest of the thesis. At the beginning of the chapter the thermodynamic laws will be presented and discussed. We will then move on to looking at some basic statistical mechanical concepts, such as partition functions and entropy. Once we have a framework from the previous sections we will then move on to a discussion of phase transitions. Finally at the end of the chapter we will discuss the topic of magnetic substances from a chemistry point of view.

Using this background theory we will move on to a development of the theory of the Ising model in Chapter \ref{ch:Ising}. We will look in detail at the development of the model, from magnetic sites, to the one-dimensional model and then extending to the two-dimensional square lattice model. At the end of the chapter the results for the triangular lattice are given. The triangular lattice is important as it will be used later for calibration of the simulation.

Having developed the theory for these lattices, we will move on in Chapter \ref{ch:UJL} to look at the Union Jack lattice model. Initially in the chapter, we will discuss the work of Vaks \etal \cite{Vaks1966}. They use an older approach to calculate the critical temperatures of the system. They also discuss the re-entrant transition, with classification of the various phases. Then we will go on to look at the work of Wu and Lin \cite{Wu1987,Wu1989}, developing prediction functions for both sublattices. After this we will discuss their classification of phases at low temperatures. Once we have finished the presentation, we will then perform analysis on interesting systems and discuss the differences between the approaches.

In Chapter \ref{ch:MFT} we will examine the alternative approach of Mean Field theory. First we will review the theory on isotropic lattices giving results for the triangular lattice. From this we will develop equations for the Union Jack lattice. These will be in two forms, a partially uncoupled set of equations which will be dependent on only one sublattice magnetisation and a coupled set of equations. We will present simulations results using our mean field predictions and compare them to those from Wu and Lin \cite{Wu1987,Wu1989}. We will also discuss the limitations of this approach and any disagreements with Wu and Lin's results. The simulation program will be discussed later in Appendix \ref{ch:ProgII}.

At this point of the thesis we briefly move away from Ising models, to discuss the required theory for the next few chapters. Chapter \ref{ch:sample} is concerned with the methods used for our numerical simulations. This section will be mainly concerned with Stochastic processes, in particular the Markov random walk. This will be simulated using a Monte Carlo method to reduce the necessary calculations. The Monte Carlo method we will be using is the Metropolis-Hasting algorithm. We will adapt this algorithm for the simulation program. The actual program code will not be discussed here, but the interested reader may find such a discussion in Appendix \ref{ch:ProgI}. 

To calibrate our simulation program, in Chapter \ref{ch:COP} we will run simulations on the triangular lattice. These simulations will range from the simplest isotropic ferromagnetic system, to the most general anisotropic antiferromagnetic system. We will compare our simulation results to the predictions of Baxter \cite{Baxter1975} and Stephenson \cite{Stephenson1964} and comment on the correlation of results.

Our numerical simulation results on the Union Jack lattice will then be presented in Chapter \ref{ch:NE}. We will follow the structure of the theoretical analysis in Chapter \ref{ch:UJL}, performing simulations on various systems of interest. These simulations will allow us to further investigate systems where there was a disagreement between the approaches on Vaks \etal \cite{Vaks1966} and Wu and Lin \cite{Wu1987,Wu1989}. Equally we will show how our simulation program allows the study of systems that are currently not able to be theoretically predicted.

To conclude the thesis, in Chapter \ref{ch:Conc} a review of the results from the theoretical, mean field theory and numerical approaches will be presented. We will then discuss how we can minimise the disagreement effects, such as additional conditions. A list of references will then follow.

\section{History of the Ising model}
\label{sec:hist}

In 1920, Wilhelm Lenz proposed a basic model of ferromagnetic substances to his then PhD student Ernst Ising. By 1925, Ising was submitting his dissertation \cite{Ising1925} which was the first exactly solved the one-dimensional case, and as such later was called the Lenz-Ising model. He discovered that there is no phase transition in this case. From this result he incorrectly extended it to say that there would be no phase transition in higher dimensional cases. Indeed during the early part of the twentieth century it was believed by some that the partition function would never show a phase transition, as the exponential function is analytic and as such the sum of these functions is also analytic. However, the logarithm of the partition function is not analytic near the critical temperature in the thermodynamic limit. The thermodynamic limit is reached as the number of particles in a system, $N$, approaches infinity. So in the infinite system the partition function is not analytic.

After Lenz and Ising proposed the model, in 1936 Peierls \cite{Peierls1936} was able to explicitly show that a phase transition occurred in  the two-dimensional model. He compared the high and low temperature limits and showed that at infinite temperatures all configurations have equal probability, while at high but not infinite temperatures there occurs clumping although the average magnetisation is still zero. At low temperatures the model resides in either a state with all the spins being positive or negative. Peierls questioned whether it was possible for the system to fluctuate between these states. In his work he managed to establish that the model defines super selection sectors. That is, domains that are not connected by finite fluctuations.

While greats of physics such as Heisenberg (1928) \cite{Heisenberg1928}, Kramers and Wannier (1941) \cite{Kramers1941,Kramers1941a} use the model to examine ferromagnetism and properties such as the Curie temperature (the critical temperature of substances), it was almost 20 years before Lars Onsager had presented a solution to the two-dimensional model with no external field \cite{Onsager1944}. His paper and solution, although algebraically complex, was a large step in understanding the model further. While other solutions have been developed that are less algebraically complex, other than in special cases, it has only been solved for a zero external magnetic field by Schultz \textit{et al.} \cite{Schultz1964}. This solution will be presented later in Chapter \ref{ch:Ising}. The derivation of the solution was published by Yang in \cite{Yang1952}, but Onsager presented the result on the 28 February 1942 at a meeting of the New York Academy of Science.

The extension to three dimensions is more complex. In the paper of Schultz \textit{et al} \cite{Schultz1964}, the method is limited by the transformation of the system into fermion annihilation and creation operators. In 1972, Richard Feynman \cite{Feynman1972} said of the three-dimensional Ising model, ``the exact solution for three dimensions has not yet been found.'' In 2000, Sorin Istrail \cite{Istrail2000}, by extending Francisco Barahona's work \cite{Barahona1982}, showed that the solution to the three-dimensional problem as well as the two-dimensional non-planar model is NP-complete. He used a method called computational intractability, which allows one to discover if a problem can be solved in a feasible timeframe\footnote{Less that the lifetime of a human}. As there are about 6000 such problems, and they are all mathematically equivalent, a solution to one would be a solution to all and this is an infeasible result. While this result does not remove the possibility of a solution (the P versus NP question is still open), it does mean that the solution is not possible using the methods known today. However, the three-dimensional problem, as with the non-planar two-dimensional model, can be solved approximately numerically and in some special cases \cite{Liu1974,Sumour2006,Suzuki1976}.

In this section, I have presented a brief history showing the major milestones in the development of the model in various dimensions, for the interested reader there are many more detailed discussions of the history, such as \cite{Brush1967} or \cite{Plischke1994}. 

\section{Applications of the Ising model}
\label{sec:AIP}
The Ising model has been applied to a great variety of other physical systems such as absorption of gases on to solid surfaces, order-disorder transitions in alloys, concentrated solutions of liquids, the helix-coil transition polypeptides and the absorption of oxygen by haemoglobin. \cite{Vecchio-Sadus1995,Mandl1988,Bray2006}

\begin{figure}[htbp]
	\centering
		\includegraphics[width=79.75mm,height=79.69mm]{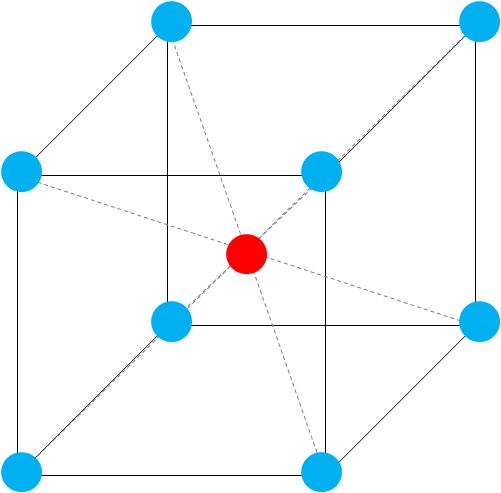}
	\caption[Diagram of the $\beta^\prime$ phase of brass]{Simplified representation of the body centred cubic of the $\beta^\prime$ phase of brass where the \textcolor{red}{red} atom is Copper and the \textcolor{cyan}{blue} atoms are Zinc.}
	\label{fig:bcent}
\end{figure}

An example, taken from \cite{Mandl1988}, of an order-disorder transition in alloys is the binary alloy of Copper and Zinc, brass. Brass has many forms depending on the atomic percentage (that is, the number of atoms) of these metals and type of lattice. For example $\alpha$-brass generally have below 35 atomic percent Zinc, a single phase and is made of a face-centred lattice. At compositions in a narrow range around 50 atomic percent Copper, 50 atomic percent Zinc, the atoms occupy the sites of a body-centred cubic lattice (as show in Figure \ref{fig:bcent}) forming $\beta$-brass. A body-centred cubic lattice consists of two interlocking simple cubic sublattices, where one site of one lattice is centred in the centre of a cube of the other lattice. This system undergoes a phase transition at the critical temperature, $T_c$, of about 740 K. The distribution of atoms on these sites is disordered above a temperature $T_c$. Below $T_c$ there is ordering with atoms of each kind preferentially distributed on one of the two simple cubic sublattices of the body-centred cubic sublattices of the body-centred cubic lattice ($\beta^\prime$ phase).

\section{Related works}
\label{sec:Inrel}
In this research we use a simulation program which uses a Metropolis-Hastings algorithm. This algorithm produces local changes in the system, which means the configuration is changed only in a small neighbourhood of each update. While there are advantages to this method, near the critical point there is a slowing down phenomenon. This means that with an increase in lattice size the computation time increases rapidly so the numerical study has been restricted to small systems. To study larger systems it has been suggested by Bl\"{o}ete \etal in \cite{Bloete2002} that cluster Monte Carlo methods can be used. In these methods, instead of flipping each individual site per iteration, clusters of sites are now flipped. This allows more efficient algorithms, such as those from Swendsen and Wang \cite{Swendsen1987}, Baillie and Coddington \cite{Baillie1991} and Wolff \cite{Wolff1989}. The Wolff single-cluster algorithm \cite{Wolff1989} stands out because of its simplicity with only one cluster being formed and flipped at a time. However, as cluster analysis is not as easy to generalise as the local interaction methods, its applicability range is limited. One possible pitfall of cluster algorithms is in the situation where the clusters tend to occupy practically the whole system, resulting in only trivial changes of the spin configuration, and in a limited efficiency.

The work of Wu and Lin has be extended by Stre\v{c}ka to mixed spin Ising models, in his papers \cite{Strecka2006} and \cite{Strecka2006a}. Mixed spin models are systems where the spins have different magnitudes. These mixed spin models have a much richer critical behaviour compared to the single spin models considered in this thesis. They are also useful as they describe the simplest ferrimagnets (see Section \ref{sec:MagSub}) and so have a wide range of practical applicability. In his solo paper \cite{Strecka2006}, Stre\v{c}ka investigates the spin-1/2 and spin-3/2 Ising model on the Union Jack
lattice. In the article he presents a formulation of the system and its equivalency to the eight-vertex model, and then performs numerical simulations to investigate the critical points. In \cite{Strecka2006a}, he and his co-authors investigate the mixed spin-$(1/2, S)$ Ising model, and compare the integer versus half-odd-integer spin-$S$ case. In the paper they extend Stre\v{c}ka's previous work and examine how critical exponents may depend on the quantum spin number. For example, the mixed spin-(1/2, 1) model exhibits quite different variations of the critical exponents when compared to the (1/2, 3/2) version.

\chapter{Theory Revision}
\label{ch:Theory}

In this chapter we will present a brief review of some of the background theory that will be used in later sections. First we will look at the topic of thermodynamics, discussing both the motivation and the laws of thermodynamics. Then we will move on to look at some important topics in statistical physics like the partition function. Finally we will present an overview of the chemistry of magnetic substances.

The topics in this section are commonly known results that form a basis for a great amount of work. As such discussions on these topics in statistical physics can be found in \cite{Mandl1988, Plischke1994} and for magnetochemistry \cite{Earnshaw1968, Orchard2003, Sharpe1992, Shriver1990}, along with many other books on both subjects.

\section{Motivation for thermodynamics}
\label{sec:MoTH}

A macroscopic system has many unmeasurable quantities that affect the overall behaviour of the system. Thermodynamics concerns itself with the relation between a small number of variables that are sufficient to describe the bulk behaviour of the system in question. In the case of a magnetic solid the variables are the magnetic
field \textbf{B}, the magnetisation \textbf{M}, and the temperature \textit{T}. If the thermodynamic variables are independent of time, the system is said to be in \textit{steady state}. If moreover there are no macroscopic currents in the system, such as a flow of heat or particles through the material, the system is in \textit{equilibrium}.

A thermodynamical transformation or process is any change in the state variables of the system. A \textit{spontaneous} process is one that takes place without any change in the external constraints on the system, due simply to the internal dynamics of the system. An \textit{adiabatic} process is one in which no heat is exchanged between the system and its surroundings. A process is \textit{isothermal} if the temperature is held fixed. A \textit{reversible} process follows a path in thermodynamic space that can be exactly reversed. If this is not possible, the process is \textit{irreversible}.

\section{Laws of thermodynamics}
\label{sec:LawTH}

The laws of thermodynamics are relations that have been found to describe the properties of a system when going through a thermodynamic process. These laws govern properties such as the direction that a process progresses in (which is stated by the second law). They also help to relate thermodynamics to mechanical ideas, such as the conservation of energy.

\subsection*{Zeroth law}
\label{subsec:Zero}

Many quantities like pressure and volume have direct meaning in mechanical systems. However some basic quantities of statistical mechanics are quite foreign to mechanical systems. One of these concepts is that of temperature. Originally temperature relates to the sensations of hot and cold of a particular system. A remarkable feature of temperature is its tendency to equilibrium. One can see this when taking a soda can out of the fridge. When it is initially out of the fridge it is cooler than the surrounding room, but it slowly warms up to be in equilibrium, that is the same temperature, with the surroundings. This type of equilibrium is often referred to as thermal equilibrium. 

The zeroth law is used to assign a measurable value to this concept of temperature. Formally the law can be stated as:

\begin{quotation}
If a system \textit{A} is in equilibrium with systems \textit{B} and \textit{C}, then \textit{B} is in equilibrium with \textit{C}.
\end{quotation}

That is if two systems are in the same temperature as a third common system then they are the same temperature as each other. In general thermometers are systems where a property that depends on its degree of hotness can be measured equally, such as the electric resistance of a platinum wire. Once calibrated with some known temperatures, such as the ice and steam points of water, and interpolating linearly for other temperatures, one can use this system to measure the temperature of other systems. Of course temperature scales depend on the particular thermometer used. This arbitrariness can be removed by developing an \textit{absolute temperature scale} using the second law of thermodynamics (\ref{subsec:Second})

\subsection*{First law}
\label{subsec:First}

The first law of thermodynamics restates the law of conservation of energy. However, it also partitions the change in energy of a system into two pieces, heat and work:

\begin{equation}
	dE=\bar{d}Q-\bar{d}W.
	\label{eq:tno1}	
\end{equation}

In (\ref{eq:tno1}) $dE$ is the change in internal energy of the system, $\bar{d}Q$ the amount of heat \textit{added} to the system, and $\bar{d}W$ the amount of work \textit{done} by the system during an infinitesimal process. Apart from the partitioning of energy into two parts, the formula distinguishes between the infinitesimals $dE$ and $\bar{d}Q$, $\bar{d}W$. The difference between the two measurable quantities $\bar{d}Q$ and $\bar{d}W$ is found to be the same for any process in which the system evolves between two given states, independently of the path. This indicates that $dE$ is an exact differential or, equivalently, that the internal energy is a function based on the state of the system. The same is not true of the differentials $\bar{d}Q$ and $\bar{d}W$, hence the difference in notation.

Consider a system whose state can be specified by the values of a set of the sate variables $x_j$ (for example the magnetisation, et cetera) and the temperature. Thermodynamics exploits an analogy with mechanics and so, for the work done during an infinitesimal process,
\[
	dW=-\sum\limits_jX_jdx_j
\]
where the $X_j$'s can be though of as generalised forces and the $x_j$'s as generalised displacements.

\subsection*{Second law}
\label{subsec:Second}
The second law of thermodynamics introduces the \textit{entropy}, $S$, as a state variable and states that for an infinitesimal reversible process at temperature $T$, the heat given to the system is
\begin{equation}
dQ_{rev} = TdS
\label{eq:tno2}
\end{equation}
while for an irreversible process
\[
dQ_{irrev} \leq TdS
\]
If the only interest is in thermodynamic equilibrium states, (\ref{eq:tno2}) can be used and then
the entropy $S$ can be treated as the generalised displacement that is coupled to the force $T$. The above formulation of the second law is due to Gibbs.

Two equivalent statements of the second law of thermodynamics are:
The \textit{Kelvin} version
\begin{quotation}
There exists no thermodynamic process whose sole effect is to extract
a quantity of heat from a system and convert it entirely to work.
\end{quotation}
and the equivalent statement of \textit{Clausius}
\begin{quotation}
No process exists in which the sole effect is that heat flows from a
reservoir at a given temperature to a reservoir at a higher temperature.
\end{quotation}

\subsection*{Third law}
\label{subsec:Third}

The third law of thermodynamics is of a more limited use, with its main applications being in chemistry and low temperature physics. It originated in Nernst's (1906) study of chemical reactions, and is sometimes called Nernst's Theorem. While it is of some interest in the report and it will briefly discussed, a more detailed treatment can be found in \cite{Wilks1961, Haar1966}. 

The third law of thermodynamics is concerned with the entropy of a system as the temperature approaches absolute zero. Stated simply it says that at absolute zero all bodies will have the same entropy. This means that at absolute zero a body will be in the only possible energy state, so would possess a definite energy (called the \textit{zero-point energy} and would have zero entropy in this state. This value is left undetermined by purely thermodynamic relations such as the previous laws, which only give entropy differences. However a postulate related to the second law, although independent from it, is that it is impossible to cool a body to absolute zero by any finite process. Although you can get as close as you desire, you can not actually reach the absolute zero.

An alternative version of the third law is given by Gilbert N. Lewis and Merle Randall in 1923
\begin{quotation}
	If the entropy of each element in some (perfect) crystalline state be taken as zero at the absolute zero of temperature, every substance has a finite positive entropy; but at the absolute zero of temperature the entropy may become zero, and does so become in the case of perfect crystalline substances.
\end{quotation}

Of course there are substances where at absolute zero there exists a residual entropy, such as carbon monoxide, but this is because there exist more than one ground state with the same zero point energy. 

\section{Partition functions}
\label{sec:Part}

The partition function ($Z$) \cite{Mandl1988} is a quantity in Statistical Mechanics that encodes the statistical properties of a system in thermodynamic equilibrium. It is a function of temperature and other parameters, such as the volume enclosing a gas. Using the partition function of a system one can derive quantities such as the average energy, free energy, entropy or pressure from the function itself and its derivatives.

There are many forms of the partition function depending on the type of statistical ensemble used. The two major versions are the \textit{Canonical partition function} and the \textit{Grand canonical partition function} \cite{Plischke1994}. The canonical partition function applies to a canonical ensemble, that is a system that is allowed to exchange heat with the environment at a fixed temperature, volume and number of particles. The grand canonical partition function applies to the grand canonical ensemble, where a system is allowed to exchange heat \textit{and particles} with the environment, at a fixed temperature, volume and chemical potential. Of course for other circumstances one can define other partition functions. In this report we will be using the canonical partition function exclusively, so we will briefly develop this here, and refer to it from now on as the partition function.

\subsection{Definition}
\label{sub:CPDef}

Assume we are looking at a thermodynamically large system that is in constant thermal contact with its surroundings, which have temperature $T$, and the system's volume and number of particles have been fixed, in other words a system in the canonical ensemble. If we label the microstates (exact states) that the system can take with $s$ ($s=1,2,\ldots$) and the energy of these microstates as $E_s$ then the partition function will be
\begin{equation}
	Z=\sum_s{e^{-\beta E_s}}
	\label{eq:part}
\end{equation}
where the ``inverse temperature'' $\beta$ is defined as
\[
	\beta \equiv \frac{1}{k_BT}.
\]

In quantum mechanics, the partition function can be written more formally as the trace over the state space
\[
	Z=\Tr \left(e^{-\beta H} \right)
\]
where $H$ is the quantum Hamiltonian operator.

\subsection{Using the partition function}
\label{sub:UsePF}

It may not be obvious why the partition function is of such importance to Statistical Mechanics from the definition above. The ability to calculated the microstates of a system using a particular model, and from those to calculate a function that can be used to calculate the thermodynamical properties of the system is a very powerful tool. The partition function has a very important statistical meaning which allows this relation to the thermodynamical properties. The probability $P_s$ that a system is in state $s$ is
\begin{equation}
	P_s = \frac{1}{Z}e^{-\beta E_s}.
	\label{eq:tno5}
\end{equation}
$e^{-\beta E_s}$ is the well known \textit{Boltzmann factor} or \textit{Boltzmann weight} \cite{Mandl1988}. As we can see the partition function is playing the role of a normalising constant for this, which ensures that the probabilities add up to one:
\[
	\sum_s{P_s}=\frac{1}{Z}\sum_s{e^{-\beta E_s}}=\frac{1}{Z}Z=1.
\]
It is for this reason $Z$ is called the ``partition function''; it weights the different microstates based on their individual energies. The reason for the partition function to be called $Z$ is because it is the first letter of the German word for sum over states, ``Zustandssumme''.

We can go on to use the partition function to calculate the average energy of the system. This is simply the expected value of the energy, which of course is the total of the energies of the microstates weighted by their relative probabilities.
\begin{equation}
	\left\langle E\right\rangle = \sum_s{E_s P_s}= \frac{1}{Z} \sum_s{E_s e^{-\beta E_s}} = -\frac{\partial \ln Z}{\partial \beta}
	\label{eq:tno6}
\end{equation}
or equivalently,
\[
	\left\langle E\right\rangle = k_B T^2 \frac{\partial \ln Z}{\partial T}.
\]

\subsection{Entropy in statistical physics}
\label{sub:EnSP}

The idea of entropy was introduced above in the second law of thermodynamics, and it was even defined as a state variable, but what is entropy? The entropy of a system is a measure of the disorder of the system and is related to the energy of the system. Each microstate, which are the states at the microscopic level, have statistical weight related to the energy of the system. The entropy of these
microstates can be defined from these statistical weights by using the following formula,
\begin{equation}
S=k_B\ln\Omega(E)
\label{eq:tno3}
\end{equation}
where $\Omega(E)$ is the number of the microstates with energy $E$. The process in which the system is heated up without any work is reversible. Hence
\[
dE=\bar{d}Q_{rev}=TdS.
\]
Therefore, by rearranging this formula
\begin{equation}
\frac{\partial S}{\partial E}=\frac{1}{T}
\label{eq:tno4}
\end{equation}
where $T$ is the temperature of the system. This shows that the entropy will increase slower as the temperature gets higher, but the entropy will never decrease as the temperature increases.

As we have seen above, one can calculate many quantities with the partition function, and the entropy of the system is one of those quantities. 
\[
	S\equiv -k_B\left(\ln Z + \beta \left\langle E\right\rangle \right)=\frac{\partial}{\partial T}\left(k_B T \ln Z\right)
\]

\section{Phase transitions}
\label{sec:Phase}

As we are studying the physics of phase transitions we should, at this point, define what we mean by a phase transition. To do this first let us define what we mean by a phase.

A phase is a homogeneous part of a system bounded by surfaces across which the properties change discontinuously. In the most general situation each phase will contain several components, that is, it will contain several different species of molecules or ions. For example a gaseous phase might consist of a mixture of gases. To specify the properties of a phase we must then specify the concentrations of the various components in it. Then we have the possibility of transfer of matter between different phases. This matter has then gone through a phase transition. 

A first order phase transition is normally accompanied by the absorption or liberation of latent heat, while a second order phase transition has no latent heat. An example of a first order phase transition is between water and steam (gaseous water), and an example of a second order phase transition that exhibited by is the two=dimensional Ising model on a square lattice (magnetisation). There is a critical temperature and energy at which the properties of the two phases of the transition are identical. A phase transition is easily seen on a graph of the spontaneous magnetism versus the temperature of the system as a discontinuity of the data, or a point where the temperature remains constant while the energy input increases. Further information on phase transitions can be found in \cite{Mandl1988}.

\section{Magnetic substances}
\label{sec:MagSub}

Substances are made up of smaller particles that cannot be divided without loss of the description of the substance, that is, if these particles were divided, the parts would not exhibit all the properties of the substance. Depending on the bonding of the substance, these particles may contain smaller particles called ions or it may be a molecule of the substance. Ions are of a similar configuration as atoms, with electrons orbiting around a nucleus, but have a different electron configuration. If the ion has more electrons than the base atom, then the ion will have a negative charge, and with less electrons will have a positive charge. Molecules are collections of atoms that share their electrons among each other by using overlapping orbits, which is called a covalent bond. As it is sufficient to describe the properties of the molecules or ions, because the substance is then in its most basic form, we shall concentrate on the description of these particles initially. Then we shall look at a crystal lattice structure of atoms or ions to investigate the possible properties that might be found as a result of the arrangement of the particles. It is sufficient to only describe a crystal lattice with ionic bonding as it can be generalised for any other structure with any type of bonding.

For a molecule or ion to be paramagnetic \cite{Earnshaw1968, Orchard2003}, that is, attracted by a magnetic field, it must contain one or more unpaired electrons. Substances that are repelled by magnetic fields are called diamagnetic and contain no unpaired electrons. The magnetic moment of a paramagnetic species increases with the increase in the number of unpaired electrons. Liquid oxygen is easily shown to be paramagnetic by pouring it between the poles of a strong magnet, when it is attracted by the field and fills the gap between the poles.

\begin{figure}[htbp]
	\centering
		\includegraphics[width= 85mm,keepaspectratio= true]{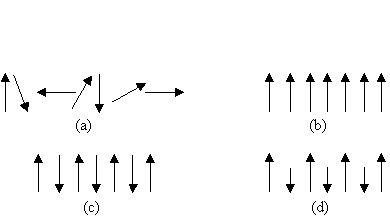}
		\caption[Diagram of the relationship between paramagnetism, ferromagnetism, antiferromagnetism and ferrimagnetism]{Pictorial representation of the relative magnetic spin directions of (a)  paramagnetism, (b) ferromagnetism, (c) antiferromagnetism (d) ferrimagnetism. Here the relative magnitude of the spins is shown by the size of the arrow. Taken from Sharpe. \cite{Sharpe1992}}
	\label{fig:paramagnetism}
\end{figure}

Substances in which the paramagnetic species are separated from one another by several diamagnetic species are said to be magnetically dilute \cite{Sharpe1992}. When the paramagnetic species are very close together (as in metals) or are separated only by an atom or monatomic ion that can transmit magnetic interactions they may interact with one another. This interaction may lead to ferromagnetism (in which large domains of magnetic dipoles are aligned in the same direction) or antiferromagnetism (in which neighbouring magnetic dipoles are aligned in opposite directions) \cite{Shriver1990}. \textit{Ferromagnetism} leads to greatly enhanced paramagnetism as in iron at temperatures up to 1041 K (the Curie temperature), above which thermal energy overcomes the alignment and normal paramagnetic behaviour occurs. \textit{Antiferromagnetism} occurs below a certain temperature called the N\'eel temperature; as less thermal energy is available with decrease in temperature the paramagnetic susceptibility falls rapidly. More complex behaviour may occur if some moments are systematically aligned so as to oppose others, but to a finite magnetic moment: this is \textit{Ferrimagnetism}. The relationship between paramagnetism, ferromagnetism, antiferromagnetism and ferrimagnetism is illustrated in Figure \ref{fig:paramagnetism}.

\section{Modelling magnetic substances}
\label{sec:MMS}
Observing the properties of magnetic substances described in Section \ref{sec:MagSub} is difficult as one has to examine the system on the quantum level with all the issues that brings. Instead, it is easier and cheaper to use a statistical model of the substance in a particular configuration\footnote{that is, a given temperature, lattice configuration and external magnetic field}, such as the Ising model. In this section, we shall concentrate on the energy that the particular configurations have, as this is crucial to the spontaneous magnetisation of the system.

The Ising model consists of a regular array of lattice sites in one, two or three dimensions, in which each site can be occupied in one of two ways. Each lattice site is occupied by an atom that can only exist in one of two spin states: $+$, with the spin in the direction of the magnetic field $B_0$, or $-$, against the field. The potential energy of a single dipole or spin is $-mB_0$ if it is oriented with the field ($+$), and $+mB_0$ if it is oriented against the field ($-$), where $m$ is the magnetic moment of an individual atom \nomenclature[ps]{$m$}{Magnetic moment of an individual atom.}. Let the state of the $j$\textsuperscript{th} lattice site be denoted by $\sigma_j$, which equals $+1$ for a $+$ state and $-1$ for a $-$ state. In terms of these $\sigma_j$, the potential energy due to the external field is $-mB_0\sum\sigma_j$. To simplify this formula, for the later sections we shall use the following definition; $h\equiv mB_0$. It is also assumed that there is an interaction $\epsilon_{ij}$ between nearest neighbour atoms, which are in sites $i$ and $j$, where $i$ and $j$ can be $+$ or $-$. In terms of $\sigma_j$, $\epsilon_{ij}$ can be written as $-J\sigma_i\sigma_j$ where $J$ is the interaction strength between spins. Note that: if the spins are parallel, $\sigma_i\sigma_j$ is positive; if there are anti parallel, $\sigma_i\sigma_j$ is negative. So the nature of the interaction is determined by the sign of $J$. If $J>0$, parallel alignments are more stable and the model will describe ferromagnetism. If $J<0$, an opposed alignment is the more stable and this will lead to antiferromagnetism. These interaction energies are similar to those in chemical bond theory \cite{Sharpe1992}.

The total energy of a given configuration of $N$ sites on a particular lattice type and variables, that is a given set of $\{\sigma_i$\} is then
\begin{equation}
E\left(\sigma_1,\sigma_2,\ldots,\sigma_N\right)=-J\sum^N_{i,j=1}\sigma_i\sigma_j-h\sum^N_{j=1}\sigma_j
\label{eq:ino1}
\end{equation}
where the first summation is over all nearest-neighbour pairs. This is the energy expression that characterises the Ising model of a magnetic system. The canonical partition function is the summation over all configurations, weighted by $\exp\left(-E/\left(k_B T\right)\right)$ or
\begin{equation}
Z_N=\sum_{\sigma_1=\pm1}\cdots\sum_{\sigma_N=\pm1}\exp\left\{-\frac{E\left(\sigma_1,\sigma_2,\ldots,\sigma_N\right)}{k_BT}\right\}
\label{eq:ino2}
\end{equation}
where $k_B$ is the Boltzmann constant. The number of terms in this equation is $2^N$, because each of the $N$ $\sigma_i$ sites can take on two values.

In a certain sense, the Ising model is a simpler system than a non-ideal gas or a liquid, because the interacting particles are allowed to be situated only at discrete lattice sites. On the other hand, the model is difficult enough that the exact solution in three dimensions has yet to be found, although the two-dimensional model has been solved exactly in the absence of a magnetic field.
\chapter{The Ising Model}
\label{ch:Ising}

In this chapter and the next we will look at the development of the solution of the Ising model for various types of lattices. We will first look at the one dimensional model, to show a simpler form of the model being exactly solved. Using this result we will then look at the two-dimensional model on a square lattice. Both the square lattice model and chain model are special in terms of Statistical Mechanics as they are examples of the few models that can be exactly solved. At the end of this chapter we will quickly cover the Ising model in higher dimensions and why these models may never be exactly solved. Also towards the end of the chapter we will quickly discuss other methods of solving and approximating solutions to the Ising model. 

\section{Calculation for one-dimension}
\label{sec:COD}
To make the development easier we shall start with the one-dimensional model, or chain, with no external magnetic field. This will allow us to see how the interactions in the chain affect the properties. Later we will explain the model in an external field so as to improve the model for its use in applications.

\begin{figure}[htbp]
	\centering
		\includegraphics[width=79.75mm]{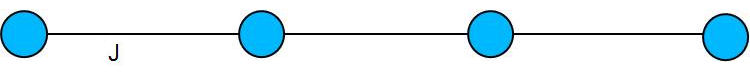}
	\caption[One-dimensional Ising chain]{Pictorial representation of the one-dimensional Ising chain. Here the circles represent the atoms in the chain, and the lines represent the interparticle interactions with strength $J$.}
	\label{fig:chain}
\end{figure}

In our model, each configuration occurs with the probability proportional to $e^{-\beta E}$, where $E$ is the energy of the system. First, we shall work out the energy, or rather what we shall call the energy which is equal to the enthalpy as there will be no work done by the system, or each configuration which is given by the following Hamiltonian formula;
\begin{equation}
	H=-J\sum_{i=1}^{N-1}{\sigma _i\sigma _{i+1}}
	\label{eq:ino3}
\end{equation}
where $J$ is a constant. The term being summed is the relative spin configuration of the neighbouring elements.

The canonical partition function $\left(Z_N\right)$ is given by
\begin{equation}
	Z_N=\sum_{\sigma _1=\pm 1}{\cdots \sum_{\sigma_N=\pm 1}{\exp \left\{ \beta J\sum_{i=1}^{N-1}\sigma _i\sigma_{i+1}\right\}}} \ \mathrm{where} \ \beta =\frac{1}{k_BT}.
	\label{eq:ino4}
\end{equation}
We can see from the equation that the spin of the last position appears only once in the equation and so we have the following formula irrespective of the value of $\sigma_{N-1}$:
\begin{equation}
	\sum_{\sigma_N=\pm 1}{\exp \left\{ \beta J\sigma_{N-1}\sigma_N\right\} }=2\cosh \beta J.
	\label{eq:ino5}
\end{equation}
Substituting this equation into (\ref{eq:part}) we get
\[
Z_N=\left[2\cosh\beta J\right]Z_{N-1}
\]
Now iterating this operation we can reduce the equation to
\begin{eqnarray}
	Z_N &=&\left( 2\cosh \beta J\right)^{N-2}Z_2 \label{eq:ino65} \\
	\mathrm{where} \ Z_2&=&\sum_{\sigma _1=\pm 1}\sum_{\sigma _2=\pm 1}\exp \left\{ \beta J\sigma _1\sigma _2\right\} =4\cosh \beta J.
	\label{eq:ino66}
\end{eqnarray}
So combining (\ref{eq:ino65}) and (\ref{eq:ino66}) we get the final formula for $Z_N$:
\begin{equation}
	Z_N = 2\left(2\cosh\beta J\right)^{N-1}
	\label{eq:ino6}
\end{equation}
The Gibbs free energy of the system is defined by
\[
G=-k_B\ln Z_N=-k_BT\left[ \ln 2+\left( N-1\right) \ln \left(2\cosh \beta J\right) \right] 
\]
In the thermodynamic limit (that is, the limit of the free energy as $N$ tends to infinity), only the term proportional to $N$ is important and so the formula becomes
\begin{equation}
	G=-Nk_BT\ln\left(2\cosh\beta J\right).
	\label{eq:ino7}
\end{equation}
These calculations give us the exact solution to the model while not in the presence of any magnetic field. We can adapt these formulae and find the solution for a model in an external magnetic field. However, one extra quantity we have to consider is the end effects of this chain when in this field. These end effects could make the equations more difficult than they would need to be, so we shall try to `fix' the model so that we can ignore the end effects. We can safely `fix' the end effects as they do not matter in the thermodynamic limit. This is because we assume the chain to be infinitely long in both directions without an end.

One way to `fix' the problem of the end effects is to consider the chain as a ring, that is, to set periodic boundary conditions. So we shall assume that the $N$\textsuperscript{th} spin is connected to the first. Then the Hamiltonian becomes
\begin{equation}
	H=-J\sum_{i=1}^{N}\sigma_i\sigma_{i+1}-h\sum_{i=1}^N\sigma_i
	\label{eq:ino8}
\end{equation}
where the spin labels run modulo $N$ (that is, $N+i=i$). This formula can then be simplified to
\[
H=-\sum_{i=1}^N\left[ J\sigma _i\sigma _{i+1}+\frac{h}{2}\left( \sigma _i+\sigma _{i+1}\right) \right].
\]
The partition function for this new Hamiltonian can be written as
\begin{eqnarray}
Z_N&=&\sum_{\sigma _1=\pm 1}\cdots \sum_{\sigma _N=\pm 1}\exp \left\{ \beta \sum_{i=1}^N\left[ J\sigma _i\sigma_{i+1}+\frac h2\left( \sigma _i+\sigma _{i+1}\right) \right] \right\} \nonumber \\
\mathrm{or } \ &=&\sum_{\sigma _i}\prod_{i=1}^N\exp\left\{ \beta \left[ J\sigma _i\sigma_{i+1}+\frac h2\left( \sigma _i+\sigma_{i+1}\right) \right] \right\}.
\label{eq:ino9}
\end{eqnarray}
We can see that this formula has the potential of being very large, so at this point we introduce a $2\times 2$ transfer matrix,
\begin{equation}
	\bf{P} = \left[ 
\begin{array}{cc}
	P_{11} & P_{1-1} \\
	P_{-11} & P_{-1-1}
\end{array}
\right]
\label{eq:ino10}
\end{equation}
where the elements are defined as follows
\begin{eqnarray}
	P_{11} &=& e^{\beta (J+h)} \nonumber \\
	P_{-1-1} &=& e^{\beta (J-h)} \nonumber \\
	P_{1-1} = P_{-11} &=& e^{-\beta J}
	\label{eq:ino11}
\end{eqnarray}
We shall now use this transfer matrix to describe our partition function in terms of a product of these transfer matrices:
\begin{equation}
	Z_N= \sum_{\left< \sigma_i \right>}{P_{\sigma_1\sigma_2}P_{\sigma_2\sigma_3}\cdots P_{\sigma_N\sigma_1}}=\Tr\mathbf P^N.
	\label{eq:ino12}
\end{equation}
We can see that the matrix P can be diagonalised and the eigenvalues $\lambda_1$ and $\lambda_2$ are the roots of the determinant
\begin{equation}
	\det\left(\mathbf{P}-\mathbf{\lambda I}\right)=0
	\label{eq:ino13}
\end{equation}
where $\mathbf{I}$ is the $2\times 2$ identity matrix. Similarly the matrix $\mathbf{P}^N$ has the eigenvalues $\lambda_1^N$ and $\lambda_2^N$, and the trace of $\mathbf{P}^N$ is the sum of the eigenvalues:
\begin{equation}
	Z_N=\lambda_1^N + \lambda_2^N.
	\label{eq:ino14}
\end{equation}
The solution of (\ref{eq:ino13}) is
\begin{equation}
	\lambda_{1,2}=e^{\beta J}\cosh\beta h\pm\sqrt{e^{2\beta J}\sinh^2\beta h+e^{-2\beta J}}
	\label{eq:ino15}
\end{equation}
We note that the eigenvalue associated with the positive root of equation (\ref{eq:ino13}), $\lambda_1$, is always larger in magnitude than the negative root eigenvalue. Now putting this partition function into the equation for the free energy we get
\begin{equation}
G=-k_BT\ln \left( \lambda _1^N+\lambda _2^N\right) =-k_BT\left\{N\ln \lambda _1+\ln \left[ 1+\left( \dfrac{\lambda _2}{\lambda _1}\right)^N\right] \right\}
\label{eq:ino67}
\end{equation}
This approaches $-Nk_BT\ln\lambda_1$ as $N$ goes to infinity. Now taking this to the thermodynamic limit,
\[
G=-Nk_BT\ln \left[ e^{\beta J}\cosh \beta h+\sqrt{e^{2\beta J}\sinh ^2\beta h+e^{-2\beta J}}\right]
\]
In the special case $h=0$ we obtain the result for the non-magnetic field case, so we have found the general formula. We may compute the magnetisation from
\[
m=\left<\sigma_0\right>=-\frac{1}{N}\frac{\partial G}{\partial h} = \frac{k_BT}{\lambda_1}\frac{\partial\lambda_1}{\partial h}.
\]
After some (straightforward) manipulations we find
\begin{equation}
	m=\frac{\sinh\beta h}{\sqrt{\sinh^2\beta h+e^{-4\beta J}}}
	\label{eq:ino16}
\end{equation}
We see that for $h=0$ there is no spontaneous magnetisation at any non-zero temperature. However, in the limit of low temperatures
\[
\sinh^2\beta h \gg e^{-4\beta J}
\]
for any $h\neq0$ and only a very small field is needed to produce near saturation of the magnetisation. The zero-field free energy will, in the limit as $T$ approaches zero, approach the value $G=-NJ$ corresponding to completely aligned spins. We could thus say that we have a phase transition at $T=0$, while for $T\neq0$ the free energy is an analytic function of its variables.

This is a very interesting result as phase transitions do not occur in the one-dimensional model at a finite temperature. If we look at the model with an initial state we should be able to intuitively illustrate why the phase transition is impossible. The model is set up so $\sigma_i=1$ if $i\leq l$ and $\sigma_i=-1$ if $i>l$, with the ground state $E_0=-(N-l)J$. There are $N-l$ such states, all with the same energy $E=E_0+2J$. At temperature $T$ the change in free energy will be
\[
\Delta G=2J-k_BT\ln(N-l).
\]		
This quantity is less than zero for all $T$ greater than zero in the limit where $N$ approaches infinity. The expectation value of the magnetism of the system is zero by using our formula given in (\ref{eq:ino16}). As the system is at equilibrium if the value of magnetism is zero, the change in free energy would be zero. So we have a contradiction and there can not be a phase transition between the equilibrium phase and the ferromagnetic phase.

\section{Square lattice two-dimensional calculation}
\label{sec:TDC}
As we have mentioned in Section \ref{sec:hist} of this chapter, the Norwegian American chemist Lars Onsager presented the first exact solution of the two-dimensional rectangular lattice Ising model without an external magnetic field \cite{Onsager1944}. While this solution is historically important, it is also quite mathematically formidable. Since Onsager's solution there have been a number of more transparent solutions presented \cite{Yang1952,Kaufman1949,Kac1952}, and in this section we will look at the method of Schultz, Mattis and Lieb \cite{Schultz1964,Plischke1994}. This method has been chosen as it follows the same form as the calculation for the one-dimensional model shown above, and it will relate well to our later development of the two-dimensional model on the Union Jack lattice.

\begin{figure}[htbp]
	\centering
		\includegraphics[width=79.75mm,height=79.69mm]{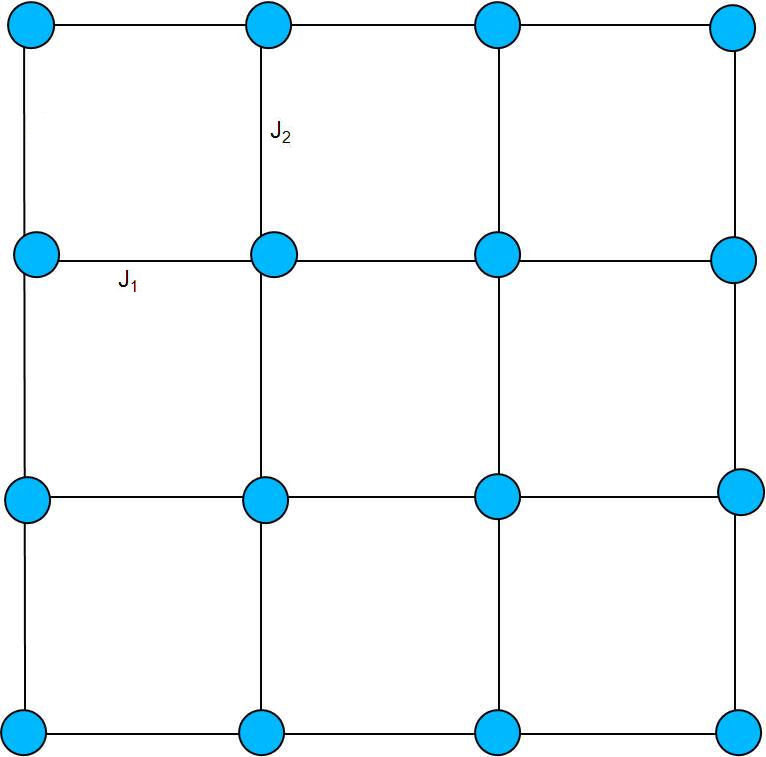}
	\caption[Two-dimensional Ising model on the square lattice]{Pictorial representation of the two-dimensional Ising model on the square lattice. Again the lines represent the interparticle interactions, with the horizontal interactions having strength $J_1$ and the vertical interactions have strength $J_2$. The sites are shown by circles.}
	\label{fig:squlat}
\end{figure}

\subsection{Transfer matrix}

In the calculation in the one-dimensional model we used a transfer matrix approach \cite{Newell1953}, and this method can be easily extended to be used for the two-dimensional model. First let us look at the one-dimensional model and this time formulate the solution in a slightly different way. First let us rewrite equation (\ref{eq:ino9}) in the following way
\begin{equation}
	Z= \sum_{\sigma_j=\pm 1}{\left(e^{\beta h\sigma_1}e^{K\sigma_1\sigma_2}\right)\left(e^{\beta h\sigma_2}e^{K\sigma_2\sigma_3}\right)\ldots\left(e^{\beta h \sigma_N} e^{K\sigma_N\sigma_1}\right)}
	\label{eq:ino27}
\end{equation}
where $K=\beta J$.

Taking the two orthonormal states $\left|+1\right\rangle$ and $\left|-1\right\rangle$ we can form a basis and rewrite the Pauli operators as,
\begin{eqnarray}
	\sigma_Z = \left(\begin{array}{cc} 1 & 0\\ 0 & -1 \end{array} \right) & \sigma^+= \left(\begin{array}{cc} 0 & 1\\ 0 & 0 \end{array}\right) & \sigma^-=\left(\begin{array}{cc} 0 & 0\\ 1 & 0 \end{array}\right)
	\label{eq:ino28}
\end{eqnarray}
with $\sigma_X = \sigma^+ + \sigma^-$ and $\sigma_Y = -i(\sigma^+ - \sigma^-)$. The Boltzmann weight $\exp\{\beta h \sigma_j\}$ can be expressed as a diagonal matrix, $\textbf{V}_1$, in this basis:
\[
	\left\langle +1\left|\textbf{V}_1\right|+1\right\rangle = e^{\beta h}, \ \left\langle -1\left|\textbf{V}_1\right|-1\right\rangle = e^{-\beta h}
\]
or
\begin{equation}
	\textbf{V}_1 = \exp{(\beta h \sigma_Z)}.
	\label{eq:ino29}
\end{equation}
We can also define the operator $\textbf{V}_2$ corresponding to the nearest-neighbour coupling by its matrix elements in this basis:
\begin{eqnarray}
	\left\langle +1\left|\textbf{V}_2\right|+1\right\rangle &=& \left\langle -1\left|\textbf{V}_2\right|-1\right\rangle = e^K \nonumber \\
	\left\langle +1\left|\textbf{V}_2\right|-1\right\rangle &=& \left\langle -1\left|\textbf{V}_2\right|+1\right\rangle = e^{-K}. \nonumber
\end{eqnarray}
Therefore,
\begin{equation}
	\textbf{V}_2 = e^K\textbf{1}+e^{-K}\sigma_X = A(K)\exp\{K^*\sigma_X\}
	\label{eq:ino30}
\end{equation}
where in the second step we have used the fact that $(\sigma_X)^{2n}=\textbf{1}$. The constants $A(K)$ and $K^*$ are determined from the equations
\begin{eqnarray}
	A\cosh K^* &=& e^K \nonumber \\
	A\sinh K^* &=& e^{-K}
	\label{eq:ino31}
\end{eqnarray}
or $\tanh K^* =\exp(-2K),\ A=\sqrt{2\sinh\{2K\}}$. From these results we can rewrite the partition function as:
\begin{eqnarray}
	Z &=& \sum_{\{\mu=+1,-1\}}{\left\langle \mu_1\left|\textbf{V}_1\right|\mu_2\right\rangle \left\langle \mu_2\left|\textbf{V}_2\right|\mu_3\right\rangle \left\langle\mu_3\left|\textbf{V}_1\right|\mu_4\right\rangle\ldots \left\langle \mu_{2N}\left|\textbf{V}_2\right|\mu_1\right\rangle } \nonumber \\
	&=& \Tr(\textbf{V}_1\textbf{V}_2)^N=\Tr(\textbf{V}_2^{1/2}\textbf{V}_1\textbf{V}_2^{1/2})^N=\lambda_1^N+\lambda_2^N
	\label{eq:ino32}
\end{eqnarray}
where $\lambda_1$ and $\lambda_2$ are the two eigenvalues of the Hermitian operator
\begin{equation}
	\textbf{V} = (\textbf{V}_2^{1/2}\textbf{V}_1\textbf{V}_2^{1/2})=\sqrt{2\sinh(2K)}e^{K^*\sigma_X/2}e^{\beta h\sigma_Z}e^{K^*\sigma_X/2}.
	\label{eq:ino33}
\end{equation}
We have managed to get to this symmetric form of the transfer matrix $\textbf{V}$ by using the property of the invariance of the trace of a product of matrices under a cyclic permutation of factors.

It is interesting to note that in this method we have taken a one-dimensional classical statistics problem and transformed it into a zero-dimensional quantum-mechanical ground state problem. This is not the only time this transformation can be achieved and the result is quite general \cite{Suzuki1976,Suzuki1985}. There is a correspondence between the ground state of quantum Hamiltonians in $d-1$ dimensions and classical partition functions in $d$ dimensions. This can sometimes be exploited, for example, in numerical simulations of quantum-statistical models. 

We can now generalise this method to the two-dimensional Ising model. This time consider an $M\times M$ square lattice with periodic boundary conditions and the Hamiltonian
\begin{equation}
	H=-J\sum_{r,c} \sigma_{r,c} \sigma_{r+1,c} - J\sum_{r,c} \sigma_{r,c} \sigma_{r,c+1}
	\label{eq:ino34}
\end{equation}
where the label $r$ refers to rows, $c$ to columns, and $\sigma_{r+M,c}=\sigma_{r,c+M}=\sigma_{r,c}$. Here the first sum contains only interactions in column $c$ and the second sum is the coupling between neighbouring columns and will lead to a non-diagonal factor in the complete transfer matrix.

Now as we did with the one-dimensional case, we introduce the $2^M$ basis states
\begin{equation}
	\left|\mu\right\rangle \equiv \left|\mu_1\right\rangle\left|\mu_2\right\rangle\ldots\left|\mu_M\right\rangle
	\label{eq:ino35}
\end{equation}
with $\mu_j = \pm 1$ and $M$ sets of Pauli operators $(\sigma_{jX},\sigma_{jY},\sigma_{jZ})$ which act on the $j$\textsuperscript{th} state in the product, that is,
\begin{eqnarray}
	\sigma_{jZ}\left|\mu_1,\mu_2,\ldots\mu_j,\ldots\mu_M\right\rangle &=& \mu_j\left|\mu_1,\mu_2,\ldots\mu_j\ldots\mu_M\right\rangle \nonumber \\
	\sigma_{j}^{+}\left|\mu_1,\mu_2,\ldots\mu_j,\ldots\mu_M\right\rangle &=& \delta_{\mu_j,-1}\left|\mu_1,\mu_2,\ldots\mu_j+2\ldots\mu_M\right\rangle \nonumber \\
	\sigma_{j}^{-}\left|\mu_1,\mu_2,\ldots\mu_j,\ldots\mu_M\right\rangle &=& \delta_{\mu_j,1}\left|\mu_1,\mu_2,\ldots\mu_j-2\ldots\mu_M\right\rangle
	\label{eq:ino36}
\end{eqnarray}
Moreover, we impose the commutation relations $[\sigma_{j\alpha},\sigma_{m\beta}]=0$, $\alpha,\beta= X,Y,Z$ for $j\neq m$. For $j=m$ the usual Pauli matrix commutation relations apply. \\ \\
\indent We can think of the index $\mu_j$ as the orientation of the $j$th spin in a given column. From this we can see that the Boltzmann factors $\exp\{K\sum_r\sigma_{r,c} \sigma_{r+1,c} \}$ are given by the matrix elements of the operator $\textbf{V}_1=\exp\{K\sum_j \sigma_{jZ} \sigma_{(j+1)Z}\}$. Similarly, the matrix element
\begin{eqnarray}
	\left\langle \left\{\mu\right\}\left|\textbf{V}_2\right|\left\{\mu^\prime\right\}\right\rangle &=& \left\langle \mu_M,\mu_{M-1},\ldots, \mu_1 \left| \prod_{j=1}^M{\left(e^K \textbf{1} + e^{-K} \sigma_{jX}\right) } \right| \mu^\prime_1, \mu^\prime_2, \ldots, \mu_M^\prime\right\rangle \nonumber \\
	&=& \exp\left\{\left(M-2n\right)K\right\}
	\label{eq:ino37}
\end{eqnarray}
where $n$ of the indices $\{\mu^\prime\}$ differ from the corresponding entries in $\{\mu\}$. Then in a zero magnetic field the partition function of the two-dimensional model is given by
\begin{eqnarray}
	Z &=& \sum_{\{\mu_1\},\{\mu_2\},\ldots, \{\mu_M\}}{\left\langle \mu_1 \left| \textbf{V}_1\right| \mu_2 \right\rangle \left\langle \mu_2 \left| \textbf{V}_2\right| \mu_3 \right\rangle \left\langle \mu_3 \left| \textbf{V}_1\right| \mu_4 \right\rangle \ldots \left\langle \mu_M \left| \textbf{V}_2\right| \mu_1 \right\rangle} \nonumber \\
	&=& \Tr\left(\textbf{V}_1\textbf{V}_2\right)^M= \Tr\left(\textbf{V}_2^{1/2}\textbf{V}_1\textbf{V}_2^{1/2}\right)^M.
	\label{eq:ino38}
\end{eqnarray}
Above the sum over each $\{\mu_J\}$ is over the entire set of $2^M$ basis states. We can then write, using \ref{eq:ino30} and \ref{eq:ino31}
\begin{equation}
	\textbf{V}_2 = \left(2\sinh2K\right)^{M/2}\exp\left\{K^*\sum^M_{j=1}{\sigma_{jX}}\right\}
	\label{eq:ino39}
\end{equation}
and as such the calculation of the partition function has been reduced to finding the largest eigenvalue of the Hermitian operator
\begin{eqnarray}
	\textbf{V} &=& \textbf{V}_2^{1/2}\textbf{V}_1\textbf{V}_2^{1/2} \nonumber \\
	&=& \left(2\sinh2K\right)^{M/2}\exp\left\{K^*\sum^M_{j=1}{\sigma_{jX}}\right\}\exp\left\{K^*\sum^M_{j=1}{\sigma_{jZ}\sigma_{j+1,Z}}\right\} \exp\left\{K^*\sum^M_{j=1}{\sigma_{jX}}\right\}. \nonumber \\
	\label{eq:ino40}
\end{eqnarray}
This is still a non-trivial task since the factors do not commute with each other, and because the matrix $\textbf{V}$ becomes infinite-dimensional in the thermodynamic limit.

\subsection{Transformation to an interacting fermion problem}

We now perform a rotation of the spin operators for future simplicity and let $\sigma_{jZ} \rightarrow -\sigma_{jX}$, $\sigma_{jX} \rightarrow \sigma_{jZ}$ for all $j=1,\ldots, M$. The eigenvalue are invariant under this rotation. Using $\sigma_{jZ} = 2\sigma^+_j\sigma^-_j-\textbf{1}$ and $\sigma_{jX} = \sigma_j^+ + \sigma_j^-$ we arrive at
\begin{eqnarray}
	\textbf{V}_1 &=& \exp\left\{K\sum^K_{j=1}\left(\sigma^+_j+\sigma^-_j\right)\left(\sigma^+_{j+1}+\sigma^-_{j+1}\right)\right\} \nonumber \\
	\textbf{V}_2 &=& \left(2\sinh2K\right)^{M/2}\exp\left\{2K^*\sum^M_{j=1}\left(\sigma^+_j\sigma^-_j-\frac{1}{2}\textbf{1}\right)\right\}.
	\label{eq:ino41}
\end{eqnarray}

Now following the method of Schultz \textit{et al} \cite{Schultz1964} we perform a Jordan-Wigner transformation, which converts the Pauli operators into fermion operators (see \cite{Plischke1994} for a discussion of second quantisation). This step is necessary as some of the subsequent canonical transformation are not possible for angular momentum operators. We write
\begin{eqnarray}
	\sigma^+_j &=& \exp\left\{\pi i \sum^{j-1}_{m=1} c^\dagger_m c_m\right\}c^\dagger_j \nonumber \\
	\sigma^-_j &=& c_j\exp\left\{-\pi i \sum^{j-1}_{m=1} c^\dagger_m c_m \right\} = \exp\left\{\pi i \sum^{j-1}_{m=1} c^\dagger_m c_m\right\}c
	\label{eq:ino42}
\end{eqnarray}
where the operators $c$, $c^\dagger$ obey the commutation relations
\begin{eqnarray}
	\left|c_j,c^\dagger_m\right|_+ &=& c_j c^\dagger_m + c^\dagger_m c_j = \delta_{jm} \nonumber \\
	\left|c_j,c_m\right|_+ &=& \left|c^\dagger_j, c^\dagger_m\right|_+ = 0. \nonumber
\end{eqnarray}
The operator $c^\dagger_m c_m$ is the fermion number operator for site $m$ with integer eigenvalues 0 and 1.

We can now express the operators $\textbf{V}_1$ and $\textbf{V}_2$ in terms of fermion operators using equation (\ref{eq:ino42}). The operator $\textbf{V}_2$ presents no difficulties and is given by
\begin{equation}
	\textbf{V}_2 = \left(2\sinh2K\right)^{M/2}\exp\left\{2K^*\sum^M_{j=1}\left(c^\dagger_j c_j -\frac{1}{2}\right)\right\}.
	\label{eq:ino43}
\end{equation}
In the case of $\textbf{V}_1$, there is a slight difficulty due to the periodic boundary condition. For $j \neq M$ we note that the term
\[
	\left(\sigma^+_j + \sigma^-_j\right)\left(\sigma^+_{j+1} + \sigma^-_{j+1}\right)= c^\dagger_j c^\dagger_{j+1} + c^\dagger_j c_{j+1} + c^\dagger_{j+1} c_j + c_{j+1} c_j.
\]
For the specific case $j=M$,
\begin{eqnarray}
	\left(\sigma^+_M + \sigma^-_M\right)\left(\sigma^+_1 +\sigma^-_1\right) &=& \exp\left\{\pi i \sum^{M-1}_{j=1} c^\dagger_j c_j\right\}c^\dagger_M\left(c^\dagger_1 + c_1\right) + \exp\left\{\pi i \sum^{M-1}_{j=1} c^\dagger_j c_j\right\}c_M\left(c^\dagger_1 + c_1\right) \nonumber \\
	&=& \exp\left\{\pi i \sum^{M}_{j=1} c^\dagger_j c_j\right\}\left[\exp{\pi i c^\dagger_M c_M}\left(c^\dagger_M + c_M\right)\left(c^\dagger_1 + c_1\right)\right] \nonumber \\
	&=& \left(-1\right)^n\left(c_M-c_M^\dagger\right)\left(c^\dagger_1+c_1\right) \nonumber
\end{eqnarray}
where $n=\displaystyle\sum_j c^\dagger_j c_j$ is the total fermion number operator. The operator $n$ commutes with $\textbf{V}_2$ but not with $\textbf{V}_1$. On the other hand, $(-1)^n$ commutes with both $\textbf{V}_1$, and $\textbf{V}_2$ as the various terms in $\textbf{V}_1$ change the total fermion number by 0 or $\pm2$. So we can write $\textbf{V}_1$ in a universal way by considering the subspaces of even and odd total number of fermions, that is,
\begin{equation}
	\textbf{V}_1 = \exp\left\{K\sum_{j=1}^M \left(c^\dagger_j - c_j\right)\left(c^\dagger_{j+1} + c_{j+1}\right)\right\}
	\label{eq:ino44}
\end{equation}
where
\begin{equation}
	\begin{array}{ccc}
		c_{M+1}\equiv -c_1, & c^\dagger_{M+1}\equiv-c^\dagger_1 & \textrm{for $n$ even} \\
		c_{M+1}\equiv c_1, & c^\dagger_{M+1} \equiv c^\dagger_1 & \textrm{for $n$ odd}.
	\end{array}
	\label{eq:ino45}
\end{equation}
This choice of boundary condition on the fermion creation and annihilation operators allows us to recover translational invariance, so we can carry out the canonical transformation
\begin{eqnarray}
	a_q &=& \frac{1}{\sqrt{M}}\sum_{j=1}^{M}{c_j e^{-iqj}} \nonumber \\
	a_q^\dagger &=& \frac{1}{\sqrt{M}}\sum_{j=1}^{M}{c_j^\dagger e^{iqj}}
	\label{eq:ino46}
\end{eqnarray}
with inverse
\begin{eqnarray}
	c_j &=& \frac{1}{\sqrt{M}}\sum_q{a_q e^{iqj}} \nonumber \\
	c_j^\dagger &=& \frac{1}{\sqrt{M}}\sum_q{a_q^\dagger e^{-iqj}}.
	\label{eq:ino47}
\end{eqnarray}
where we take $q=j\pi/M$ to reproduce the boundary conditions in (\ref{eq:ino45}). Substituting this into the equations for $\textbf{V}_2$ and $\textbf{V}_1$ we get for $n$ even,
\begin{eqnarray}
	\textbf{V}_2 &=& \left(2\sinh 2K\right)^{M/2} \exp\left\{2K^*\sum_{q>0}{\left(a_q^\dagger a_q + a^\dagger_{-q} a_{-q} -1\right)}\right\} \nonumber \\
	&=& \left(2\sinh 2K\right)^{M/2}\prod_{q>0}\textbf{V}_{2q}
	\label{eq:ino48}
\end{eqnarray}
and
\begin{eqnarray}
	\textbf{V}_1 &=& \exp\left\{2K\sum_{q>0}{\left[\cos q\left(a_q^\dagger a_q + a_{-q}^\dagger a_{-q} \right) - i\sin q\left(a_q^\dagger a_{-q}^\dagger + a_q a_{-q} \right) \right]} \right\} \nonumber \\
	&=& \prod_{q>0}\textbf{V}_1q.
	\label{eq:ino49}
\end{eqnarray}
In (\ref{eq:ino48}) and (\ref{eq:ino49}), the terms corresponding to $q$ and $-q$ have been combined. The resultant operators can be written as products and we can then see that bilinear operators with different wave vectors commute. This in turn allows great simplification and the eigenvalues of the transfer matrix can be written as a product of eigenvalues of at most $4\times4$ matrices. For odd $n$ we need the operators $\textbf{V}_{1q}$ and $\textbf{V}_{2q}$ for $q=\pi$ and $q=0$. These are given by
\begin{equation}
	\begin{array}{ll}
		\textbf{V}_{10} = \exp\left\{2Ka_0^\dagger a_0\right\} & \textbf{V}_{20} = \exp\left\{2K^*\left(a^\dagger_0 a_0 - \frac{1}{2}\right)\right\} \\
		\textbf{V}_{1\pi} = \exp\left\{-2Ka^\dagger_\pi a_\pi\right\} & \textbf{V}_{2\pi} = \exp\left\{2K^*\left(a^\dagger_\pi a_\pi - \frac{1}{2}\right)\right\}
	\end{array}
	\label{eq:ino50}
\end{equation}
which are already in diagonal form and commute with each other.

\subsection{Calculation of eigenvalues}
\label{sub:CoE2D}

We now go on to calculate the eigenvalues of the operator
\[
	\textbf{V}_q = \textbf{V}^{1/2}_{2q}\textbf{V}_{1q}\textbf{V}^{1/2}_{2q}
\]
for $q \neq 0$ and $q \neq \pi$. As we are dealing with fermions, we only have four possible states: $\left|0\right\rangle$, $a^\dagger_q\left|0\right\rangle$, $a^\dagger_{-q} \left|0\right\rangle$ and $a^\dagger_q a^\dagger_{-q}\left|0\right\rangle$, where $\left|0\right\rangle$ is the zero particle state defined by $a_q\left|0\right\rangle = a_{-q}\left|0\right\rangle = 0$. These states are eigenstates of $\textbf{V}_2$. Since the operator $\textbf{V}_1$ has non-zero off-diagonal matrix elements only when two states differ by two fermions, the problem reduces to finding the eigenvalues of $\textbf{V}_q$ in the basis $\left|0\right\rangle$ and $\left|2\right\rangle = a_q^\dagger a_{-q}^\dagger\left|0\right\rangle.$ We note that
\begin{equation}
	\textbf{V}_{1q}a_{\pm q}^\dagger\left|0\right\rangle = \exp\left\{2K\cos q\right\}a_{\pm q}^\dagger\left|0\right\rangle
	\label{eq:ino51}
\end{equation}
and
\begin{eqnarray}
	\textbf{V}_{2q}^{1/2}\left|0\right\rangle &=& \exp\left\{-K^*\right\}\left|0\right\rangle \nonumber \\
	\textbf{V}_{2q}^{1/2}\left|2\right\rangle &=& \exp\left\{K^*\right\}\left|2\right\rangle.
	\label{eq:ino52}
\end{eqnarray}
To obtain the matrix elements of $\textbf{V}_{1q}$ in the basis $\left|0\right\rangle$, $\left|2\right\rangle$, we let
\[
	\textbf{V}_{1q}\left|0\right\rangle = \alpha(K)\left|0\right\rangle + \beta(K)\left|2\right\rangle.
\]
When we differentiate this with respect to $K$, we get
\begin{eqnarray}
	\frac{d\alpha}{dK} &=& 2i\beta(K)\sin q \nonumber \\
	\frac{d\beta}{dK} &=& 4\beta(K)\cos q - 2i\alpha(K)\sin q.
	\label{eq:ino53}
\end{eqnarray}
Subject to the boundary conditions $\alpha(0) = 1$, $\beta(0) = 0$, we solve these equations to find
\begin{eqnarray}
	\left\langle 0\left|\textbf{V}_{1q}\right|0\right\rangle &=& \alpha(K) = e^{2K\cos q}\left(\cosh 2K - \sinh 2K \cos q\right) \nonumber \\
	\left\langle 2\left|\textbf{V}_{1q}\right|0\right\rangle &=& \beta(K) = -ie^{2K\cos q}\sinh 2K\sin q.
	\label{eq:ino54}
\end{eqnarray}
Using this method the matrix elements for $\left\langle 2\left|\textbf{V}_{1q}\right|2\right\rangle$ and $\left\langle 0\left|\textbf{V}_{1q}\right|2\right\rangle = \left\langle 2 \left|\textbf{V}_{1q}\right|0\right\rangle^*$ and obtain the matrix
\begin{equation}
	\textbf{V}_{1q} = e^{2K\cos q} \left[\begin{array}{cc}
		\cosh 2K - \sinh 2K\cos q & i\sinh 2K \sin q \\
		-i\sinh 2K \sin q & \cosh 2K + \sinh 2K \cos q
	\end{array} \right]
	\label{eq:ino55}
\end{equation}
and
\begin{equation}
	\textbf{V}_q = \left[\begin{array}{cc}
		\exp\left\{-K^*\right\} & 0 \\
		0 & \exp\left\{K^*\right\}
	\end{array} \right] \left[ \textbf{V}_{1q}\right]\left[\begin{array}{cc}
		\exp\left\{-K^*\right\} & 0 \\
		0 & \exp\left\{K^*\right\}
	\end{array} \right].
	\label{eq:ino56}
\end{equation}
The eigenvalues of this matrix are easily determined and can be given in the form
\begin{equation}
	\lambda^\pm_q = \exp\left\{2K \cos q \pm \epsilon(q)\right\}
	\label{eq:ino57}
\end{equation}
where after a bit of algebra $\epsilon(q)$ can be defined by
\[
	\cosh\epsilon(q) = \cosh 2K \cosh 2K^* + \cos q \sinh 2K \sinh 2K^*.
\]
By convention we choose $\epsilon(q)\geq 0.$ It can be seen that the minimum of the right hand side of this equation occurs as $q\rightarrow \pi$ and that for all $q$
\begin{equation}
	\epsilon(q) > \epsilon_{\textrm{min}}=\lim_{q\rightarrow\pi}\epsilon(q) = 2\left|K-K^*\right|
	\label{eq:ino58}
\end{equation}
and also
\begin{equation}
	\lim_{q\rightarrow 0}\epsilon(q) = 2\left(K+K^*\right).
	\label{eq:ino59}
\end{equation}

Now we combine this information and consider the subspace on an even number of fermions. The wave vectors in this case do not include $q=0$ or $q=\pi$, and it can be seen that the largest eigenvalue of $\textbf{V}_q$ for each $q$ is $\lambda_q^+$. Thus the largest eigenvalue in this subspace is given by
\begin{eqnarray}
	\Lambda_e &=& \left(2\sinh 2K\right)^{M/2}\prod_{q>0}\lambda_q^+ \nonumber \\
	&=& \left(2\sinh 2K\right)^{M/2}\exp\left\{\sum_{q>0}\left[2\cos q + \epsilon(q)\right]\right\} \nonumber \\
	&=& \left(2\sinh 2K\right)^{M/2}\exp\left\{\frac{1}{2}\sum_q\epsilon(q)\right\}
	\label{eq:ino60}
\end{eqnarray}
where we have used that $\sum_q \cos q =0$ and have extended the summation over the entire range, $-\pi<q<\pi$.

It is slightly more difficult to work with the odd subspace. As before for $q\neq 0$ and $q\neq \pi$ the maximum possible eigenvalue is $\lambda_q^+$. The corresponding eigenstates are all states with $(-1)^n = -1$. So to make the overall state have the property $(-1)^n = -1$, we occupy the $q=0$ state and leave the $q=\pi$ state empty, and as such obtain a contribution of $(2\sinh 2K)^{M/2}\exp\left\{2K\right\}$ to the eigenvalue $\Lambda_o$. So the largest eigenvalue in the odd subspace is
\begin{equation}
	\Lambda_o = \left(2\sinh 2K\right)^{M/2}\exp\left\{2K+\frac{1}{2}\sum_{q\neq 0 ,\pi}\epsilon(q)\right\}.
	\label{eq:ino61}
\end{equation}
With further consideration of these results, it can be shown that $\Lambda_e$ and $\Lambda_o$ are degenerate. The critical temperature of the two-dimensional Ising model is given by the equation $K=K^*$ or by expression
\begin{equation}
	\sinh\frac{2J}{k_B T_c} = 1
	\label{eq:ino62}
\end{equation}
which is approximately $k_B T_c/J = 2.269185\ldots$

The degeneracy of the two largest eigenvalues is negligible in the dimensionless free energy and only contributes an additive term of $\ln 2$. So for any temperature the free energy is given by
\begin{eqnarray}
	\frac{\beta G(0,T)}{M^2} &=& \beta g(0,T) = -\frac{1}{2}\ln\left(2\sinh 2K\right) - \frac{1}{2M}\sum_q\epsilon(q) \nonumber \\
	&=& -\frac{1}{2}\ln\left(2\sinh 2K\right)-\frac{1}{4\pi}\int_{-\pi}^{\pi} dq~\epsilon(q)
	\label{eq:ino63}
\end{eqnarray}
where the sum over the wave vectors has been converted to an integral.

\subsection{Spontaneous magnetisation}
\label{sub:SM2D}
While other thermodynamic functions can be derived, now we have the free energy of the system, which can be simplified with a bit more algebra. A full derivation of this result along with the derivation of the specific heat capacity of the model can be found in \cite{Plischke1994}. It is also possible to extend the theory discussed in this section to calculate the spontaneous magnetisation, as in \cite{Schultz1964}. The result is
\begin{eqnarray}
	m_0(T) &=&
		-\lim_{h\rightarrow 0}\frac{\partial}{\partial h} g(h,T) \nonumber \\
		&=& \left\{
	\begin{array}{cc}
		\left[1-\frac{\left(1-\tanh^2\beta J\right)^4}{16\tanh^4\beta J}\right]^\frac{1}{8} & T<T_c \\
		0 & T>T_c
	\end{array}
	\right.
	\label{eq:ino18}
\end{eqnarray}
As $T\rightarrow T_c$, the limiting form of the spontaneous magnetisation is given by
\begin{equation}
	m_0(T)\approx (T_c-T)^{1/8}\equiv(T_c-T)^\beta.
	\label{eq:ino64}
\end{equation}
At the critical point, the order parameter has a power law singularity. 

\section{Triangular lattice two-dimensional model}
\label{sec:ITri}

\begin{figure}[htbp]
	\centering
		\includegraphics[width=79.75mm,height=79.69mm]{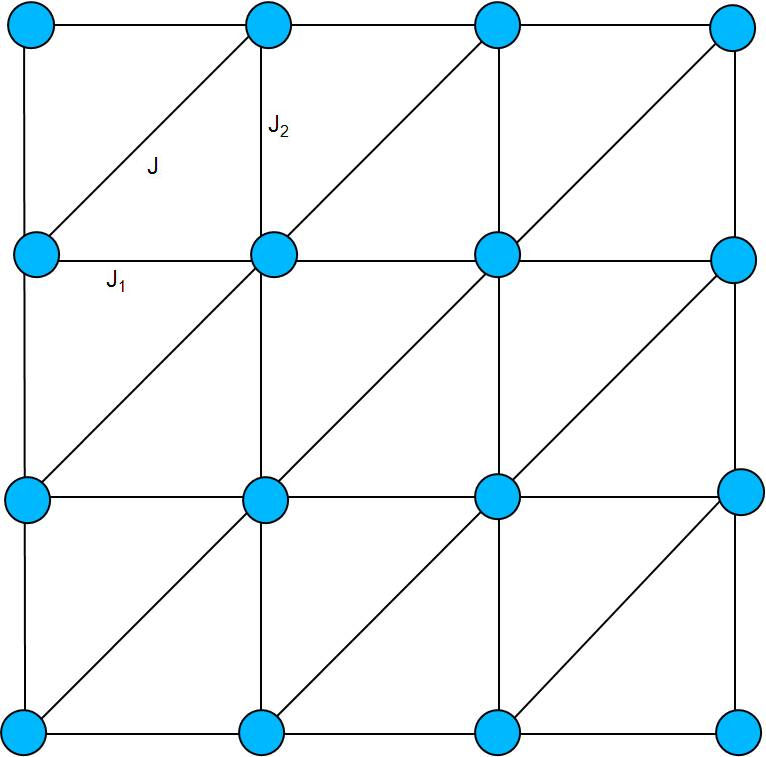}
	\caption[Two-dimensional Ising model on the triangular lattice]{Pictorial representation of the two-dimensional Ising model on the Triangular lattice. Interparticle interactions are shown here with lines, and have interactions strength $J_1$ horizontally, $J_2$ vertically and $J$ diagonally. Sites are shown with circles.}
	\label{fig:trilat}
\end{figure}

Now we add diagonals in one direction to the square lattice to obtain the triangular lattice seen in Figure \ref{fig:trilat}. This lattice configuration will be used as our ``test'' lattice in the later chapters. In this section we will state the results for the two site interaction spontaneous magnetisation and then go on to look at the three site interaction result.

\subsection{Two-site interactions on a triangular lattice}
\label{sub:2Sitetri}
Stephenson in his paper \cite{Stephenson1964} shows the development of the two site interaction result. As this development is similar to the development for the square lattice result, here we will state the final result for the spontaneous magnetisation in (\ref{eq:ino72}):
\begin{equation}
	M= \left\{
	\begin{array}{cc}
		\left(1-k_1^2\right)^\frac{1}{8} & T < T_c \\
		0 & T \geq T_c
	\end{array} \right. , 
	\label{eq:ino72}
\end{equation}
where
\begin{eqnarray}
	k_1^2 = \frac{\left[\left(1-v_1^2\right)\left(1-v_2^2\right)\left(1-v_3^2\right)\right]^2}{16\left(1+v_1 v_2 v_3\right)\left(v_1+v_2 v_3\right)\left(v_2+v_3 v_1\right)\left(v_3+v_1v_2\right)}, \nonumber \\
	\begin{array}{ccc}
		v_1= \tanh{\beta J_1}, & v_2= \tanh{\beta J_2}, & v_3= \tanh{\beta J}.
	\end{array}
	\label{eq:ino73}
\end{eqnarray}
Similar derivations can be seen in \cite{Potts1952,Green1962}. 

\subsection{Three-site triplet interactions on a triangular lattice}
\label{sub:3sitetri}
Having a result for the two-site nearest neighbour interactions, we now move on to look at the three site triplet interactions. When we consider these interactions the Hamiltonian (\ref{eq:ino34}) becomes
\begin{equation}
	H= -J^\prime \sum{\sigma_x\sigma_y\sigma_z} - J \sum{\sigma_x\sigma_y} - B \sum{\sigma_x},
	\label{eq:ino74}
\end{equation}
where the first summation is over all $2N$ three-site triplet interactions, the second is over all $3N$ two-site nearest neighbour interactions and the third is over all $N$ sites of the lattice. When any of the two of $B$, $J$, $J^\prime$ are zero, the free energy of the system can be evaluated exactly. In the case where $B=J=0$ we obtain the pure three-spin model, whose free energy was obtained by Baxter and Wu in \cite{Baxter1973}.

In his approach to calculate the spontaneous magnetisation of the three-site interactions, Baxter \cite{Baxter1975} considers the case when $H=J^\prime=0$. This allows a return to the normal two-site interaction triangular Ising model. For this he introduces a new parameter for the three-site spontaneous magnetisation, $M_3$, and defines $\mathcal{R}_\infty$ to be the ratio between the three-site correlator and the magnetisation in the infinite lattice case, i.e.
\[
	\mathcal{R}_\infty=\frac{M_3}{M}.
\]
Here the three-site correlator is the interaction between a triplet of sites $\sigma_x,~\sigma_y~\sigma_z$ on a triangular face, $\left\langle \sigma_x~\sigma_y~\sigma_z\right\rangle$. From (62) of \cite{Baxter1975} we have that
\begin{eqnarray}
	\mathcal{R}_\infty&=& \frac{1}{2}\left(v_1+v_1^{-1}+v_2+v_2^{-1}+v_3+v_3^{-1}\right) \nonumber \\
	&& -\frac{1}{2}\left(v_1 v_2 v_3\right)^{-1}\left[\left(1+v_1 v_2 v_3\right)\left(v_1+v_2 v_3\right)\left(v_2+v_3 v_1\right)\left(v_3+v_1v_2\right)\right]^{\frac{1}{2}}.
	\label{eq:ino75}
\end{eqnarray}
This then means we can evaluate the spontaneous magnetisation of the three-site interactions as
\begin{equation}
	M_3= \left\{
	\begin{array}{cc}
		M\mathcal{R} & T<T_c, \\
		0 & T \geq T_c.
	\end{array} \right.
	\label{eq:ino76}
\end{equation}

\subsection{Three-site triplet interactions on a square lattice}
\label{sub:3sitesq}
Now consider the case when $J=v_3=0$. By removing the diagonal interactions we have the square lattice. $M_3$ now becomes the three-spin magnetisation around a corner of the square lattice. Using the equations above we can obtain
\begin{equation}
	M_3=M\left[1-\frac{4e^{-4\beta J_1-4\beta J_2}}{\left(1-e^{-4\beta J_1}\right)\left(1-e^{-4\beta J_2}\right)}\right],
	\label{eq:ino77}
\end{equation}
	where $M$ is the spontaneous magnetisation of the square lattice. Other three-spin magnetisations have been evaluated by Pink \cite{Pink1968}.
\chapter{The Union Jack Lattice}
\label{ch:UJL}
\begin{figure}[htbp]
	\centering
		\includegraphics[width=79.75mm,height=79.69mm]{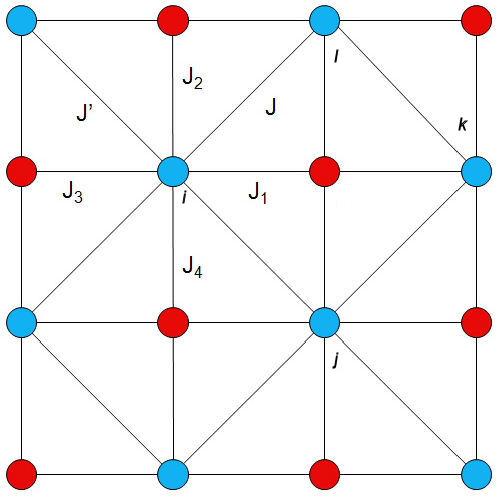}
	\caption[Two-dimensional model on the Union Jack lattice.]{Pictorial representation of the two dimensional Ising model on the Union Jack lattice. Here, as before, the interparticle interactions are shown with lines. The \textcolor{red}{red} circles represent $\tau$-sites which interact with four particles. The \textcolor{cyan}{blue} circles represent $\sigma$-sites with eight interparicle interactions. The various interaction strengths are shown on the different interactions.} 
	\label{fig:union}
\end{figure}

Having discussed the two-dimensional model both on the square and triangular lattices in the last chapter, in this chapter we will focus on the Union Jack lattice or centred squared lattice. We obtain the Union Jack lattice by adding alternate diagonals to the squares of the square lattice, as shown in Figure \ref{fig:union}.
Here the the nearest neighbour interaction strengths can be one of six values $J_1$, $J_2$, $J_3$, $J_4$, $J$, $J'$. The Hamiltonian for this lattice is
\begin{eqnarray}
		H &=& \sum \left[\sigma_{i,j}\left(-J_1\sigma_{i+1,j} -J_2\sigma_{i,j+1} -J_3\sigma_{i-1,j} -J_4\sigma_{i,j-1} \right) \right] \nonumber \\
		&+&\sum\left[\sigma_{2i-1,2j-1} \left(-J\left(\sigma_{i+1,j+1}+\sigma_{i-1,j-1}\right)-J^\prime\left(\sigma_{i-1,j+1}+\sigma_{i+1,j-1}\right)\right)\right] \nonumber\\
		&-& B \sum \sigma_{i,j}.
		\label{eq:uno8}
\end{eqnarray}
It is clear to see that this lattice ($\mathcal{L}_l$) can be made of two sublattices, indicated by red and blue circles. The sublattice $\mathcal{L^{\prime\prime}}$ contains the $\tau$ sites, red circles, with four nearest neighbour interactions and the sublattice $\mathcal{L^\prime}$ contains the $\sigma$ sites, blue circles, with eight nearest neighbour interactions. The partition function on $\mathcal{L}_l$ is
\begin{equation}
	Z_l=\sum_{\sigma _1=\pm 1}{\cdots \sum_{\sigma_N=\pm 1}{\exp \left( \beta \sum_{i,j=1}^{N-1}{J_r\sigma _i\sigma_j}\right) }}.
	\label{eq:ino20}
\end{equation}

Vaks, Larkin and Ovchinnikov \cite{Vaks1966} first considered the Union Jack lattice Ising model as a system exhibiting a re-entrant transition in the presence of competing interactions. They considered a system where the interactions were symmetric, and obtained its free energy and a sublattice two-spin correlation function. This was later generalised by Sacco and Wu \cite{Sacco1975} as a 32-vertex model on a triangular lattice.

Wu and Lin presented a simple formulation for the general Union Jack lattice model \cite{Wu1987, Wu1989} and showed that it is equivalent to a free fermion model \cite{Fan1970}. Using this equivalence they obtained the free energy and the eight site interaction sublattice spontaneous magnetisation. This was extended for symmetric interactions by Lin and Wang \cite{Lin1987} to produce a result for four interaction sublattice. Wu and Lin in their 1989 paper \cite{Wu1989} then extend the result further to produce results for both sublattices in the general interaction model.

\section{The Valks \textit{et al} result}
\label{sec:UJSR}
\begin{figure}[htbp]
	\centering
		\includegraphics[width=0.8\textwidth]{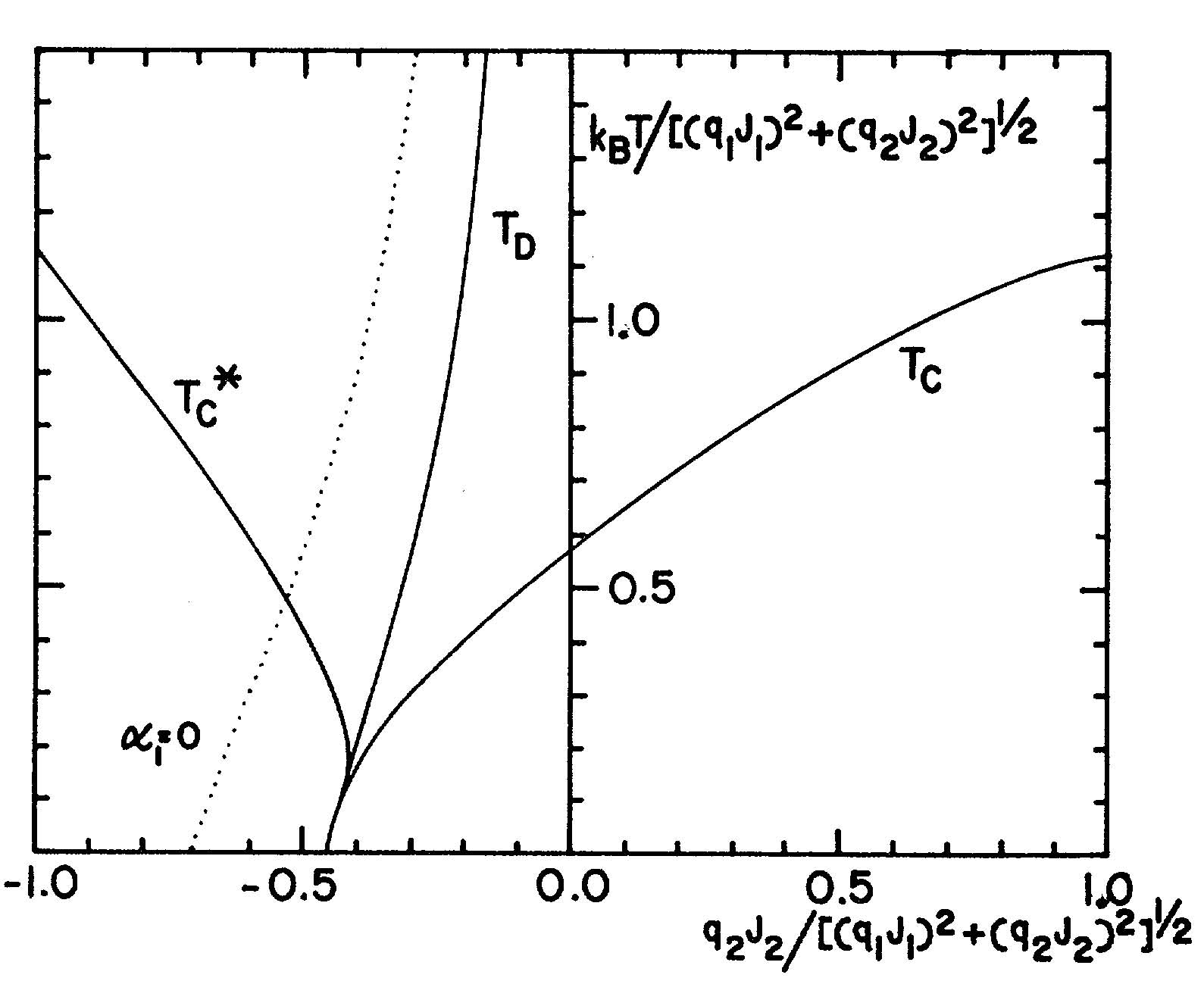}
		\caption[Phase diagram for the Union Jack lattice Ising model, taken from \cite{Stephenson1970}]{Phase diagram of the Vaks \etal result on the Union Jack lattice Ising model showing the graphs of the critical temperature $T_c$, disorder point $T_D$, and second critical temperature $T_c^*$. Below $T_c$ the system is in a ferromagnetic phase. Below $T_c^*$ the system is in an ordered antiferromagnetic phase. Above these critical temperatures the system is in a disorder phase. Taken from \cite{Stephenson1970}.}
	\label{fig:stevephase}
\end{figure}

Although we will be presenting the general Union Jack lattice result from Wu and Lin \cite{Wu1987, Wu1989} in this chapter, it is useful to look at the results from the paper by Valks \textit{et al} \cite{Vaks1966}. In their paper, they use a symmetric interaction model where the $J_r$, $r=1,2,3,4$ are equal and $J=J^\prime$. They showed that the phase of the system on the $\sigma$-sublattice can be determined by two parameters $\alpha_1$ and $\alpha_2$, given by (\ref{eq:uno2}).
\begin{eqnarray}
	\alpha_1=\frac{e^{2K}\left(\cosh{4K_1}-e^{-2K}\right)}{\left(1+e^{-2K}\right)}, \nonumber \\
	\alpha_2=\frac{e^{-2K}\left(1-e^{-2K}\right)}{\left(\cosh{4K_1}+e^{-2K}\right)}
	\label{eq:uno2}
\end{eqnarray}
where $K=J/k_BT$ and $K_1=J_1/k_BT$. The critical temperature $T_c$ is determined when $\alpha_1=1$, which has solutions when $-\left|J_1\right|<J$. A second critical temperature $T_c^*$ is determined by $\alpha_2=-1$, which has solutions when $J_2<-0.907\left|J_1\right|$. This is shown in Figure \ref{fig:stevephase}.
With further analysis of the formulas given by Valks \textit{et al} one can classify the phases of the system. When $-\left|J_1\right|<J_2<0$, there is an ordered ferromagnetic phase below $T_c$. Above $T_c$, there is disordered ferromagnetic phase up to a disorder temperature $T_D$, determined by $\alpha_1=-\alpha_2$. At temperatures above $T_D$, there is a disordered antiferromagnetic phase. This region extends to $T=\infty$, if $J$ satisfies $-0.907\left|J_1\right|<J<0$. However, if $-\left|J_1\right|<J<-0.907\left|J_1\right|$, there is an intervening ordered antiferromagnetic phase between the lower and upper critical temperatures $T_c^*$.

\section{Computation on the Union Jack lattice}
\label{sec:CUJL}
Now we move on to look at the exact result for the general anisotropic Ising model on the Union Jack lattice from Wu and Lin \cite{Wu1987,Wu1989}. As the development is similar to that of the two-dimensional square lattice model shown in section \ref{sec:TDC}, we will present the result only. 

By summing over all the $N/2$ $\tau$ sites on $\mathcal{L}^{\prime\prime}$ we can rewrite the partition function $Z_l$ as limit variables
\begin{equation}
	Z_l=\sum_{\tau_1 =\pm 1}{\cdots\sum_{\tau_{\frac{N}{2}} =\pm 1}{\left[\prod_{i,j,k,l}{\omega \left( \sigma _i,\sigma _j,\sigma _k,\sigma _l\right)}\right]}}
	\label{eq:ino21}
\end{equation}
where the product is over all $N/2$ faces of the sublattice $\mathcal{L'}$, each surrounded by four spins $i$, $j$, $k$, $l$. Here the Boltzmann factor $\omega(\sigma_1, \sigma_2, \sigma_3, \sigma_4)$ is obtained by dividing each of the diagonal interactions $-J$ and $-J'$ into halves, one for each of the two adjacent faces. This leads to
\begin{equation}
	\omega \left( a,b,c,d\right) =2\exp\left[ \frac{\beta J\left( ab+cd\right) }{2}+\frac{\beta J^\prime \left(ad+bc\right) }{2}\right] \cosh \left( a\beta J_1+b\beta J_2+c\beta J_3+d\beta J_4\right).
	\label{eq:ino22}
\end{equation} 
This equation with its four spin variables gives us sixteen separate states which by symmetry can be reduced to eight distinct expressions:
\begin{eqnarray}
	\omega _1&=&\omega \left( ++++\right)=2e^{\beta J+\beta J^\prime }\cosh \left( \beta \left( J_1+J_2+J_3+J_4\right)\right) \nonumber \\
	\omega _2&=&\omega \left( +-+-\right) =2e^{-\beta J-\beta J^\prime }\cosh \left( \beta \left( J_1-J_2+J_3-J_4\right) \right) \nonumber \\
	\omega _3&=&\omega \left( +--+\right) =2e^{-\beta J+\beta J^\prime }\cosh \left( \beta \left( J_1-J_2-J_3+J_4\right) \right) \nonumber \\
	\omega _4&=&\omega \left( ++--\right) =2e^{\beta J-\beta J^\prime }\cosh \left( \beta \left( J_1+J_2-J_3-J_4\right) \right) \nonumber \\
	\omega _5&=&\omega \left( +-++\right) =2\cosh \left( \beta \left( J_1-J_2+J_3+J_4\right) \right) \nonumber \\
	\omega _6&=&\omega \left( +++-\right) =2\cosh \left( \beta \left( J_1+J_2+J_3-J_4\right) \right) \nonumber \\
	\omega _7&=&\omega \left( ++-+\right) =2\cosh \left( \beta \left( J_1+J_2-J_3+J_4\right) \right) \nonumber \\
	\omega _8&=&\omega \left( -+++\right) =2\cosh \left( \beta \left( -J_1+J_2+J_3+J_4\right) \right).
	\label{eq:ino23}
\end{eqnarray}
There can only be eight possible arrangements of the four spin variables and so we can consider it to be an eight-vertex model with weights (\ref{eq:ino22}). This eight-vertex model was proved to satisfy the free fermion condition by Fan and Wu \cite{Fan1970}.
\begin{equation}
	\omega_1\omega_2 + \omega_3\omega_4 = \omega_5\omega_6 + \omega_7\omega_8
	\label{eq:ino24}
\end{equation}
and as such is a free fermion model with the partition function:
\[
Z=\frac{1}{2}Z_l.
\]
Here the factor $1/2$ takes account of the two-to-one mapping of the spin and vertex configurations. The spontaneous magnetisation of a free fermion model was given by Baxter in \cite{Baxter1986} and is
\begin{eqnarray}
	\left\langle \sigma \right\rangle &=&\left\{
	\begin{array}{cc} 
		\left(1-\Omega^{-2}\right)^{1/8}, & \Omega^{-2}\geq 1 \\
		0, & \Omega^{-2} \leq 1,
	\end{array}
	\right.
	\label{eq:ino25} \\ \nonumber \\
	\Omega^2 &=&1-\frac{\gamma_1 ~ \gamma_2 ~  \gamma_3 ~ \gamma_4}{16\omega_5 ~\omega_6 ~\omega_7 ~\omega_8}.
	\label{eq:uno3}
\end{eqnarray}
where
\begin{eqnarray}
	\gamma_1 &=& -\omega_1+\omega_2+\omega_3+\omega_4 \nonumber \\
	\gamma_2 &=& \omega_1-\omega_2+\omega_3+\omega_4 \nonumber \\
	\gamma_3 &=& \omega_1+\omega_2-\omega_3+\omega_4 \nonumber \\
	\gamma_4 &=& \omega_1+\omega_2+\omega_3-\omega_4
	\label{eq:uno1}
\end{eqnarray}
The system exhibits an Ising transition at the critical point(s)
\begin{equation}
	\Omega^2=1
	\label{eq:uno4}
\end{equation}
or equivalently,
\begin{equation}
	\omega_1+\omega_2+\omega_3+\omega_4=2\ \mathrm{max}\left\{\omega_1,\omega_2,\omega_3,\omega_4\right\}.
	\label{eq:uno5}
\end{equation}

We can now go on to finding the spontaneous magnetisation for the other sublattice, $\mathcal{L^\prime}$. While Lin and Wang presented a result for the symmetric case in 1988 \cite{Lin1988a}, here we will quickly review the result for the general case given by Wu and Lin in \cite{Wu1989}. Their approach uses the three site triplet interactions and again, as in section \ref{sub:3sitetri}, we will simply present the result here. The $\tau$-sublattice magnetisation is given by
\begin{equation}
	\left\langle \tau \right\rangle = \left\langle \sigma \right\rangle \left[A_{1234}(K)(F_+ + F_-)+A_{2341}(K)(F_+ - F_-)\right].
	\label{eq:ino68}
\end{equation}
We can see that this equation has a similar form to (\ref{eq:ino76}), and can be seen to be a multiple of the $\sigma$-sublattice value. In (\ref{eq:ino68}),
\begin{eqnarray}
	A_{1234}(K)&=& \frac{\sinh{2(\beta J_1+\beta J_3)}}{\sqrt{2G_{-}(\beta J)\sinh{2\beta J_1}\sinh{2\beta J_3}}} \nonumber \\
	A_{2341}(K)&=& \frac{\sinh{2(\beta J_2+\beta J_4)}}{\sqrt{2G_{-}(\beta J)\sinh{2\beta J_2}\sinh{2\beta J_4}}} \nonumber
\end{eqnarray}
and
\begin{equation}
	G_{-}(\beta J)= \cosh{2(\beta J_1+\beta J_3)}+\cosh{2(\beta J_2-\beta J_4)}. \nonumber \\
\end{equation}
The calculation for $F_+$ and $F_-$ is a little more involved. We start by calculating
\begin{equation} 
	F_{\pm}= \sqrt{\frac{A+2\sqrt{BC}}{D+2E\sqrt{B}}}
	\label{eq:ino69}
\end{equation}
where
\begin{eqnarray}
	A&=& 2\omega_5\omega_6\omega_7\omega_8\left(\omega^2_1+\omega^2_2+\omega^2_3\omega^2_4\right)-\left(\omega_1\omega_2+\omega_3\omega_4\right)\left(\omega_1\omega_3 + \omega_2\omega_4\right)\left(\omega_1\omega_4+\omega_2\omega_3\right) \nonumber \\
	B&=& \omega_5\omega_6\omega_7\omega_8\left(\omega_5\omega_6\omega_7\omega_8-\omega_1\omega_2\omega_3\omega_4\right) \nonumber \\
	C&=& \left(\omega_1^2+\omega_2^2+\omega_3^2+\omega_4^2\right)^2-4\left(\omega_5\omega_6-\omega_7\omega_8\right)^2 \nonumber \\
	D&=& \left(\omega_1^2+\omega_2^2\right)\left(2\omega_5\omega_6\omega_7\omega_8 - \omega_1\omega_2\omega_3\omega_4\right) - \omega_5\omega_6\omega_7\omega_8\left(\omega_3^2+\omega_4^2\right) \nonumber \\
	E&=& \omega_1^2-\omega_2^2. \nonumber
\end{eqnarray}
We can relate $F_+$ and $F_-$ with the following formula, allowing us to get values for each variable,
\begin{equation}
	F_+F_- = \frac{\omega_5\omega_6-\omega_7\omega_8}{\omega_1\omega_2}.
	\label{eq:ino71}
\end{equation}

We can compute the overall nearest neighbour magnetisation by taking the mean of the two sublattice magnetisations
\[
	M_0= \frac{1}{2}(\left\langle \sigma \right\rangle + \left\langle \tau \right\rangle).
\]

\subsection{Classification of phases}
\label{sub:CoPh}
At low temperatures the phase of the $\sigma$-sublattice can be classified based on the following energy value,
\begin{eqnarray}
	-E_1 &=& J + J^\prime + \left|J_1+J_2+J_3+J_4\right| \nonumber \\
	-E_2 &=& -J - J^\prime + \left|J_1-J_2+J_3-J_4\right| \nonumber \\
	-E_3 &=& -J + J^\prime + \left|J_1-J_2-J_3+J_4\right| \nonumber \\
	-E_4 &=& J - J^\prime + \left|J_1+J_2-J_3-J_4\right|
	\label{eq:uno6}
\end{eqnarray}
The sublattice is in a ferromagnetic phase when
\[
	E_1 < E_2,~E_3,~E_4;
\]
is antiferromagnetic when
\[
	E_2 < E_1,~E_3,~E_4;
\]
and finally is metamagnetic when
\[
	E_3 < E_1,~E_2,~E_4 \ \mathrm{ or } \ E_4 < E_1,~E_2,~E_3.
\]
As the temperature rises, depending on the relative strengths of the interactions $J_r$, the occurrence of phase change is signified by one or more of the following equations being realised
\begin{equation}
	\omega_1+\omega_2+\omega_3+\omega_4 = 2\max\left\{\omega_1,~\omega_2,~\omega_3,~\omega_4\right\}.
	\label{eq:uno7}
\end{equation}
A re-entrant transition occurs if any one equation admits two solutions.

\section{Theoretical analysis}
\label{sub:TAnal}
In this section we will plot the theoretical predictions of Wu and Lin, and compare those against the critical temperatures from the results of Stephenson. To do this we will plot results from each of the possible systems, along with a plot of the $\gamma$ from (\ref{eq:uno3}). Wu and Lin's predictions will be plotted against our simulation results later in Chapter \ref{ch:NE}. Here we are only concerned with identifying systems that need further investigation, due to conflict between the predictions of Vaks \textit{et al.} and, Wu and Lin.

\begin{figure}[htbp]
	\centering
		\includegraphics[width= 0.5\textwidth]{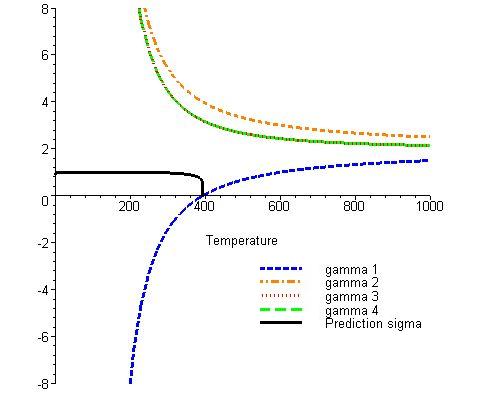}
	\caption[Separation of $\gamma$ terms for a ferromagnetic system]{Graph of the separate  $\gamma$ terms from (\ref{eq:uno1}) and the $\sigma$-sublattice prediction for the ferromagnetic system with interactions strengths $J_n=J=J^\prime=100k_B$. Here $\gamma_3$ is overlaid on $\gamma_4$ as they are equal. Note that at low temperatures $\gamma_1$ is negative and the other terms are positive. At the critical temperature $\gamma_1=0$.}
	\label{fig:ferrosep}
\end{figure}

For our first system we will look at the isotropic ferromagnetic case. This is the simplest system as all the interactions are the same, $J_n=100k_B$. The graph of this system is shown in  Figure \ref{fig:wulinferro}:
\begin{figure}[htbp]
	\centering
		\includegraphics[width= 0.5\textwidth]{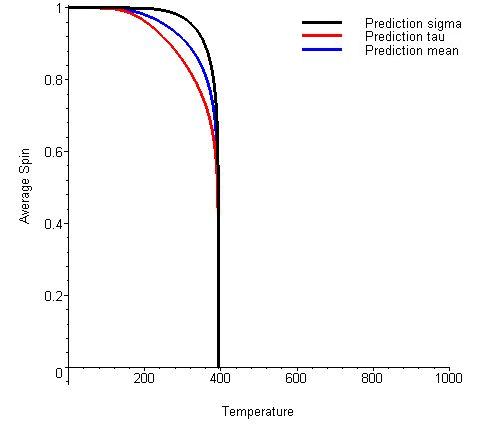}
	\caption[Wu and Lin prediction for a ferromagnetic system]{Theoretical predictions of Wu and Lin for a ferromagnetic system with interaction strengths $J_n=J=J^\prime=100k_B$. Here the prediction for the $\sigma$-sublattice (eqn. (\ref{eq:ino25})) is shown in black, the prediction for the $\tau$-sublattice (eqn. (\ref{eq:ino68})) is shown in red and the prediction for the overall lattice is shown in blue. A phase transition can be seen just below 400 Kelvin.}
	\label{fig:wulinferro}
\end{figure}
We can see that the prediction of this system moves from having a non-zero average magnetic spin to a zero average magnetic spin. The system has this phase transition at the critical temperature, $T_c$ around 400 Kelvin. This prediction agrees with the value obtained from $\alpha_1$ in \ref{eq:uno2}, and a solution to $\alpha_2$ does not exist. Intuitively this is the sort of graph we would expect. Although for this system both theories agree, we will go on to examine the changes in the gamma terms. This will set the groundwork for our later analysis. This plot is shown in Figure \ref{fig:ferrosep}.
Here we can see that at low temperatures $\gamma_1$ is the only negative term, and that it equals zero at $T_c$ and then moves to being positive and converges to the level of the other $\gamma$. 

\begin{figure}[htbp]
	\centering
		\subfloat[]{
		\includegraphics[width= 0.48\textwidth]{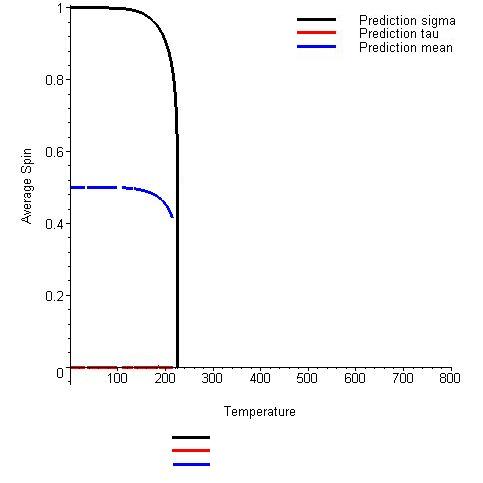}
		\label{subfig:metares}
		}
		\subfloat[]{
		\includegraphics[width= 0.48\textwidth]{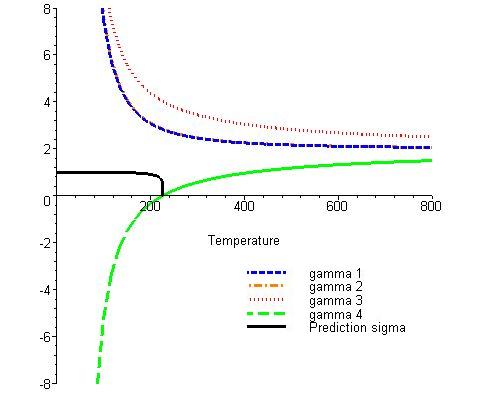}
		\label{subfig:metasep}
		}
	\caption[Wu and Lin prediction for a metamagnetic system]{Results for a metamagnetic system with interaction strengths $J_n=10k_B$, $J=-J^\prime=100k_B$. (a) shows the theoretical predictions of Wu and Lin. We see that the result for the $\tau$-sublattice is zero across the temperature range while the $\sigma$-sublattice shows a transition at just above 200 Kelvin. (b) shows the values of the separate $\gamma$ terms. Here we see that $\gamma_4$ is negative at low temperatures and crosses the x-axis at the transition temperature. In this graph $\gamma_1$ is overlaid on $\gamma_2$ as they are equal.}
	\label{fig:wulinmeta}
\end{figure}

Now we will move on to looking at a system that starts in a metamagnetic phase. For this system we need to set the diagonals to be opposite polarities, that is $J=-J^\prime=100k_B$. Due to this the Vaks \textit{et al.} result does not apply to this system. We also set the horizontal and vertical interactions to be $J_n=10k_B$. The plots for this system are shown in Figure \ref{fig:wulinmeta}
As a metamagnetic system only has an average magnetic spin when in the presence of an external magnetic field, we can see that something is incorrect here. From Figure \ref{subfig:metasep} we can see that $\gamma_4$ is negative at low temperatures and behaves similarly to $\gamma_1$ in the ferromagnetic case. As
\[
	\gamma_4= \omega_1 + \omega_2 + \omega_3 - \omega_4 < 0,
\]
we note that $\omega_4=\omega(++--)$ is the dominant term. As this term would suggest an antiferromagnetic phase we would intuitively expect $\left\langle \sigma\right\rangle=0$.

The system we next consider is that on an anisotropic antiferomagnetic system. For this system we set the vertical and horizontal interactions to be $J_n=100k_B$ and the diagonal interactions to be $J=J^\prime=-100k_B$. As the system is antiferromagnetic we expect it to have a zero average spin across the temperature range. The plots are shown in Figure \ref{fig:wuantipredict} below.
\begin{figure}[htbp]
	\centering
		\subfloat[]{
		\includegraphics[width= 0.48\textwidth]{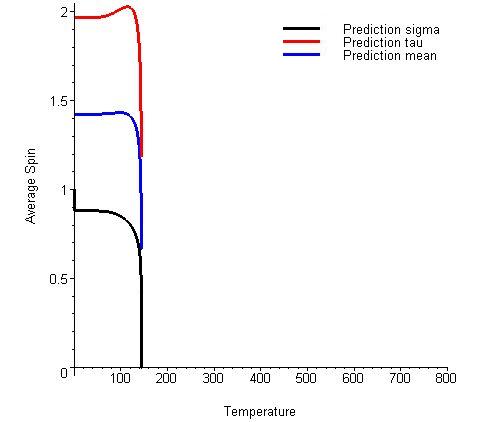}
		\label{subfig:antipred}
		}
		\subfloat[]{
		\includegraphics[width= 0.48\textwidth]{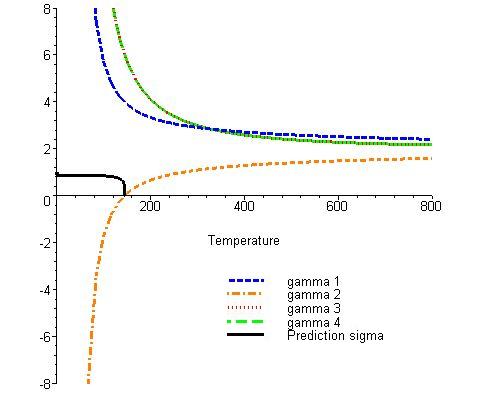}
		\label{subfig:antisep}
		}
	\caption[Wu and Lin prediction for a antiferromagnetic system]{Results for the antiferromagnetic system with interactions $J_n=100k_B$, $J=J^\prime=-100k_B$. (a) shows the predictions from the results of Wu and Lin. Note that the graph for the $\tau$-sublattice is above $+1$ which is physically impossible in our model. There is a transition below 200 Kelvin in all predictions. (b) shows the separate $\gamma$ values for the system. Here $\gamma_2$ is identified as the dominant term for the system, and is negative while the predictions are non-zero. Again in this plot, $\gamma_3$ is overlaid on $\gamma_4$.}
	\label{fig:wuantipredict}
\end{figure}
From Figure \ref{subfig:antipred} we can see that the $\left\langle \tau\right\rangle$ prediction produces physically impossible results. In our system the values of the spin can only be $\pm 1$ and so a result greater than one is impossible. As the value of $\left\langle \tau\right\rangle$ is determined as a multiple of the value of $\left\langle \sigma\right\rangle$ we should also look at this plot. We note that although our system is antiferromagnetic the prediction graph shows a non zero value for the magnetisation. As before, when we plot the individual $\gamma$ in Figure \ref{subfig:antisep} we see that this time it is $\gamma_2$ that is negative before the critical temperature. This implies that $\omega_2=\omega(+-+-)$ dominates, and so again should have average spin of zero. When we compare these predictions to those of Vaks \etal we can see that the critical temperature here is $T_D$, that is the disorder temperature where the system goes from being an ordered antiferromagnetic phase to a disordered antiferromagnetic phase.

As one of the motivations for studying the Union Jack lattice is the property of re-entrant phase transitions, we will now move on to looking at one such system. The system we will now study is the anisotropic ferromagnetic case. For this type of model we will look at the system where the square lattice has interactions with strength $J_n=100k_B$ while the diagonals will have interaction strengths of $J=J^\prime=-92k_B$. When we calculate the possible critical temperatures from Vaks \etal, we see that all three are possible, and there exists a re-entrant phase transition. The graph of Wu and Lin's prediction is shown in Figure \ref{fig:wulinferrononiso}:

\begin{figure}[htbp]
	\centering
		\subfloat[]{
			\includegraphics[width= 0.48\textwidth]{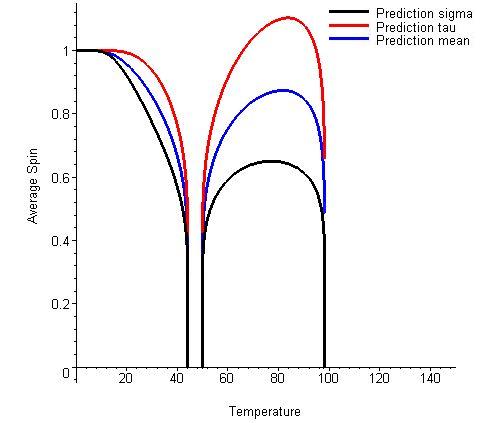}
			\label{subfig:ferrononiso}
		}
		\subfloat[]{
			\includegraphics[width= 0.48\textwidth]{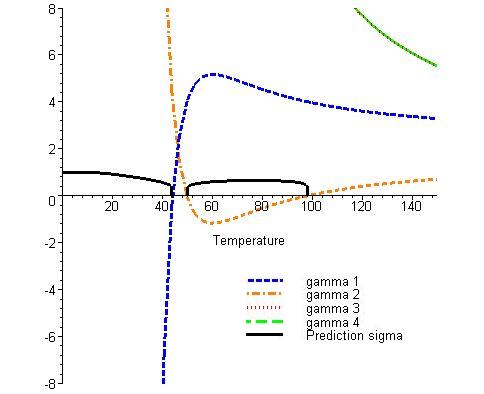}
			\label{subfig:ferrononisosplit}
		}
	\caption[Wu and Lin prediction for a anisotropic ferromagnetic system with equal horizontal and vertical interactions]{Results for the anisotropic ferromagnetic system with interaction strengths $J_n=100k_B$, $J=J^\prime=-92k_B$. This means that the horizontal and vertical interactions are equal. (a) shows the predictions from Wu and Lin for this system. Note after the first phase transition there is a second area of non-zero magnetisation. Here the prediction for the $\tau$-sublattice again has physically impossible results given the conditions of our model. (b) shows graphs of the separate $\gamma$ terms. Here we see that initially the $\gamma_1$ term is negative up to the first critical temperature. In the range of the second area of non-zero magnetisation we see that $\gamma_2$ is negative. As $\gamma_3$ is equal to $\gamma_4$ their plots are overlaid in this graph.}
	\label{fig:wulinferrononiso}
\end{figure}
Looking at Figure \ref{subfig:ferrononiso} we see that in the set of curves between 45 and 100 Kelvin we have physically impossible results again for the $\left\langle \tau\right\rangle$ variable. Up to the first critical temperature we can see that the system behaves as we would expect a ferromagnetic system to, and then between the second and third critical temperatures we have the re-entrant phase. Examining Figure \ref{subfig:ferrononisosplit} we see that the first of these curves is determined when $\gamma_1$ is negative, suggesting a ferromagnetic system. During the re-entrant phase however, it is $\gamma_2$ which is negative, suggesting that its is an ordered antiferromagnetic region. In the paper of Vaks \etal \cite{Vaks1966} the re-entrant phase transition is defined to be between a disordered antiferromagnetic phase and an ordered antiferromagnetic phase. From only looking at the average magnetisation of the system it is not possible to see the transition from a disordered phase to an ordered phase. However the critical temperatures here are consistent with those predicted from (\ref{eq:uno2}). 

However, the results from some anisotropic ferromagnetic systems do have agreement between both theories. To show an example of this, we will look at the system where the interactions on the square lattice are $J_n=-100k_B$ and the diagonal interactions are $J=J^\prime=92k_B$. This is illustrated below in Figure \ref{fig:anisoferrowork}.

\begin{figure}[htbp]
	\centering
		\subfloat[]{
		\includegraphics[width= 0.48\textwidth]{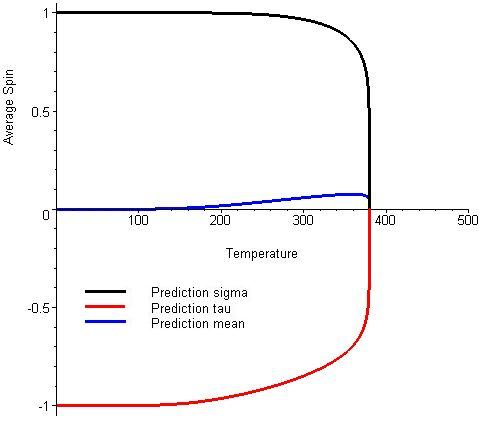}
		\label{subfig:aifwork}
		}
		\subfloat[]{
		\includegraphics[width= 0.48\textwidth]{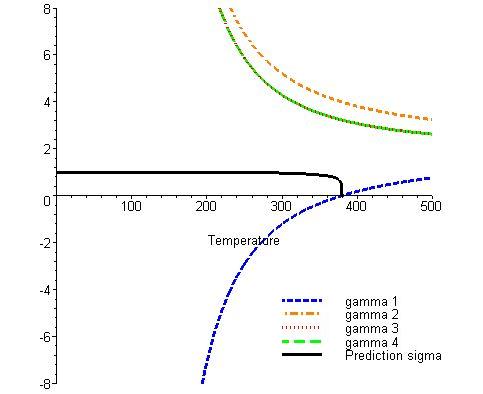}
		\label{subfig:aifworksplit}
		}
	\caption[Wu and Lin prediction for a anisotropic ferromagnetic system with equal negative horizontal and vertical interactions and equal positive diagonal interactions. ]{Results for the anisotropic ferromagnetic system with interactions strengths $J_n=-100k_B$, $J=J^\prime=92k_B$. These are the same interaction strengths as the last example but with the opposite sign. (a) shows the predictions of Wu and Lin for this system. Note that the overall system is antiferromagnetic with a zero average magnetisation but the sublattices have spins of different signs. As the temperature rises, due to the shape of the different curves, an overall magnetisation increases up to the critical temperature of the phase transition. (b) shows the graphs of the individual $\gamma$ terms. Here we see that only $\gamma_1$ is negative in the temperature range, becoming positive at the critical temperature. In this graph $\gamma_3$ and $\gamma_4$ are equal and as such their plots are overlaid.}
	\label{fig:anisoferrowork}
	
\end{figure}
Looking at Figure \ref{subfig:aifwork} we can see that the curves are of opposite signs, but both display a ferromagnetic shape. Using the calculations from Vaks \etal we would expect the critical temperature to be at the value it is, and of the type it is. From examining Figure \ref{subfig:aifworksplit} we can see that it is only $\gamma_1$ which is negative at low temperatures. Equally as $\omega_1=\omega(++++)=\omega(----)$, the result for $\left\langle \tau\right\rangle$ is acceptable, even though the overall system is antiferromagnetic.

However there are some anisotropic systems, with non-uniform square lattice interactions where we see some interesting results. We shall now go on to look at a system where the horizontal interactions are $J_1=J_3=100k_B/0.9^2$, the vertical interactions are $J_2=J_4=100k_B/0.9$ and the diagonal interactions are $J=J^\prime=100k_B$. The graph of this prediction is shown in Figure \ref{fig:UJFunky} below.

\begin{figure}[htbp]
	\centering
		\subfloat[Prediction results]{
		\includegraphics[width= 0.48\textwidth]{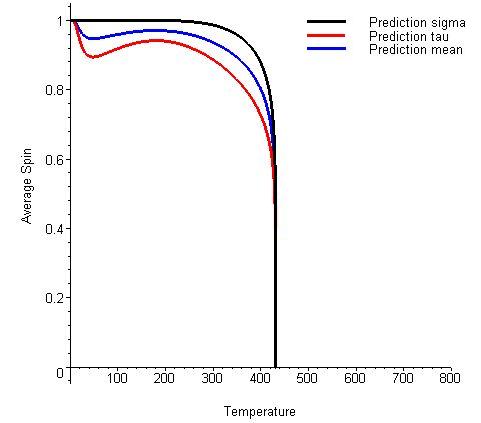}
		\label{subfig:funky}
		}
		\subfloat[Separation of $\gamma$]{
		\includegraphics[width= 0.48\textwidth]{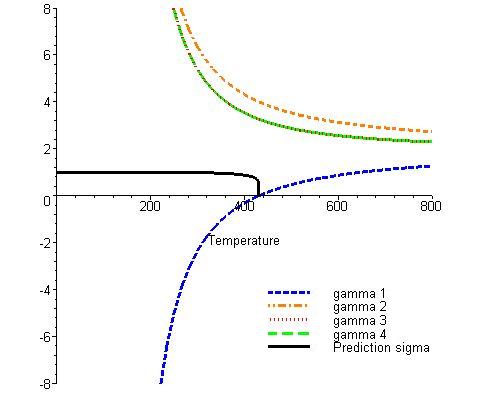}
		\label{subfig:funkysep}
		}
	\caption[Wu and Lin prediction of an anisotropic ferromagnetic system where the horizontal interactions are not equal to the vertical interactions.]{Results of the anisotropic ferromagnetic system with interactions horizontally of $J_1=J_3=100k_B/0.9^2$, vertically of $J_2=J_4=100k_B/0.9$ and diagonally of $J=J^\prime=100k_B$. (a) shows the predictions of Wu and Lin for the system. Note that the predictions are all physically plausible being less than or equal to $+1$. In the $\tau$-sublattice prediction it can be seen that there is a little dip in the value before returning to the curve similar to the prediction function of the $\sigma$-sublattice. (b) shows the separate $\gamma$ terms. As with previous ferromagnetic systems $\gamma_1$ is the dominant term, and the graphs of $\gamma_3$ and $\gamma_4$ are overlaid.}
	\label{fig:UJFunky}
	
\end{figure}
\begin{figure}[htbp]
	\centering
		\subfloat[]{
		\includegraphics[width= 0.48\textwidth]{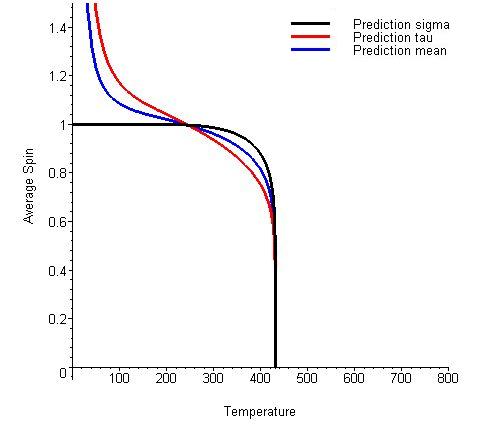}
		\label{subfig:funky2}
		}
		\subfloat[]{
		\includegraphics[width= 0.48\textwidth]{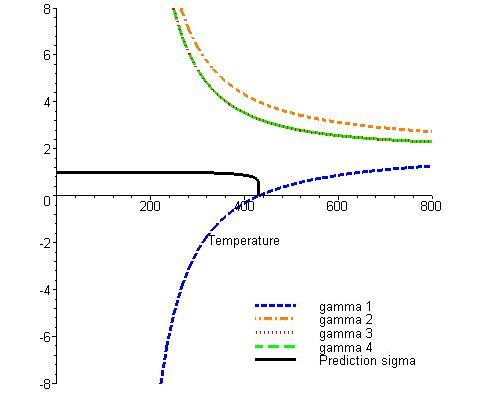}
		\label{subfig:funky2sep}
		}
	\caption[Wu and Lin prediction of the anisotropic ferromagnetic system with interactions as in Figure \ref{fig:UJFunky} rotated 90 degrees]{Results for the anisotropic ferromagnetic system with interaction strengths $J_1=J_3=100k_B/0.9$, $J_2=J_4=100k_B/0.9^2$ and $J=J^\prime=100k_B$. Note these interactions are those of Figure \ref{fig:UJFunky} rotated by 90 degrees. (a) shows the predictions of Wu and Lin for the system. Here we see that the prediction for the $\tau$-sublattice is physically impossible in our model below around 250 Kelvin. (b) shows the graphs of the separate $\gamma$ terms. Note that the result is identical to Figure \ref{subfig:funkysep}.}
	\label{fig:UJFunky2}	
\end{figure}

As we can see from Figure \ref{subfig:funky}, the curves for $\tau$ and subsequently the mean predictions deviate from the shape of the $\sigma$ curve at temperatures below 200 Kelvin. Above this temperature, the graphs show the expected ferromagnetic curves. As there are no values over $+1$ this is physically plausible. When we look at the separated $\gamma$ in Figure \ref{subfig:funkysep} the only term that crosses the x-axis is $\gamma_1$. This again suggests a purely ferromagnetic system, and shows the critical temperature where we would expect. We now look at a similar system where we swap around the interaction strengths of the vertical and horizontal interactions. The graph of this system is shown in Figure \ref{fig:UJFunky2}.

From Figure \ref{subfig:funky2} we can see that again, the $\tau$ and so the mean magnetic spin results are physically impossible given our system. However we can see from this figure and Figure \ref{subfig:funky2sep} that the prediction for $\sigma$ remains the same. Intuitively we would expect this, as we have only rotated the system from Figure \ref{fig:UJFunky} by 90 degrees. Again we see that the predictions become more like we would expect above 200 Kelvin. 

\section{Summary}
\label{sec:conuj}
We have seen in this section that there are a number of systems for which the predicted results from Vaks \etal and Wu and Lin disagree. Equally we note that the Wu and Lin prediction only works when either $\gamma_1<0$ or $\gamma_1~\gamma_2~\gamma_3~\gamma_4>0$. These added conditions were not stated in either Wu and Lin's 1987 \cite{Wu1987} or 1989 \cite{Wu1989} papers. This suggests that these results need to be investigated further with other methods, such as mean field theory in Chapter \ref{ch:MFT} and numerical simulations in Chapter \ref{ch:NE}.

\chapter{Mean Field Theory}
\label{ch:MFT}
In this chapter we will briefly discuss alternative ways of investigating the Ising model through the use of approximations. While exact solutions are useful for investigating systems under special conditions, such as in the absence of a magnetic field, approximations allow us to look at systems where an exact solution may not be possible. After this discussion we will look at one particular method of approximation, mean field theory, in more detail. Initially we will look at how mean field theory is defined for the triangular lattice model \cite{Baxter1989}, which is a specific case of the mean field theory method for the isotropic lattice. After this we will develop our own equations for the Union Jack lattice. At the end of the chapter we will the perform some numerical analysis to investigate how well the approximation models our systems.

\section{Approximations}
\label{sec:OSA}
Mean field theory is an older approach to the subject of phase transitions and generally give us a qualitative description of the phenomena of interest. There are many different approaches that can be use with mean field theory, but common to all is the identification of an order parameter. For this section we will be using the Curie-Weiss model \cite{Plischke1994}. In this method we approximate a system where nearest neighbours interact with each other, to a system where the particles do not interact with each other but instead with the mean spin of the entire lattice. This means that the approximation is affected less by the dimensionality or lattice configuration of the system and so we are able to avoid the issues discussed in Section \ref{sec:hist} (work by \cite{Barahona1982,Istrail2000}). This will allow us to investigate systems with external magnetic field in two dimensions. Alternatively, a similar approach where the free energy is expressed, then minimised with respect to the order parameter, can be used. An example of this free energy minimisation approach is the Bragg-Williams approximation \cite{Bragg1934,Yew1970}. 

These methods can be improved to gain numerical values for the properties of the Ising model, by such approximations as those presented by Bethe in \cite{Bethe1935} and its extension by Guggenheim \cite{Fowler1940}.  However,  the asymptotic critical behaviour of mean field theories is always the same. The most serious fault of mean field theories lies in the neglect of long-range fluctuations of the order parameter. The importance of this omission depends very much on the dimensionality of the problem. In problems involving one and two-dimensional systems, the results predicted by mean field theory are often quantitatively wrong, although they do give a qualitative idea of the behaviour.

\section{The triangular lattice model}
\label{sec:RLM}
First we will look at the development of mean field theory on the triangular lattice, which we will use as a basis for our development on the Union Jack lattice. This is a well studied area, with many textbooks on the subject, including \cite{Landau1980}, \cite{Kubo1965} (summarised in \cite{Plischke1994}) and \cite{Baxter1989}.

Let us consider the triangular lattice Ising model in an external magnetic field, $B$. The Hamiltonian of such a system is given by
\begin{equation}
	H= -J\sum_{\left\langle i,j\right\rangle}^{N} \sigma_{i,j}\sigma_{i+1,j} + \sigma_{i,j}\sigma_{i,j+1} + \sigma_{i,j}\sigma_{i+1,j+1} - B\sum_{i,j=1}^{N} \sigma_{i,j}.
	\label{eq:mno9}
\end{equation}
Now we focus on one site and consider its expectation value, which at a certain temperature has value $m$,
\begin{equation}
	\left\langle \sigma_i\right\rangle = m
	\label{eq:mno10}
\end{equation}
for all $i$. This $m$ is referred to as order parameter of the system. For the next step, we will focus on a particular spin of the lattice $\sigma_0$. As in our final model all spins will be interacting with the parameter $m$. We can first develop the method for one site with $j$ nearest neighbours and then extend it to the entire lattice. For this we take the local version of the Hamiltonian in (\ref{eq:mno9}), which is
\begin{eqnarray}
	h\left(\sigma_0\right) &=& -\sigma_0 \left(J\sum_j{\sigma_j} +B\right) \nonumber \\
	&=& -\sigma_0\left(qJm+B\right)-J\sigma_0\sum_j{\left(\sigma_j-m\right)}
	\label{eq:mno11}
\end{eqnarray}
where $q$ is half the number of nearest neighbours of site 0. This definition of $q$ is such that the duplication of counting of nearest neighbour spins can be eliminated. If we disregard the second term in this equation we obtain a non-interacting system. In a non-interacting system each spin is an effective magnetic field composed of the applied field and an average exchange field due to the neighbours. This means that it will work in any number of dimensions, since the lattice configuration only effects the value of $q$. For example $q=2$ for a square lattice, $q=3$ for both the triangular and simple cubic lattice. The magnetisation has to be determined self-consistently from the condition
\[
m=\left< \sigma_0 \right>=\left\langle \sigma_j \right\rangle, \ j=1,2,\ldots N.
\]
Extending the local Hamiltonian for an non-interacting system to all sites in the matrix we obtain the following equation,
\begin{equation}
	H=-Jqm\sum_{i=1}^N{\sigma_i}-B\sum_{i=1}^N{\sigma_i}.
	\label{eq:mn12}
\end{equation}
As before, the partition function can be found from the Hamiltonian
\begin{eqnarray}
	Z_N &=& \sum_\sigma {\exp \left[\left( \beta Jqm+\beta B\right) \sum_{i=1}^N{\sigma_i}\right]} \nonumber \\
	&=& 2\cosh \left( \beta Jqm+\beta B\right)Z_{N-1} \nonumber \\
	\mathrm{where} \ Z_{N-1}&=&\sum_{\sigma^\prime}{\exp \left[ \left( \beta Jqm+\beta B\right) \sum_{i=1}^{N-1}{\sigma_i}\right] }
	\label{eq:mno13}
\end{eqnarray}
and $\sigma^\prime$ denotes a configuration over $N-1$ sites, $\beta=1/(k_B T)$.

Now we return to looking at the specific case for $\sigma_0$
\begin{eqnarray}
	m&=&\left\langle \sigma _0\right\rangle \nonumber \\ 
	&=&Z_N^{-1}\sum_\sigma{ \sigma_0\exp \left[ \left( \beta Jqm+\beta B\right) \sum_{i=1}^N{\sigma _i}\right]} \nonumber \\
	&=&Z_N^{-1}2\sinh \left( \beta Jqm+\beta B\right) \sum_{\sigma^\prime}{\exp
\left[ \left( \beta Jqm+\beta B\right) \sum_{i=1}^{N-1}{\sigma_i}\right]} \nonumber \\
	 &=&Z_N^{-1}\sinh \left( \beta Jqm+\beta B\right)Z_{N-1} \nonumber \\
	 &=&\tanh \left( \beta Jqm+\beta B\right).
	 \label{eq:mno14}
\end{eqnarray}
To find $m$, the above equation must be solved numerically for $B$ and $T$. This equation is invariant under the map $B\rightarrow-B$, $m\rightarrow-m$
\begin{eqnarray}
	-m &=& \tanh \left( -\beta Jqm+-\beta B\right) \nonumber \\
	&=& -\tanh (-\left( \beta Jqm+\beta B\right)) \nonumber \\
	m &=& \tanh \left( \beta Jqm+\beta B\right).
	\label{eq:mno19}
\end{eqnarray}
It can be seen from this that for a nonzero external field there is at least one solution and sometimes three. For a system with no external field there is always the solution $m=0$, and two further solutions may exist at $\pm m_0$ where $m_0$ signifies the spontaneous magnetisation. As $T\rightarrow 0$, $\tanh{(\beta q J m)}\rightarrow \pm 1$ for $m\neq 0$ and $m_0\rightarrow \pm 1$. As $T$ approaches the critical temperature $T_c$ from $T < T_c$, $\left|m_0(T)\right|$ decreases and we can obtain its asymptotic dependence from a low order Taylor series expansion of the hyperbolic tangent. That is,
\[
	m_0= \beta q J m_0 - \frac{1}{3}\left(\beta q J\right)^3 m_0^3 + \cdots
\]
or
\begin{equation}
	m_0(T)= \pm \sqrt{3}\left(\frac{T}{T_c}\right)^{3/2}\left(\frac{T_c}{T}-1\right)^{1/2}.
	\label{eq:mno15}
\end{equation}
From this we can see that the order parameter $m$ approaches zero in a singular fashion as $T$ approaches $T_c$, vanishing asymptotically as
\begin{equation}
	m_0(T) \propto \left(\frac{T_c}{T}-1\right)^{1/2}.
	\label{eq:mno16}
\end{equation}
It should be noted at this stage that when we compare this with (\ref{eq:ino64}) that the power is $\frac{1}{2}$ rather than $\frac{1}{8}$, so will only give a qualitative result, as will be seen later in Figure \ref{fig:Meantriangleferro}.

\section{Extending to the Union Jack}
\label{sec:EUJ}

As we have stated before in Chapter \ref{ch:UJL}, the Union Jack lattice can be thought of as two square lattices superimposed on each other. This means that we can develop the mean field approximation for this lattice in one of two ways, both of which will be shown in this section. First, in the most simple development, we can consider these sublattices independently of each other and thus develop a set of partially uncoupled equations. In a slightly more complex development, we can approximate the complete lattice with a set of coupled equations. That is, equations where there is a dependence on both sublattices. In the next section we will perform some analysis of the quality of the approximation of both systems.

\subsection{Partially uncoupled equations}
\label{sub:UCEQ}
The partially uncoupled equations can be quickly developed from the equation for a triangular lattice, that is the sublattices have sites with the same number of nearest neighbours. As in Chapter \ref{ch:UJL} we denote the sites with four nearest neighbours as $\tau_k$ and the sites with eight nearest neighbours as $\sigma_j$. We now simplify the system further by taking the interaction strengths along the vertical and horizontal bonds as $J$, and along the diagonals as $K$. We also include an applied external magnetic field of $B$. Using the above theory for triangular lattices, we can construct equations for each type of site:
\begin{eqnarray}
	\left\langle \sigma \right\rangle &=& \tanh{\left[\beta\left(2K\left\langle \sigma\right\rangle+B\right)\right]}, \nonumber \\
	\left\langle \tau \right\rangle &=& \tanh{\left[\beta\left(4J\left\langle \sigma\right\rangle+B\right)\right]}.
	\label{eq:mno17}
\end{eqnarray}
In these equations we can see that the number of nearest neighbours is perhaps not what we would expect from the previous section. In the equation for $\left\langle \sigma \right\rangle$, we are only concerned with interactions between one $\sigma$ and another $\sigma$, so to avoid duplication we have to divide by two. Meanwhile in the equation for $\left\langle \tau \right\rangle$, we are concerned with the number of interactions between a $\tau$ and a $\sigma$, so we do not need to worry about duplication as there are no $\tau-\tau$ interactions. From this point forward we will refer to equations (\ref{eq:mno17}) as the \textit{partially uncoupled equations}, since both equations are dependent on $\left\langle \sigma\right\rangle$ and so only the equation for $\left\langle \sigma\right\rangle$ is uncoupled.

\subsection{Coupled equations}
\label{sub:CEQ}
While the partially uncoupled equations can be useful, to make the equations uncoupled we purposefully ignored the interactions with the $\left\langle \tau\right\rangle$ in the equation for $\left\langle \sigma\right\rangle$. Now we will develop a system of equations where we consider all the interactions on the $\sigma$, which will be referred to as the \textit{coupled equations}. Considering the lattice as before, we can rewrite the Hamiltonian for the system as
\[
	H=-J\sum_{\left\langle j,k\right\rangle}{\sigma_j\tau_k}-K\sum_{\left\langle j,l\right\rangle }{\sigma_j\sigma_l}-B\sum_j{\sigma_j}-B\sum_k{\tau_k}.
\]
Again we focus on the expectation values of the sites and put them into our Hamiltonian to get
\[
H=-2J\left(\sum_k{\left\langle \sigma \right\rangle \tau_k}+\sum_j{\left\langle \tau \right\rangle \sigma_j}\right)-2K\sum_j{\left\langle \sigma \right\rangle \sigma_j} +B\sum_j{\sigma_j}+B\sum_k{\sigma _k}.
\]
The partition function is then
\[
Z_N=\sum_{\sigma _j,\tau _k=\pm 1}{\exp \left[ \beta \left(2K\left\langle \sigma \right\rangle +2J\left\langle \tau \right\rangle +B\right)\sum_j\sigma_j+\beta \left( 2J\left\langle \sigma \right\rangle+B\right) \sum_k\tau _k\right] }.
\]
The partition function is defined as the sum over all possible values of $\sigma_j$ and $\tau_k$, and there are $N=L^2$ sites in total. Let us pick two such spins, say $\sigma_P$ and $\tau_Q$, and perform that part of the sum involving these sites.
\begin{eqnarray}
	\sum_{\sigma _P=\pm 1}{\exp \left[ \beta \left(2K\left\langle \sigma \right\rangle +2J\left\langle \tau \right\rangle +B\right)\sigma_P\right]} &=&\exp \left[ \beta \left( 2K\left\langle \sigma \right\rangle +2J\left\langle \tau \right\rangle +B\right) \right] \nonumber \\
	&&+\exp \left[ -\beta \left( 2K\left\langle \sigma \right\rangle+2J\left\langle \tau \right\rangle +B\right) \right] \nonumber \\
	&=&2\cosh \left[ \beta \left( 2K\left\langle \sigma \right\rangle+2J\left\langle \tau \right\rangle +B\right) \right]. \nonumber
\end{eqnarray}
Likewise
\[
\sum_{\tau _Q=\pm 1}{\exp \left[ \beta \left( 2J\left\langle \sigma\right\rangle +B\right) \tau _Q\right]} =2\cosh \left[ \beta \left(2J\left\langle \sigma \right\rangle +B\right) \right].
\]
If $Z_{N-2}$ is the partition function for the same system but with $N-2$ lattice sites, then
\[
Z_N=4\cosh \left[ \beta \left( 2K\left\langle \sigma \right\rangle
+2J\left\langle \tau \right\rangle +B\right) \right] \cosh \left[ \beta \left(
2J\left\langle \sigma \right\rangle +B\right) \right] Z_{N-2}.
\]	
Now we compute the magnetisation for $\sigma_P$ and $\tau_Q$:
\begin{eqnarray}
	\left\langle \sigma_P\right\rangle &=&Z_N^{-1}\sum_{\sigma j,\tau _k=\pm 1}{\sigma _P\exp \left[ \beta \left(2K\left\langle \sigma \right\rangle +2J\left\langle \tau \right\rangle +B\right)\sum_j\sigma_j+\beta \left(2J\left\langle \sigma \right\rangle +B\right)
\sum_k\tau_k\right] }, \nonumber \\
	\left\langle \tau_Q\right\rangle &=& Z_N^{-1}\sum_{\sigma_j,\tau_k=\pm 1}{\tau _Q\exp \left[ \beta \left(2K\left\langle \sigma \right\rangle +2J\left\langle \tau \right\rangle +B\right)
\sum_j\sigma_j+\beta \left( 2J\left\langle \sigma \right\rangle +B\right)
\sum_k\tau_k\right] }. \nonumber
\end{eqnarray}	
Now we use the fact that
\begin{eqnarray}
	\sum_{\sigma _P=\pm 1}{\sigma_P\exp \left[ \beta\left( 2K\left\langle \sigma \right\rangle +2J\left\langle \tau \right\rangle+B\right) \sigma_P\right]} &=& 2\sinh \left[ \beta \left( 2K\left\langle \sigma\right\rangle +2J\left\langle \tau \right\rangle +B\right) \right], \nonumber \\
	\sum_{\tau _Q=\pm 1}{\tau _Q\exp \left[ \beta \left( 2J\left\langle \sigma \right\rangle +B\right) \tau _Q\right]}  &=&2\sinh \left[ \beta \left( 2J\left\langle \sigma \right\rangle +B\right) \right] \nonumber
\end{eqnarray}
to deduce
\begin{eqnarray}
	\left\langle \sigma _P\right\rangle&=&Z_N^{-1}\sum_{\sigma_j,\tau_k=\pm 1}{\sigma_P\exp \left[ \beta \left(2K\left\langle \sigma \right\rangle +2J\left\langle \tau \right\rangle +B\right)\sum_j{\sigma_j}+\beta \left(2J\left\langle \sigma \right\rangle +B\right)
\sum_k{\tau _k}\right] } \nonumber \\
	&=&\frac{4}{Z_N}\sinh \left[ \beta\left( 2K\left\langle \sigma \right\rangle +2J\left\langle \tau \right\rangle+B\right) \right] \cosh \left[ \beta \left( 2J\left\langle \sigma \right\rangle
+B\right) \right] Z_{N-2} \nonumber \\
	&=&\tanh \left[ \beta \left(2K\left\langle \sigma \right\rangle +2J\left\langle \tau \right\rangle +B\right)\right] \nonumber, \\
	\left\langle \tau _Q\right\rangle &=& Z_N^{-1}\sum_{\sigma_j,\tau_k=\pm 1}{\tau_Q\exp \left[ \beta \left(2K\left\langle \sigma \right\rangle +2J\left\langle \tau \right\rangle +B\right)
\sum_j{\sigma_j}+\beta \left( 2J\left\langle \sigma \right\rangle +B\right) \sum_k{\tau _k}\right] } \nonumber \\
	&=&\frac{4}{Z_N}\sinh \left[ \beta \left( 2J\left\langle \sigma \right\rangle +B\right) \right] Z_{N-2}\cosh\left[ \beta \left( 2K\left\langle \sigma \right\rangle +2J\left\langle \tau
\right\rangle +B\right) \right] \nonumber \\
	&=&\tanh \left[ \beta\left( 2J\left\langle \sigma \right\rangle +B\right) \right]. \nonumber
\end{eqnarray}
Since we have $\left\langle \sigma _P\right\rangle =\left\langle \sigma \right\rangle $ for all $P$ and $\left\langle \tau _Q\right\rangle =\left\langle \tau \right\rangle $ for all $Q$ we need to analyse the equations
\begin{eqnarray}
	\left\langle \sigma \right\rangle &=&\tanh \left[ \beta\left( 2K\left\langle \sigma \right\rangle +2J\left\langle \tau \right\rangle+B\right) \right],  \nonumber \\
	\left\langle \tau \right\rangle &=&\tanh\left[ \beta \left( 2J\left\langle \sigma \right\rangle +B\right) \right].
	\label{eq:mno30}
\end{eqnarray}

\subsection{Wu and Lin adapted}
\label{sub:WLA}
To test our mean field approximations against the known results of Wu and Lin \cite{Wu1987,Wu1989} we will have to rewrite the conditions that they use to classify the type of system. To quickly review, for a ferromagnetic system the following conditions hold:
\begin{equation}
	E_1 < E_2,E_3,E_4;
	\label{eq:mno1}
\end{equation}
for antiferromagnetic systems:
\begin{equation}
	E_2 < E_1,E_3,E_4;
	\label{eq:mno2}
\end{equation}
and for the metamagnetic system:
\begin{eqnarray}
	E_3 &<& E_1,E_2,E_4 \nonumber \\
	\mathrm{or} \ E_4&<&E_1,E_2,E_3;
	\label{eq:mno3}
\end{eqnarray}
where
\begin{eqnarray}
	-E_1&=&J+J^\prime +\left| J_1+J_2+J_3+J_4\right|, \nonumber \\
	-E_2&=&-J-J^\prime +\left| J_1-J_2+J_3-J_4\right|, \nonumber \\
	-E_3&=&-J+J^\prime +\left| J_1-J_2-J_3+J_4\right|, \nonumber \\
	-E_4&=&J-J^\prime +\left| J_1+J_2-J_3-J_4\right|.
	\label{eq:mno4}
\end{eqnarray}
Now we rewrite (\ref{eq:mno4}) with the horizontal and vertical interactions being set to $J$ and the diagonal interactions set to 	$K$, as before. This produces the following	equations:
\begin{eqnarray}
	-E_1&=&K+K+\left| J+J+J+J\right| = 2K+4\left| J\right|, \nonumber \\
	-E_2&=&-K-K+\left| J-J+J-J\right| = -2K, \nonumber \\
	-E_3&=&-K+K+\left| J-J-J+J\right| = 0, \nonumber \\
	-E_4&=&K-K+\left| J+J-J-J\right| = 0.
	\label{eq:mno5}
\end{eqnarray}
Using these equations we can rewrite our conditions for the phase of system as:\\
to be ferromagnetic:
\begin{equation}
	K+\left| J\right| > 0;
	\label{eq:mno6}
\end{equation}
to be antiferromagnetic:
\begin{equation}
	K+\left|J\right| < 0 \ \mathrm{so} \ K<-\left|J\right|;
	\label{eq:mno7}
\end{equation}
and finally to be metamagnetic
\begin{equation}
	K+2\left| J\right| < 0 \ \mathrm{and} \ K > 0 .
	\label{eq:mno8}
\end{equation}
However $\left| J \right|$ is always positive, so the conditions for metamagnetic systems cannot be satisfied. This is due to our simplification to two variables, and means that mean field theory will not be able to model metamagnetic systems. At $K=0$ it can be seen that the Union Jack lattice reduces to the square lattice.

\section{Analysis of the magnetisation}
\label{sec:AOTM}
Having now developed the mean field approximation for the Union Jack lattice, we will now analyse the equations (\ref{eq:mno17}) and (\ref{eq:mno30}). We will look at the two types of equations separately, looking at such quantities as the predicted critical temperatures. Then we will go on to look at the graphs of the functions at a constant temperature. As the equations for $\left\langle \tau\right\rangle$ rely on $\left\langle \sigma\right\rangle$ we will not be discussing them in this section, as they will follow those of $\left\langle \sigma\right\rangle$. Equally as the partially uncoupled equations are adapted forms of those from the square lattice, the analysis will be similar and so not included here (see \cite{Baxter1989}).

\subsection{Partially uncoupled equations}
\label{sub:SWFB}
As the equation for $\left\langle \sigma\right\rangle$ in (\ref{eq:mno17}) only depends on itself no further rearrangement is required. For this analysis, and simplicity, we consider a system with no external magnetic fields applied, that is $B=0$. For any temperature $T$, the corresponding value of $\left\langle \sigma\right\rangle$ is determined by the point of intersection of the curves $y=x$ and $y=\tanh \left[ 4\beta K\left\langle \sigma \right\rangle \right]$. The slope of the latter curve varies from the initial value of $4\beta K$ at the origin and asymptotically approaches the final value of zero. This can be seen graphically in Figure \ref{fig:4sitemagnetism} below.
\begin{figure}[htbp]
	\centering
		\includegraphics[width=8.191cm]{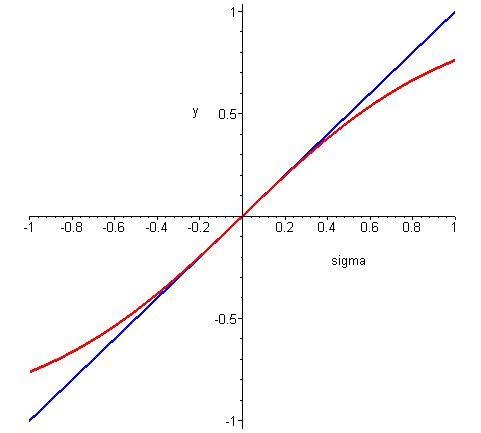}
	\caption[Partially uncoupled mean field prediction on the Union Jack lattice for $\sigma$ at $T_c$]{Graph of the partially uncoupled mean field prediction on the Union Jack lattice for $\sigma$ (from eqn. (\ref{eq:mno17})), shown in red, against the graph of $y=x$, shown in blue. The intersection of the two lines shows the value of $\left\langle \sigma\right\rangle$ at temperature $T$. Here $T=T_c$ and so the graphs intersect at exactly zero.}
	\label{fig:4sitemagnetism}
\end{figure}
It can be seen that there exists a critical temperature above which will have no spontaneous magnetisation. That is the magnetisation will be zero. It follows that an intersection of the two curves (other than at the origin) can only eventuate if
\begin{equation}
	4\beta K > 1 \; \; \; \mathrm{or} \; \; \; T < T_c = \frac{4K}{k_B},
	\label{eq:mno20}
\end{equation}
where we call $T_c$ the critical temperature. At temperatures below $T_c$ the model acquires a non-zero spontaneous magnetisation but at temperature $T>T_c$ no spontaneous magnetisation is possible.

\subsection{Coupled equations}
\label{sub:SWEB}
Now we move to the coupled equations and perform a similar analysis. However, we first need to rearrange the equations to make the equation for $\sigma$ dependent only on the value of $\sigma$. We can do this easily as the equation for $\tau$ only depends on $\sigma$ so we can simply substitute it in as shown below:
\begin{eqnarray}
	\left\langle \sigma \right\rangle &=&\tanh \left[ \beta \left( 2K\left\langle
\sigma \right\rangle +2J\left\langle \tau \right\rangle +B\right) \right], \nonumber \\
	\left\langle \tau \right\rangle &=&\tanh \left[ \beta \left( 2J\left\langle
\sigma \right\rangle +B\right) \right], \nonumber \\
	\mathrm{thus }\ \left\langle \sigma \right\rangle &=&\tanh \left[ \beta \left(
2K\left\langle \sigma \right\rangle +2J\tanh \left[ \beta \left( 2J\left\langle
\sigma \right\rangle +B\right) \right] +B\right) \right].
\label{eq:mno21}
\end{eqnarray}
While this substitution has reduced the number of variables, it has also made the analysis more complicated. However at the origin the equation still behaves like the linearised version. We can find a critical temperature equivalently by solving
\[
	2\beta K+4\beta ^2J^2>1,
\]
or equivalently by solving
\begin{equation}
	4\beta ^2J^2+2\beta K-1=0.
	\label{eq:mno22}
\end{equation}
for $\beta$. We put the coefficients into the solution for a quadratic equation
\begin{eqnarray}
	\beta &=&\frac{-2K\pm \sqrt{4K^2+16J^2}}{8J^2} \nonumber \\
	&=&\frac{-K\pm \sqrt{K^2+4J^2}}{4J^2}.
	\label{eq:mno24}
\end{eqnarray}
The critical temperature can be found with the following formula
\begin{equation}
	T_c=\frac{4J^2}{k_B\left( -K\pm \sqrt{K^2+4J^2}\right) }.
	\label{eq:mno23}
\end{equation}
This formula produces two roots, and we need to perform some further analysis to find which root we should take. We can rewrite (\ref{eq:mno24}) as a Taylor series expansion for small $\left| J\right|$ where the first term will be
\begin{equation}
	\beta \sim \frac{-K \pm \left|K\right|\left(1+\frac{2J^2}{K^2}\right)}{4J^2}.
	\label{eq:mno25}
\end{equation}
So for $K>0$ we first take the positive sign, getting
\begin{equation}
	\beta \sim \frac{-K+K\left(1+\frac{2J^2}{K^2}\right)}{4J^2} \sim \frac{1}{2K};
	\label{eq:mno26}
\end{equation}
and if we take the negative sign we get
\begin{equation}
	\beta \sim \frac{-K-K\left(1+\frac{2J^2}{K^2}\right)}{4J^2} \sim \frac{-K}{2J^2}-\frac{1}{2K}.
	\label{eq:mno27}
\end{equation}
Now we look when $K<0$, and again first take the positive sign:
\begin{equation}
	\beta \sim \frac{-K-K\left(1+\frac{2J^2}{K^2}\right)}{4J^2} \sim \frac{-K}{2J^2}-\frac{1}{2K};
	\label{eq:mno28}
\end{equation}
and finally taking the negative sign:
\begin{equation}
	\beta \sim \frac{-K+K\left(1+\frac{2J^2}{K^2}\right)}{4J^2} \sim \frac{1}{2K}.
	\label{eq:mno29}
\end{equation}
For ferromagnetic systems, when $K>-\left|J\right|$, we require a positive critical temperature. So we take the positive sign, for $K>0$ using (\ref{eq:mno26}) and for $K<0$ using (\ref{eq:mno28}). However, as $J$ approaches zero, the $\sigma$ sublattice becomes a square lattice with the critical temperature defined by $T_c=2K/k_B$. For antiferromagnetic systems we require a zero or negative temperature, so we can see for consistency as $J$ approaches zero, we must take the negative sign for $K<-\left|J\right|$ (\ref{eq:mno29}). Here $K$ will always be negative, so we only use one equation. 

Again we plot the graph of this function at the critical temperature, shown below in Figure \ref{fig:8sitemagnetism}.
\begin{figure}[htbp]
	\centering
		\includegraphics[width=8.191cm]{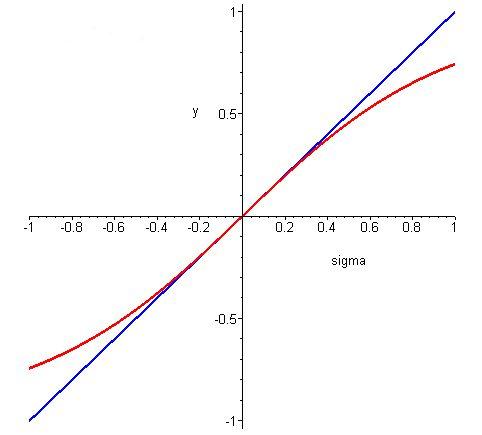}
	\caption[Coupled mean field prediction on the Union Jack lattice for $\sigma$ at $T_c$]{The graph of the coupled mean field prediction on the Union Jack lattice for $\sigma$ (from eqn. (\ref{eq:mno21})) is shown in red against the graph of $y=x$ shown in blue. For this prediction we set the temperature of the system to be again $T_c$. Again the intersection of the two lines shows the final value of $\sigma$, here at zero.}
	\label{fig:8sitemagnetism}
\end{figure}
We can see at the critical point of the system the function follows the line near the origin. As we have seen before with the sites with four bonds, as the system passes the critical temperature it falls below this line.

\section{Numerical results}
\label{sec:MFNE}

Having developed our mean field approximations we can test them against our known results. In this section we will qualitatively compare results from a numerical simulation of the mean field approximation (program code given in Appendix \ref{ch:ProgII}) to the known results of Wu and Lin \cite{Wu1987}. While we do not expect the quantitative correlation to be very close, qualitatively the approximation should bring a general idea of the character of the magnetisation. We will start by looking at the ferromagnetic square lattice and then we will move on to looking at a similar system on the Union Jack lattice.

\subsection{Triangular ferromagnetic lattice}
\label{sub:MFSFL}
\begin{figure}[htbp]
	\centering
		\includegraphics[width=8.191cm]{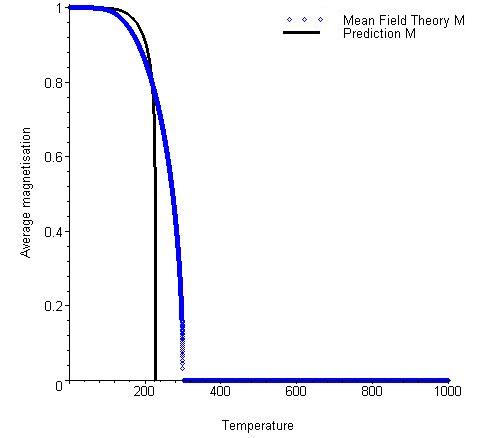}
	\caption[Numerical results for the mean field prediction of the ferromagnetic triangular lattice Ising model.]{Graphs for the triangular lattice Ising model for a ferromagnetic system with interaction strength $J=100k_B$. Here the solid line shows the prediction of Stephenson for nearest neighbour interactions (eqn. (\ref{eq:ino72})). The data points show the numerical results from the mean field simulation of the system. Note that the critical temperature for the mean field curve is higher than the prediction temperature. However qualitatively the curves are similar.}
	\label{fig:Meantriangleferro}
\end{figure}

First we will look at a triangular lattice where the interactions have strength one hundred times the Boltzmann constant and there is no external magnetic field. In Figure \ref{fig:Meantriangleferro} our data points are plotted against the known result.
As we can see, the mean field result is approximately the same shape, and shows a phase transition along with a critical temperature, of 300 K. That is a critical temperature higher than the known result. However the mean field result starts to move away from the positive ground state earlier than the analytical result and so has a slightly less steep curve. As a qualitative approximation it is not bad, though quantitatively it is quite far from the mark.

\subsection{Union Jack ferromagnetic lattice}
\label{sub:MFUFL}
Next we move on to the Union Jack lattice. In general we have seen a good correlation between the theories of Stephenson \cite{Stephenson1970}, Wu and Lin \cite{Wu1987}, and our intuitive expectation in Section \ref{sub:TAnal}. So as our first example of the results we get from our mean field approximation, we will look at the isotropic ferromagnetic case with interaction strengths of one hundred times the Boltzmann constant. The graph of the coupled equations versus the Wu and Lin's result is shown in Figure \ref{fig:Meanferro} below.
\begin{figure}[htbp]
	\centering
		\includegraphics[width=8.191cm]{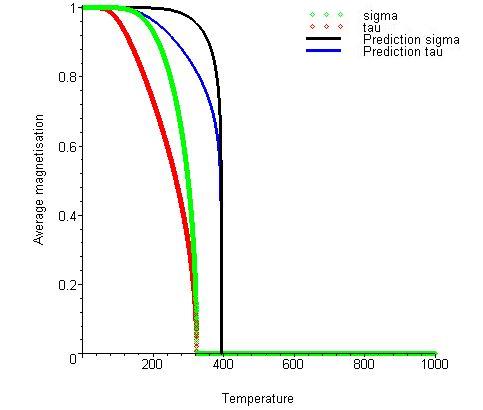}
	\caption[Numerical results for the mean field prediction of an isotropic ferromagnetic Union Jack lattice Ising model using coupled equations]{Graphs for the isotropic ferromagnetic system on the Union Jack lattice with interaction strengths $J=K=100k_B$. The theoretical predictions of Wu and Lin are shown by the solid lines and the numerical results from the coupled mean field equations are shown by the points. Note that the curves are qualitatively the same though quantitatively the critical temperature of the mean field results is lower than the predictions.}
	\label{fig:Meanferro}
\end{figure}
Again we have a good qualitative correlation between the two sets of results, although quantitatively the approximation is slightly under the known result. When we compare to Figure \ref{fig:Meantriangleferro} we see that it is of about the same quality as the triangular lattice predictions.

When we look at the partially uncoupled equations we obtain Figure \ref{fig:Meanunferro} below.
\begin{figure}[htbp]
	\centering
		\includegraphics[width=8.191cm]{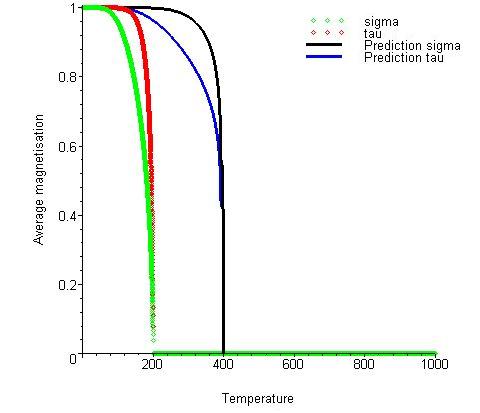}
	\caption[Numerical results for the mean field prediction of an isotropic ferromagnetic Union Jack lattice Ising model using partially uncoupled equations]{Graphs for the isotropic ferromagnetic system on the Union Jack lattice with interaction strengths $J=K=100k_B$. Again the theoretical predictions of Wu and Lin are shown by the solid lines. The data points show the numerical results from the mean field partially uncoupled equations. Note that the order of the curves for the mean field predictions is opposite to the theoretical predictions. Also the critical temperature of the mean field is much lower than the theoretical prediction.}
	\label{fig:Meanunferro}
\end{figure}
As we can see the correlations of this mean field theory approximation with the known results is quantitatively not good. Equally the values for $\tau$ are larger than those of the $\sigma$, which is in the opposite order to the known result. It is a very rough prediction and even when looking at it qualitatively we can see that the curves of the magnetism for the two sublattices are different. In general it would be of more use to use the coupled equations as those results are closer to the known theory, so for the rest of the section we will concentrate on those equations in (\ref{eq:mno30}).

\begin{figure}[htbp]
	\centering

		\includegraphics[width=8.19cm]{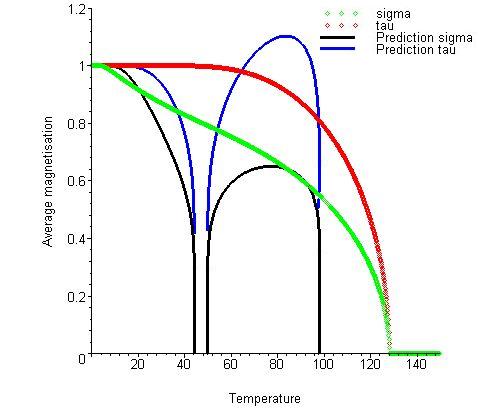}
	\caption[Numerical results for the mean field prediction of an anisotropic ferromagnetic Union Jack lattice Ising model using the coupled equations]{Graphs for the anisotropic ferromagnetic system on the Union Jack lattice with interactions $J=100k_B$ and $K=-92k_B$. The data points are calculated using the coupled mean field equations and the solid lines are the Wu and Lin theorectical predictions. Note that the mean field curves do not see the second range of non-zero magnetisation. Equally the temperature is higher than all the critical temperatures.}
	\label{fig:92negcouple}
\end{figure}

Here we see that the results from the couple equations are more accurate than those from the uncoupled equations on isotropic systems. Next we shall look a anisotropic system. Of course this allows us to look at a system with a re-entrant phase transition. So we will look at the system with horizontal and vertical interactions of $J_n=100k_B$ and diagonal interactions of $J=J^\prime=92k_B$. The graph we obtain is shown in Figure \ref{fig:92negcouple}.
The result from the mean field has a poor correlation with the prediction of Wu and Lin. Qualitatively it can seen that while the sublattice magnetisations are ordered in the same way as the predictions, but do not see the re-entrant phase transition at all. Interestingly the mean field result here quantitatively is above the critical temperature. Also we can see in the graph for the $\left\langle \sigma\right\rangle$ that there is a deviation from a smooth curve. 

\section{Summary}
\label{sec:MFC}
In this section, we have shown that the mean field approximation is a powerful tool for qualitative prediction of a given system in some instances. It allows us to study systems outside the range of the exact solutions as it is not limited by dimensionality. This allows us to study systems with for example an external magnetic field. It does have its limitation however as it is not able to model metamagnetic systems due to the simplification used. Also we have seen that to model the Union Jack lattice models we require a coupled set of equations to produce results that more closely follow those of the known predictions. In addition we see that when the diagonal interactions are negative our mean field results do not closely follow the predictions. In the exact solution (\ref{eq:ino25}) the overall power is $1/8$, where in the mean field equations (\ref{eq:mno30}) the overall power is $1/2$. This limits the accuracy of the approximation. The approximation is an iterative process, and this could lead to any error being compounded. Thus a defect might lead to an instability, resulting in the accuracy of the approximation being poor.
\chapter{Sampling Methods}
\label{ch:sample}
As we have seen from the calculations of Section \ref{sec:TDC} and Chapter \ref{ch:UJL}, it is not a practical possibility to calculate every state of the Ising model on a lattice of a reasonable size as it is NP complete. While one can use an approximation method like the mean field theory approach seen in the previous Chapter \ref{ch:MFT}, these generally only have a qualitative use and we would like a method that could produce reasonably accurate qualitative results. With this in mind in this chapter we will briefly discuss the area of Markov Processes \cite{Grimmett1992} and how these can be sampled using Monte Carlo Methods to produce our results in a reasonable timeframe. While this is going to be a short presentation, the reader is directed to books such as \cite{Landau2005,Binder1995} for a more in depth development of this theory. Our development however will follow the flow of the development in Section 7 of \cite{Plischke1994}. In Chapter \ref{ch:NE} we will use these methods to preform some numerical experiments on various Ising model systems.

\section{Monte Carlo methods}
\label{sec:MCM}
In a Monte Carlo simulation \cite{Metropolis1949} one does not attempt to simulate the dynamics of the system; instead the idea is to generate states $i,j,\ldots$ by a stochastic process such that the probability $\pi_i$ of state $i$ is that given by an approximate distribution (in our case the canonical distribution). In a production run of a simulation, $N$ states are generated and the desired quantity $x_i$ (energy, magnetisation, pressure, etc.) is calculated for each state. If the probabilities are correct, then
\begin{equation}
\left\langle x\right\rangle=\lim_{N\rightarrow\infty}\frac{1}{N}\sum_ix_i.
\label{eq:sno1}
\end{equation}
In our case we will calculate the canonical ensemble, so the probability will be given by
\[
\pi_i=\frac{1}{Z}e^{-\beta E_i} \ \mathrm{where}\  Z=\sum_i{e^{-\beta E_i}}.
\]

In the general case a Monte Carlo simulation consists of the following steps:
\begin{enumerate}
	\item Choose an initial condition
	\item Select a move
	\item Accept or reject the move using a criterion based on a detailed balance
	\item Repeat steps (2) and (3) until enough data is collected.
\end{enumerate}
Typically the data from the early part of a run is discarded since the system will not have had enough time to reach equilibrium.

From this definition we now have two questions:
\begin{enumerate}
	\item How does the computer generate the states?
	\item How can we make sure the probabilities are correct?
\end{enumerate}

\section{Markov processes}
\label{sec:MP}
Now let us consider our Ising model. If it has $N$ positions then it will have $2^N$ microstates. Using the Hamiltonians given in (\ref{eq:ino34}), (\ref{eq:ino74}) and (\ref{eq:uno8}), it is easy to calculate the energy of one of these states. If we were to calculate the expected value with the canonical ensemble, we would use a random number generator to assign the values of spin to each position, weight the contribution of that microstate by $e^{-\beta E_i}$ and repeat until the expectation value had converged. This is inefficient since all the states would appear with equal probability, including those with such small weight that they do not contribute to the thermodynamical average.

Instead to make the process more efficient, we will be sampling the transition between the states. This is because it will be more likely to occur between those of more dominance to the system. The process we shall use to generate these states will be a \textit{Markov Process}.

If the model starts in a given microstate $i$, it will move to state $j$ with transition probability $P_{j\leftarrow i}$ that does not depend on the previous history of the model. If we assume the model to be under some fairly general conditions, processes after the passage of a transient would produce states with a unique steady-state probability distribution. This steady-state probability $\pi_j$ is an eigenvector, with eigenvalue one, of the transition matrix:
\begin{equation}
	\pi_j = \sum_i{P_{j\leftarrow i}\pi_i}.
\label{eq:sno2}
\end{equation}
The steady-state probabilities are unique if the matrix $P_{j\leftarrow i}$ is regular, which means that for
some integer $n$ all elements of $(P_{j\leftarrow i})^n$ are positive and non-zero. Physically, this restriction
implies that it is always possible to go from one state to any other state in a finite number of steps. Exceptions are matrices that are block diagonal, for example
\[
\left[	
\begin{array}{cccc}
	P_{1\leftarrow 1} & P_{1\leftarrow 2} & 0 & 0 \\
	P_{2\leftarrow 1} & P_{2\leftarrow 2} & 0 & 0 \\
	0 & 0 & P_{3\leftarrow 3} & P_{3\leftarrow 4} \\
	0 & 0 & P_{4\leftarrow 3} & P_{4\leftarrow 4}
\end{array}.
\right]
\]
Since there is no way of going from states 1 or 2 to 3 or 4, the stationary probability distribution will depend on whether one started with one of the first two states or one of the last two.

\section{The Metropolis-Hastings algorithm}
\label{sec:Met}
Suppose we wish to determine the transition matrix $P_{i\leftarrow j}$ for our Ising model so that the steady-state distribution is
\begin{equation}
	\pi(i)=\frac{\exp-\beta E_i}{Z}
	\label{eq:sno3}
\end{equation}
where $\beta= 1/\left(k_B~T\right)$ and $Z$ is the partition function. A possible method of generating a sequence of states from an initial state is to pick a site $\alpha$ randomly and attempt to flip (change the sign of) its spin. The resulting state (which may be the same state $i$ if the attempt to flip $\sigma_\alpha$ fails) we call $j$. Let $P_{j\leftarrow i}$ be the transition probability from $i$ to $j$. After $n$ steps the transition probability $P_{f\leftarrow i}(n)$ is given by
\[
P_{f\leftarrow i}(n)= \sum_{i_1,i_2,\ldots,i_{n-1}}{P_{f\leftarrow i_{n-1}}P_{i_{n-1}\leftarrow i_{n-2}}\ldots P_{i_1\leftarrow i}}.
\]
After many steps the system will approach a limiting distribution
\[
\frac{\pi(m)}{\pi(j)}=\exp\left[-\beta\left\{E(m)-E(j)\right\}\right]
\]
for all pairs of states $m$,$j$. We now also require the transition probabilities to be normalised
\begin{equation}
	\sum_j{P_{j\leftarrow m}}=1
	\label{eq:sno4}
\end{equation}
and to obey
\begin{equation}
	\frac{P_{j\leftarrow m}}{P_{m\leftarrow j}}=\frac{\pi(j)}{\pi(m)}=\exp\left[-\left\{E(j)-E(m)\right\}\right].
	\label{eq:sno5}
\end{equation}
We find that
\begin{equation}
	\pi(m)=\sum_j{P_{m\leftarrow j}\pi(j)}.
	\label{eq:sno6}
\end{equation}
This relation holds, by definition, for any Markov process. The first step above follows from the normalisation in (\ref{eq:sno4}), while the second step involves substituting (\ref{eq:sno5}). From (\ref{eq:sno1}) we see that (\ref{eq:sno6}) implies that $\pi(m)$ is a stationary probability distribution of the process. Equation (\ref{eq:sno5}) is called the \textit{principle of detailed balance}. It can be shown that it is a sufficient condition for arriving at the correct limiting probability distribution, provided that the process for selecting moves does no contain any traps. That is, it should always be possible to get from any given microstate to any other microstate.

The simplest and most frequently used method of achieving a detailed balance is the \textit{Metropolis algorithm} \cite{Metropolis1953}:
\begin{enumerate}
	\item Pick a site $\alpha$ randomly
	\item Compute the energy change $\Delta E=E(f)-E(i)$ that would occur if the spin at site $\alpha$ was flipped
	\item If $\Delta E<0$, flip the site at site $\alpha$; if $\Delta E > 0$ flip this site with probability $\exp(-\beta\Delta E)$
	\item Repeat steps (1) to (3) until enough data has been collected.
\end{enumerate}
An alternative to (3) that is sometimes used is to flip the spin $\sigma_\alpha$ with probability $[\exp(\beta\Delta E)+1]^{-1}$ regardless of the sign of $\Delta E$. It is easy to see that (\ref{eq:sno4}) is satisfied in both cases for all possible pairs of states. The allowed states therefore will occur with the correct frequency if the simulation is run long enough to reach the steady state.

\chapter{Calibration}
\label{ch:COP}

In this chapter we will be calibrating the program written for the project (code given in Appendix \ref{ch:ProgI}) by running simulations on a triangular lattice and comparing them to the known results given in \cite{Stephenson1964} and \cite{Baxter1975}. This is an important step as it will allow us to see how accurately the program is able to simulate a system quantitatively, rather than just qualitatively as seen with mean field theory in Chapter \ref{ch:MFT}.

\section{Isotropic system}
\label{sec:ITS}
\subsection{Ferromagnetic}
\label{sub:FITS}
\begin{figure}[htbp]
	\centering
		\includegraphics[width= 0.6\textwidth]{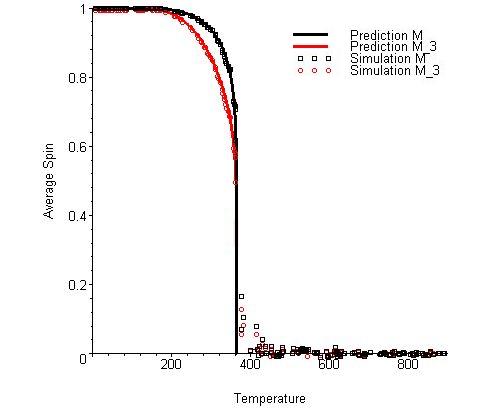}
	\caption[Plot of simulation results for an isotropic ferromagnetic system on the triangular lattice.]{Graph of numerical simulation results plotted against the theoretical predictions of Stephenson \cite{Stephenson1964} and Baxter \cite{Baxter1975} for a isotropic ferromagnetic system on the triangular lattice with interaction strengths $J_1= 100k_B$, $J_2=100k_B$, $J= 100k_B$. The theoretical predictions are shown with solid lines. The numerical simulations are shown with points of the colour corresponding to the relative prediction.}
	\label{fig:triferro}
\end{figure}

In this first simulation we will investigate an isotropic system with all nearest neighbour interactions as one hundred times the Boltzmann constant and a three site interaction of zero. This is the simplest system on a triangular lattice, and contains a phase transition. This will give us a rough guide as to whether we can continue with more complicated models. The resultant graph is shown in Figure \ref{fig:triferro}.
We can see from this graph that our simulation shows a good correlation with the known results that we described earlier. Our simulation tracks the phase transition at under 400 Kelvin quite well. The correlation is good for both the magnetisation data and the three-site correlator data. There is little noise after the phase transition and the simulation results quickly move to a zero average magnetisation. While visible, the noise is of a magnitude that is much lower than the rest of the data.

\subsection{Antiferromagnetic}
\label{sub:AITS}
\begin{figure}[htbp]
	\centering
		\includegraphics[width= 0.6\textwidth]{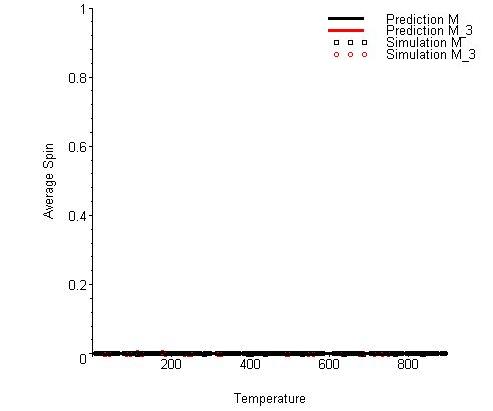}
	\caption[Plot of simulation results for an isotropic antiferromagnetic system on the triangular lattice.]{Graph of numerical simulation results plotted against the theoretical predictions of Stephenson and Baxter for the isotropic antiferromagnetic system on the triangular lattice with interaction strengths $J_1= -100k_B$, $J_2=-100k_B$, $J= -100k_B$.}
	\label{fig:trianti}
\end{figure}

Having looked at an isotropic ferromagnetic system, the next step is to look at an isotropic antiferromagnetic system. According to the conditions set out in the papers \cite{Stephenson1964,Baxter1975}, we should expect results of zero average magnetisations. The graph we obtain from the simulations is shown in Figure \ref{fig:trianti}.
From the graph we see that the simulation data shows a good correlation with the predicted zero magnetisation, and appears to have little noise. We also note that there is no evidence of a phase transition across our temperature range. This is however a frustrated system. That is, not all of the interactions can be in their preferred states, one interaction must always be in a parallel state. To further investigate antiferromagnetic systems we have to move on to looking at anisotropic systems.

\section{Anisotropic systems}
\label{sec:ATS}
\subsection{Ferromagnetic}
\label{sub:FATS}
Our investigation of anisotropic systems will start by trying to match the results given in Baxter's paper \cite{Baxter1975} of the three-site interactions on the square lattice. We do this by setting the diagonal interaction strength to zero, while choosing values for the other two interactions. In our first simulation of these systems we will look at the ferromagnetic system where the horizontal and vertical interactions are one hundred times the Boltzmann constant. The graph we obtain is show in Figure \ref{fig:squareferrotrip}:
\begin{figure}[htbp]
	\centering
		\includegraphics[width= 0.6\textwidth]{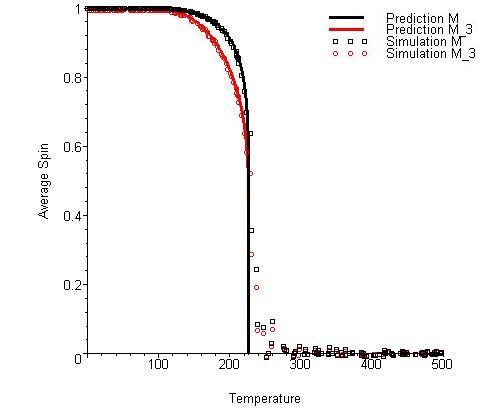}
	\caption[Plot of simulation results for an isotropic ferromagnetic system on the square lattice.]{Graph of numerical simulation results plotted against Stephenson and Baxter theoretical predictions on the isotropic ferromagnetic system on the square lattice with interactions $J_1= 100k_B$, $J_2=100k_B$, $J= 0$. }
	\label{fig:squareferrotrip}
\end{figure}
Once again our simulation has a good correlation with the know results. There is noise around the critical temperature, though the results quickly return to the expected value. The phase transition appears to be consistent with the known results. 

\subsection{Antiferromagnetic}
\label{sub:AATS}
Now we move to a more general anisotropic system in this set of systems, an antiferromagnetic system where the vertical interactions are of equal magnitude as the horizontal interactions but negative. Again we expect the results to show a zero average magnetisation across the entire temperature range and no phase transitions. The graph of this simulation is show in Figure \ref{fig:squareantitrip}:
\begin{figure}[htbp]
	\centering
		\includegraphics[width= 0.6\textwidth]{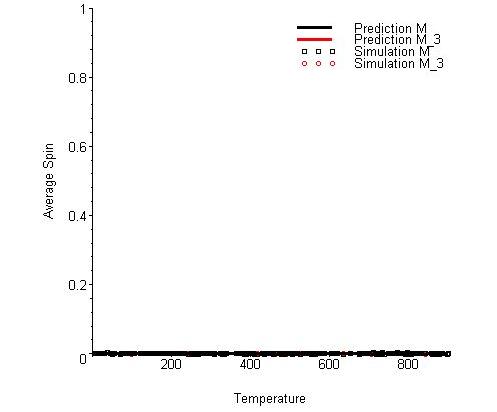}
	\caption[Plot of simulation results for an isotropic antiferromagnetic system on the square lattice.]{Graph of numerical simulation results plotted against the theoretical predictions from Stephenson and Baxter for the isotropic antiferromagnetic system on the square lattice with interactions $J_1= 100k_B$, $J_2=-100k_B$, $J= 0$.}
	\label{fig:squareantitrip}
\end{figure}
As expected we can see that the simulation data is approximately zero average magnetisation and there is no sign of a phase transition in the temperature range. The noise seen is the data is low and seems constant across the data set. In this system we do not have the issue of frustrated bonds as the lack of the diagonal interactions allows all the interactions to be in their preferred state.

As an extension to the last simulation, we will now perform a simulation of a system of an antiferromagnetic system where the vertical and horizontal interactions are negative one hundred times the Boltzmann constant and the diagonal interactions are positive one hundred times the Boltzmann constant. The graph of this simulation is shown in Figure \ref{fig:antitriangle}:
\begin{figure}[htbp]
	\centering
		\includegraphics[width= 0.6\textwidth]{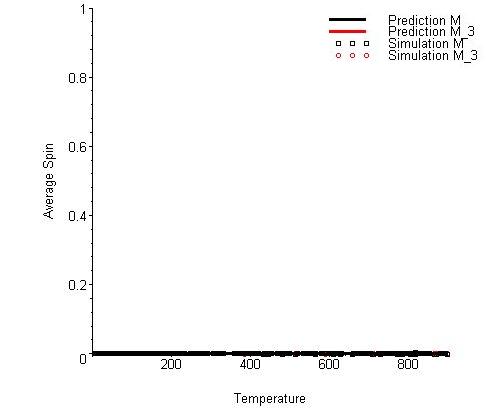}
	\caption[Plot of simulation results for as anisotropic antiferromagnetic system on the triangular lattice.]{Graph of numerical simulation results plotted against the theoretical predictions from Stephenson and Baxter for the isotropic antiferromagnetic system on the square lattice with interactions $J_1= -100k_B$, $J_2=-100k_B$, $J= 100k_B$.}
	\label{fig:antitriangle}
\end{figure}
Once again, as with the previous antiferromagnetic systems, we see that the data is around the zero average magnetisation. This shows a strong correlation to the known results for the triangular lattice, and the noise seems to be at a low level. There appears to be no phase transition across the data range, which is as expected. In this simulation the interactions will be in their preferred state. That is, along the horizontal and vertical bonds the spins will be in an anti-parallel arrangement, due to the ferromagnetic coupling of the diagonal bond.

\section{Overall result}
\label{sec:TOR}
In the simulations carried out during this chapter we have seen that the data produced shows a high correlation with the known results of Baxter and Stephenson. In producing accurate results for these systems, confidence can now be taken in future results as to their accuracy of modelling the system. In general we can say that the calibration of the program has gone well and a useful tool has been developed for the next chapter.
\chapter{Numerical Simulations}
\label{ch:NE}
Having calibrated our program against the triangular lattice, we move on to look at the Union Jack lattice. As we saw previously in Chapter \ref{ch:UJL}, the theoretical predictions from Wu and Lin \cite{Wu1989} sometimes give results that intuitively we do not expect. In this chapter we will perform numerical simulations on various systems and compare the results against Wu and Lin's predicted results, and the critical temperatures again against those from Vaks \etal \cite{Vaks1966}. As in the triangular lattice simulations, we will be using a lattice of 100 sites by 100 sites in our program. The simulation results obtained will be for a finite lattice, while the theoretical predictions are for an infinite lattice. This lattice size is chosen as it is small enough to have a reasonable run time, while being large enough to suppress the finite size effects.

\section{Isotropic ferromagnetic}
\label{sec:IF}
Following the structure of our analysis in Section \ref{sub:TAnal}, we will start our numerical simulations with the isotropic ferromagnetic system. Here the interactions for our system will be $J=J^\prime=J_n=100k_B$. When we plot our simulation results against the predicted results of Wu and Lin we obtain Figure \ref{fig:ferro} below.
\begin{figure}[htbp]
	\centering
		\includegraphics[width= 0.6\textwidth]{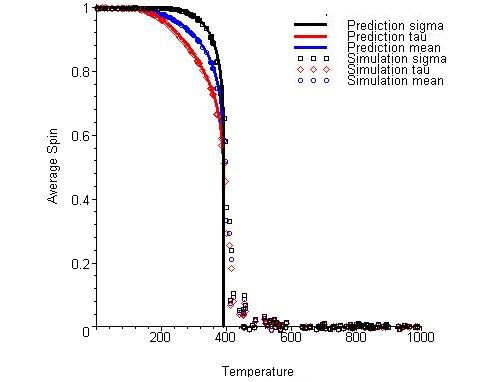}
	\caption[Plot of simulation results for an isotropic ferromagnetic system on the Union Jack lattice.]{Graph of numerical simulation results plotted against theoretical predictions of Wu and Lin for an isotropic ferromagnetic system on the Union Jack lattice with interactions $J_1= 100k_B$, $J_2=100k_B$, $J_3=100k_B$, $J_4=100k_B$, $J= 100k_B$, $J^\prime= 100k_B$. Here the theoretical predictions are shown with the solid lines and the numerical results are shown with the points.}
	\label{fig:ferro}
\end{figure}
As we can see, our simulation data has a high correlation with the prediction functions. As we have discussed before, Wu and Lin's result follows that of Vaks \textit{et al}., and so we can say that our simulation data also agrees with their phase predictions. There is some noise around the critical temperature, though is of a small magnitude when compared to the other results. After the noise part of the data, the points again follow the prediction with a high correlation.

\section{Anisotropic metamagnetic}
\label{sec:NSmeta}
We move on to the first system in which we saw a disagreement between the results of Wu and Lin and the configuration we expect. Our next simulation will be on an anisotropic metamagnetic system where $J_n=10k_B$, $J=100k_B$ and $J^\prime=100k_B$. The graph we obtain when we plot our simulation results against the predictions of Wu and Lin is shown in Figure \ref{fig:meta} below.
\begin{figure}[htbp]
	\centering
		\includegraphics[width= 0.6\textwidth]{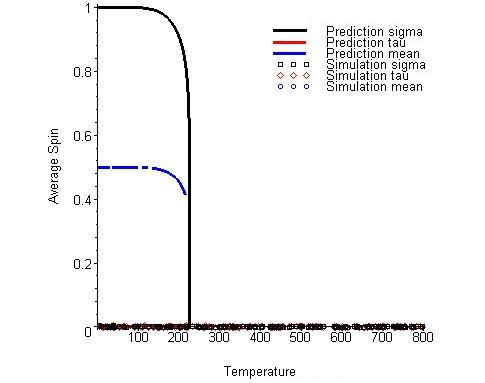}
	\caption[Plot of simulation results for an anisotropic metamagnetic system on the Union Jack lattice.]{Graph of numerical simulation results plotted against theoretical predictions of Wu and Lin for the anisotropic metamagnetic system on the Union Jack lattice with interactions $J_n=10k_B$, $J=100k_B$ and $J^\prime=-100k_B$. Note that the simulation results do not follow the curves of the prediction functions.}
	\label{fig:meta}
\end{figure}
Our simulation results show a poor correlation to the prediction functions, with the average spin across the temperature being zero. From our analysis in Section \ref{sub:TAnal} we concluded that the system should be antiferromagnetic when there is no external magnetic field. The results have a high correlation to an antiferromagnetic system, and so back up this analysis.

\section{Anisotropic antiferromagnetic}
\label{sec:NSanti}
Next we look at the anisotropic antiferromagnetic system. The system will have horizontal and vertical interactions of $J_n=100k_B$ and diagonal interactions of $J=J^\prime=-100k_B$ as before. The results are plotted against Wu and Lin's predictions in Figure \ref{fig:antiferro} below.
\begin{figure}[htbp]
	\centering
		\includegraphics[width= 0.6\textwidth]{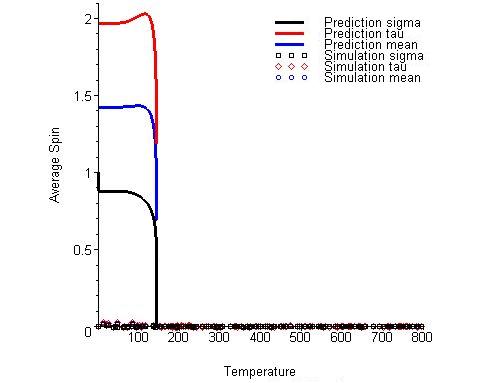}
	\caption[Plot of simulation results for an anisotropic antiferromagnetic system on the Union Jack lattice.]{Graph of numerical simulation results plotted against theoretical predictions of Wu and Lin for an anisotropic antiferromagnetic system on the Union Jack lattice with interactions $J_1= 100k_B$, $J_2=100k_B$, $J_3=100k_B$, $J_4=100k_B$, $J= -100k_B$, $J^\prime= -100k_B$. Note that the simulation results do not follow the curves of the theoretical predictions and are physically possible.}
	\label{fig:antiferro}
\end{figure}
The correlation between the simulation data and Wu and Lin's prediction is again poor. Vaks \etal predict that this system should have average spin of zero. As we can see this confirms the results of the simulation. Overall the noise in the results is low and does not show any signs of detecting any non-zero magnetisation. 

\section{Anisotropic ferromagnetic}
\label{secNSanferro}
\begin{figure}[htbp]
	\centering
		\includegraphics[width= 0.6\textwidth]{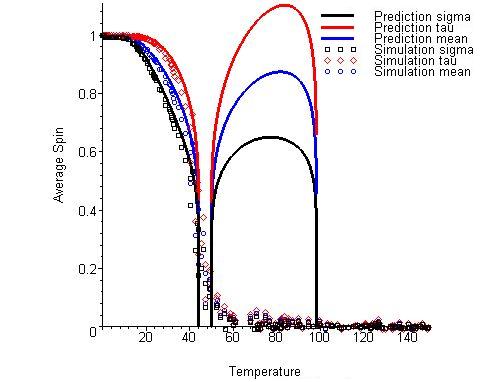}
	\caption[Plot of simulation results for an anisotropic ferromagnetic system on the Union Jack lattice with equal horizontal and vertical interactions.]{Graph of numerical simulation results plotted against theoretical predictions of Wu and Lin for an isotropic ferromagnetic system on the Union Jack lattice with interactions  $J_1= 100k_B$, $J_2=100k_B$, $J_3=100k_B$, $J_4=100k_B$, $J= -92k_B$, $J^\prime= -92k_B$. Note that the simulation results have good correlation up to the first critical temperature, but do not show the re-entrant phase.}
	\label{fig:anti92150}
\end{figure}

So far in simulations we have focussed on systems with only one dominant system. In Section \ref{sub:TAnal} we saw that we can have an anisotropic ferromagnetic system where there is a re-entrant phase transition in the temperature range. The system we studied there was one with horizontal and vertical interactions of $J_n=100k_B$ and diagonal interactions of $J=J^\prime=-92k_B$. The graph of our simulation results plotted against the prediction functions of Wu and Lin is shown in Figure \ref{fig:anti92150}.
We can see that our simulation results show good correlation with the predictions of Wu and Lin up to the first phase transition. Note that our simulation results do not show the re-entrant phase transition. As we discussed in Section \ref{sub:TAnal}, the re-entrant phase transition is between a disordered antiferromagnetic phase and an ordered antiferromagnetic phase. Due to the set up of the simulation program we are not able to see the difference between the different antiferromagnetic phases. Compared to the prediction functions of Wu and Lin, we can see that our simulation results do not follow their predictions for the ordered antiferromagnetic phase, although do show an average magnetisation of zero.

\begin{figure}[htbp]
	\centering
		\includegraphics[width= 0.6\textwidth]{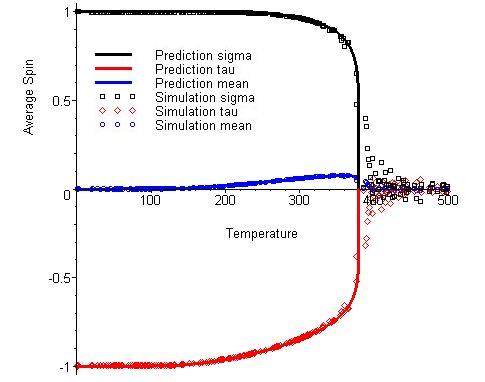}
	\caption[Plot of simulation results for an anisotropic ferromagnetic system on the Union Jack lattice with equal negative horizontal and vertical interactions and positive diagonal interactions.]{Graph of numerical simulation results plotted against theoretical predictions of Wu and Lin for an anisotropic ferromagnetic system on the Union Jack lattice with interactions $J_1= -100k_B$, $J_2=-100k_B$, $J_3=-100k_B$, $J_4=-100k_B$, $J= 92k_B$, $J^\prime= 92k_B$. Note that the simulation results shows high correlation with both prediction curves.}
	\label{fig:antia92500}
\end{figure}

Next in our study of the systems in Section \ref{sub:TAnal}, we will now go on to look at the anisotropic ferromagnetic system with horizontal and vertical interactions of $J_n=-100k_B$ and diagonal interactions of $J=J^\prime=92k_B$. The graph of this simulation is show in Figure \ref{fig:antia92500}.
Here we see that our simulation results have good correlation with the predicted functions of Wu and Lin. Our simulation also shows the slight overall magnetisation after 200 Kelvin, up to the critical temperature. There is noise after the critical temperature as the two sets of data converge to the average zero spin. We have seen this in all our ferromagnetic systems and it is not significant given the range of the rest of the data. So we have agreement with Wu and Lin's prediction.

Following on in our studies, we will look at the anisotropic systems where rotation produces different graphs. We first look at the system with horizontal interactions of $J_1=J_3=100k_B/0.9^2$, vertical interactions of $J_2=J_4=100k_B/0.9$ and diagonal interactions of $J=J^\prime=100k_B$, and the similar system rotated through 90 degrees. The graphs of our simulation results are shown in Figure \ref{fig:NSfunkypos} below.
\begin{figure}[htbp]
	\centering
		\subfloat[]{
		\includegraphics[width= 0.48\textwidth]{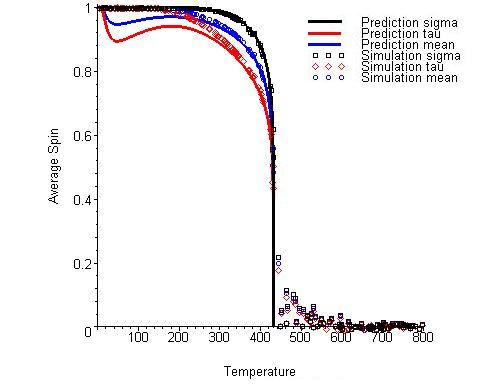}
		\label{subfig:funkygraphpos}
		}
		\subfloat[]{
		\includegraphics[width= 0.48\textwidth]{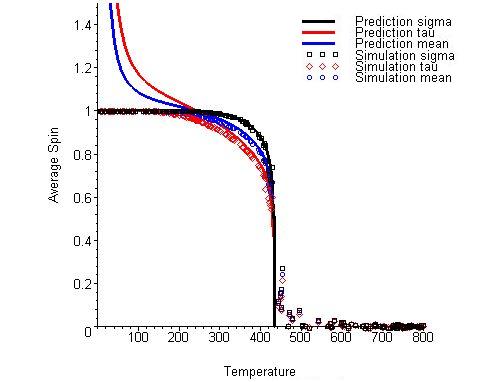}
		\label{subfig:funkyposgraph}
		}
	\caption[Plot of simulation results for an anisotropic ferromagnetic system on the Union Jack lattice with horizontal and vertical interactions that are not equal and positive diagonals.]{Graphs of numerical simulation results plotted against theoretical predictions of Wu and Lin for an anisotropic ferromagnetic system on the Union Jack lattice. (a) shows the system with interactions $J_1=J_3=100k_B/0.9^2$, $J_2=J_4=100k_B/0.9$ $J=J^\prime=100k_B$. (b) shows the rotated system with interactions $J_1=J_3=100k_B/0.9$, $J_2=J_4=100k_B/0.9^2$ $J=J^\prime=100k_B$. Note that although the prediction functions are different, the simulation results are identical. Also there is poor correlation with the $\tau$-sublattice prediction.}
	\label{fig:NSfunkypos}
\end{figure}
As we see when we compare the simulations results, the data forms similar curves and has a similar phase transition at equal critical temperatures. In comparison to the predictions of Wu and Lin, we see that at the higher temperatures the data follows all three curves with good correlation. At lower temperatures, below about 200 Kelvin, we see that the $\sigma$ prediction still has good correlation for both systems. This confirms that the prediction for $\tau$ (and subsequently the mean magnetisation) is not correct below this temperature. It also shows that rotating the lattice should not have an effect on the results of the system. 

\begin{figure}[htbp]
	\centering
		\subfloat[]{
		\includegraphics[width= 0.48\textwidth]{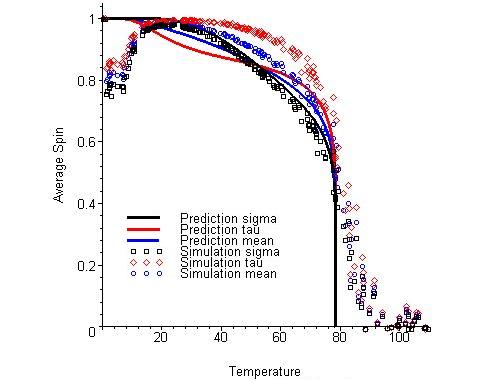}
		\label{subfig:funkydata1}
		}
		\subfloat[]{
		\includegraphics[width= 0.48\textwidth]{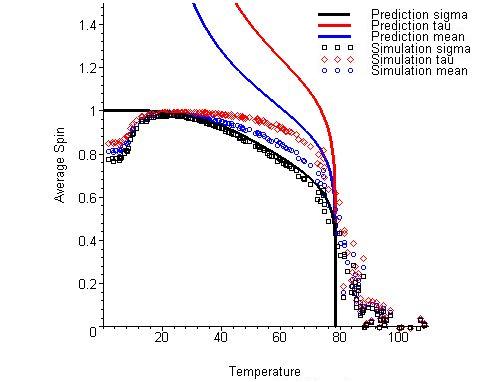}
		\label{subfig:funkydata2}
		}
	\caption[Plot of simulation results for an anisotropic ferromagnetic system on the Union Jack lattice with horizontal and vertical interactions that are not equal and negative diagonals.]{Graphs of numerical simulation results plotted against theoretical predictions of Wu and Lin for an anisotropic ferromagnetic system on the Union Jack lattice. (a) shows the system with interactions $J_1=J_3=100k_B/0.9^2$, $J_2=J_4=100k_B/0.9$ $J=J^\prime=-100k_B$. (b) shows the rotated system with interactions $J_1=J_3=100k_B/0.9$, $J_2=J_4=100k_B/0.9^2$ $J=J^\prime=-100k_B$. Note that although the prediction functions are different, the simulation results are identical. Also there is slight deviation at low temperatures of the simulation results.}
	\label{fig:NSFunky}
\end{figure}

A further result can be seen if we now take a system similar to the previous example but with negative diagonal interactions. For an example of this type of system we will look at the system with horizontal interactions of $J_1=J_3=100k_B/0.9^2$, vertical interactions of $J_2=J_4=100k_B/0.9$ and diagonal interactions of $J=J^\prime=-100k_B$, and the same system rotated through 90 degrees. The graphs of our simulation results are shown in Figure \ref{fig:NSFunky}.
Again we see as in Figure \ref{fig:NSfunkypos} that the simulation results are very similar to each other. However, in this case we see that all three simulation results have a range of lower results at low temperatures. After these lower results the simulations then move up again to the prediction curves, following the $\sigma$ prediction in both cases. We do see that in both cases the simulation results and predicted results show the same critical temperature and phase transition. This oddity maybe due to the simulation being performed on a finite system, while the theoretical predictions are for an infinite lattice.

\section{Magnetic fields}
\label{sec:MagFields}
\begin{figure}[htbp]
	\centering
		\includegraphics[width= 0.6\textwidth]{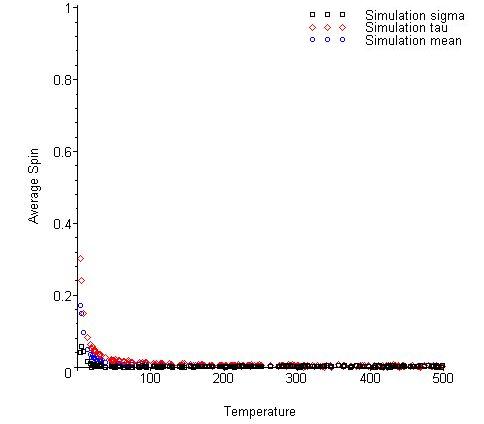}
	\caption[Plot of the simulation results of an anisotropic metamagnetic system on the Union Jack lattice with external magnetic field strength $k_B$]{Graph of the simulation results for the anisotropic metamagnetic system on the Union Jack lattice in an external magnetic field of strength $k_B$. Note that at low temperatures there is non-zero magnetisation of greater magnitude than the external magnetic field.}
	\label{fig:metabolt}
\end{figure}

One advantage of using a numerical simulation, is that we can study systems in external magnetic fields. Metamagnetic systems have the property that when we add a external magnetic field we obtain results that show average magnetisation of a higher magnitude than that of the applied field. In all the following systems, the base system is one where the square lattice interactions are $J_n=10k_B$ and the diagonal interactions are $J=-J^\prime=100k_B$, our standard metamagnetic system. In our first simulation we will apply a magnetic field of strength equal to the Boltzmann constant. The graph is show in Figure \ref{fig:metabolt}.
With this magnetic field we can see already that at low temperatures that there is now a non-zero average magnetic spin. Also the sublattices now have distinct curves at these temperatures, rather than being in the disorder antiferromagnetic state. The average magnetisation is small in comparison with the possible values for the spin states. So our next step is to increase the magnetic field to a strength ten times the Boltzmann constant. The graph we obtain is shown in Figure \ref{fig:metabolt10} below.
\begin{figure}[htbp]
	\centering
		\includegraphics[width= 0.6\textwidth]{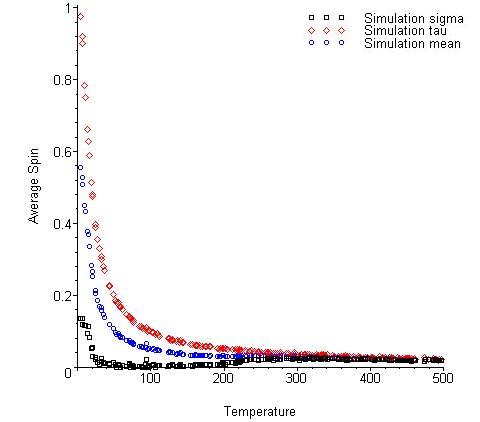}
	\caption[Plot of the simulation results of an anisotropic metamagnetic system on the Union Jack lattice with external magnetic field strength $10k_B$]{Graph of the simulation results for the anisotropic metamagnetic system on the Union Jack lattice in an external magnetic field of strength $10k_B$.}
	\label{fig:metabolt10}
\end{figure}
We can see again that a non-zero average magnetisation has resulted. We see also that the $\sigma$ prediction is now showing more of a sign of a ferromagnetic curve, while the $\tau$ prediction is a less sharp curve. Above 200 Kelvin we see a slight increase in the magnetisation suggesting that this is the critical temperature where the system is going into a disordered phase. When we look back to Figure \ref{fig:wulinmeta} we see that this is also the point where Wu and Lin's prediction hits the x-axis.

\begin{figure}[htbp]
	\centering
		\includegraphics[width= 0.6\textwidth]{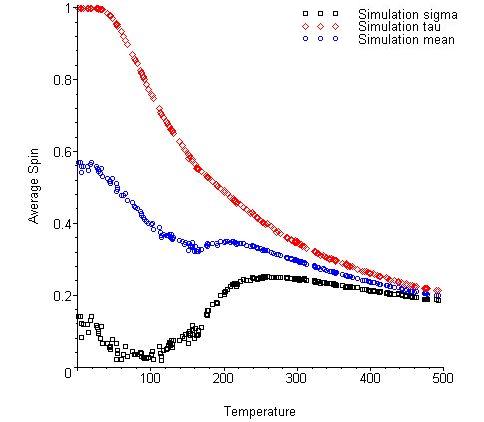}
	\caption[Plot of the simulation results of an anisotropic metamagnetic system on the Union Jack lattice with external magnetic field strength $100k_B$]{Graph of the simulation results for the anisotropic metamagnetic system on the Union Jack lattice in an external magnetic field of strength $100k_B$.}
	\label{fig:metabolt100}
\end{figure}

For completeness we will now look at the same metamagnetic system in an external magnetic field of $100k_B$. The graph of the results we obtain is shown in Figure \ref{fig:metabolt100}.
Here we see that the results are showing more of a ferromagnetic shape. Again above 200 Kelvin the sublattice results are converging to a central value, and all the curves seem to be related to each other. Below this temperature we see that the $\tau$-sublattice is showing more of a ferromagnetic phase, although the curve is not sharp. Meanwhile the $\sigma$-sublattice has a shape suggesting a ferromagnetic phase at low temperatures but moves up to a maximum at 200 Kelvin.

\section{Summary}
\label{NSSum}
We have seen in this chapter that our simulation results help support our findings in Chapter \ref{ch:UJL}. We have good correlation between our simulation results in isotropic ferromagnetic and most anisotropic ferromagnetic systems and Wu and Lin's predictions \cite{Wu1987,Wu1989} while in a ferromagnetic phase. However in terms of the $\tau$-sublattice prediction we saw that our simulations support that the results should not change by a 90 degree lattice rotation. We also saw that our simulation agrees with Valks \etal predictions \cite{Vaks1966} for the antiferromagnetic systems. In addiction we showed that the re-entrant phase transition can not be seen from average magnetisation alone.

In addition we showed that our simulation can be used to further investigate system properties that occur in external magnetic fields. This shows that we have a powerful tool for numerical simulation of the Union Jack lattice.
\chapter{Conclusion}
\label{ch:Conc}
Over the course of the thesis we have seen two main results. The re-entrant phase transition can not be seen when looking at the average magnetisation results. In addition we have seen that the prediction functions of Wu and Lin in their papers \cite{Wu1987,Wu1989} produce some anomalous results. In this section we will review the results from the different approaches we undertook and discus the disagreements from Wu and Lin's results that were found. At the end of the section we will discuss the additional conditions that when applied allow the prediction functions of Wu and Lin to show a greater agreement with known theories.

Our aim was to investigate the properties of the model and specifically the re-entrant phase transitions. From our investigations we have seen that these re-entrant phase transitions can not be seen when only looking at the average magnetisation results. This is due to the transition being from an unordered antiferromagnetic phase to an ordered antiferromagnetic phase. While this can been seen in the Vaks \etal paper \cite{Vaks1966}, this can be seen from other order parameters, the average magnetisation of an unordered antiferromagnetic phase and an ordered antiferromagnetic phase is identically zero. This was further confirmed by our numerical simulations, which also did not see these re-entrant phase transitions due to being focused on the average magnetisation of the system. However we have seen that the predictions of Wu and Lin can be used to find the critical temperatures of these phase transitions, and agree with the calculations of Vaks \etal

In addition, we have seen that Wu and Lin's prediction for the $\sigma$-sublattice given in \cite{Wu1987} requires additional conditions to agree with our numerical simulations. It is possible to classify the phases of the system by examining the $\gamma$ terms of equation (\ref{eq:uno3}). Wu and Lin's predictions under the current conditions produce non-zero magnetisations for non-ferromagnetic systems. However if we impose the conditions that we only use the prediction formula (\ref{eq:ino25}) when $\gamma_1<0$ or $\gamma_1~\gamma_2~\gamma_3~\gamma_4~>0$ and it is zero outside those conditions, their results now work for all systems.

When we look at Wu and Lin's prediction for the $\tau$-sublattice, given in \cite{Wu1989}, we see for the general anisotropic lattice there are some additional issues. When the interactions on the square lattice are equal, by applying the conditions above we can eliminate the physically impossible results that we saw in the antiferromagnetic and metamagnetic phases. This is in agreement with the earlier work of Lin and Wang \cite{Lin1988}. When we consider unequal vertical and horizontal interactions on the square lattice we see that physically plausible results are only possible when the horizontal interactions are larger than the vertical interactions. From Figure \ref{fig:NSfunkypos} we see however that our simulation results show that there is still some disagreement at low temperatures. While this suggests that further conditions are required for this prediction, they would be more involved than those for the $\sigma$-sublattice. For the current prediction, imposing the condition that $J_1=J_2=J_3=J_4$, together with the conditions imposed on $\sigma$ would bring agreement with the simulation results.

From our investigation of mean field theory, we saw that it is a poor quantitative predictor for the systems. In addition it has the limitation of not being able to look at metamagnetic systems. Qualitatively however, it does allows a prediction of the general behaviour of symmetric systems without a re-entrant phase transition.Again the mean field prediction does not see re-entrant phase transitions in terms of the magnetisation. In these systems its prediction of critical temperature is generally in line with the third phase transition critical temperature. The results were useful for confirming those the results we found in our theoretical analysis.

In conclusion, in the thesis we have not been able to investigate the re-entrant phase transition using the average magnetisation. However, we have developed an accurate simulation program for finite systems for both the triangular and Union Jack lattices. This simulation program has allowed us to investigate inconsistencies in the theoretical predictions of Wu and Lin \cite{Wu1987,Wu1989}. We have also shown that it is possible with additional conditions to make these predictions model the systems more accurately.

Future research could extend this work to look more closely at the rotational variance that is observed in the $\tau$-sublattice prediction. Intuitively the predictions should be invariant under rotation of the lattice, as shown in the simulation results, and this disagreement requires further investigation. Equally as we saw in Figure \ref{fig:NSFunky} that there is an oddity in the results at low temperatures, and this should be investigated to determine if it is affected by the lattice size.

As well as looking to the current lattice configuration, further research could take place to see if the results found occur in other configurations. Research into the hexagonal and double hexagonal lattices may be interesting to determine if a re-entrant phase is present. In addition in terms of the Union Jack lattice, an investigation into mixed spin lattices such as those of Stre\v{c}ka \etal \cite{Strecka2006, Strecka2006a} may bring insight into both the re-entrant phase transitions and the metamagnetic systems.

In terms of the simulation program, other Monte Carlo methods could be investigated to improve its efficiently. After a conversation with Dr Tim Garoni of the University of Melbourne at the Australian Mathematics Society meeting in 2010, the algorithm of Swendsen and Kotecky \cite{Swendsen1987}, which has been improved by Wang, Swendsen and Kotecky in \cite{Wang1990}, was suggested to be a more efficient method. As this is a cluster method, its use on a Union Jack lattice would need to be investigated. If the algorithm could be used on this lattice type, it would overcome the problem of non-ergodic data in low temperature antiferromagnetic systems. It would also allow the simulation to not be affected by the critical slowing down effects around the second-order phase transitions.
\bibliographystyle{amsalpha}
\bibliography{bibliography}
\appendix
\chapter{Constants and Conversion Factors}
\label{ch:CCF}

\begin{flushright}
Boltzmann Constant ($k_B$) \dotfill $8.617343\times 10^{-5}\mathrm{eVK}^{-1}$ \\
Avogadro's number \dotfill $6.02\times 10^{23}\mathrm{mol}^{-1}$ \\
1 electron-volt (eV) \dotfill $1.60\times 10^{-19}\mathrm{J}$ \\
or $1.1605\times 10^4\mathrm{K}$ \\
0 Kelvin \dotfill $-273.16^{\circ }\mathrm{C}$ \\
\end{flushright}
\chapter{Programming I}
\label{ch:ProgI}

\section{Programmer's guide}

The broad purpose of this code is to simulate the Ising Model on a two-dimensional lattice with specific configurations. It does this by first asking the user for the initial conditions of the system along with the configuration of the lattice, the temperature range required and how many initial points to ignore. The program then simulates the model over randomly chosen temperature points, and outputs the data to a text file. The program has the ability to be split the output into sublattices if required. 

\subsection{Flow}
Initially the user is asked for the type of lattice, the interaction strengths, external magnetic field strength they want the system to be running with, the number of points to ignore and  the temperature range they are interested in. After all the conditions have been entered the system is initialised with all the sites being set to the "UP" (+1) position. The initial energy is then calculated for the system. Then a temperature is picked at random within the range given by the user, and the system is simulated at this temperature. Initially the first set of data points are ignored up to the level required by the user. Then the next points are averaged and stored as the average spin value for that temperature. The user is then told the overall average spin. The system is reset to the "UP" position for every site and is simulated again at a random temperature, not equal to the past temperatures. After 200 data points have been stored in the array, it is then outputted in a suitable format to a text file. The program then ends, telling the user how long the simulation took.

\begin{figure}[htbp]
	\centering
		\includegraphics[width=150mm]{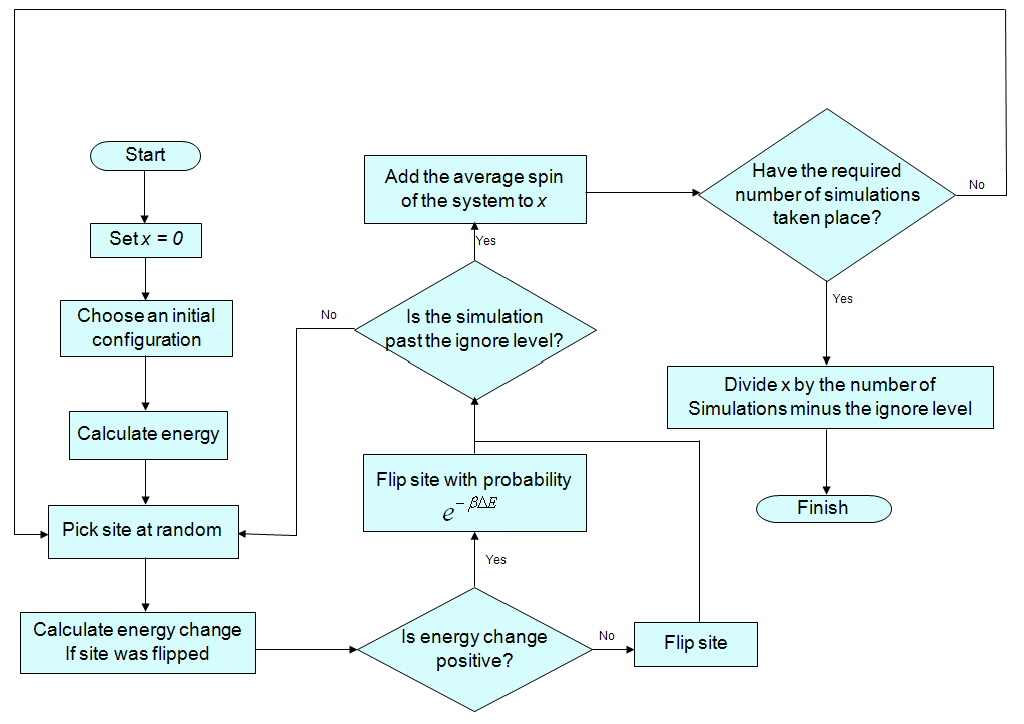}
	\caption{Flowchart of the numerical simulation program}
	\label{fig:flow}
\end{figure}

\newpage
\section{Code listings for main program}
\subsection{isling2dt.h}
\begin{lstlisting}
%auto-ignore
// standard C headers
#include <stdio.h>
#include <math.h>
#include <stdlib.h>
#include <time.h>

// Some useful system variables
enum {
    UP= 1,
    DOWN= -1,
    ITERATIONS= 10000,
    MODELS= 1000,
    COND= 1000,
    length= 100,
    height= 100,
    array_len= 200
};

#define	BOLT 8.617343*pow(10,-5)   /*The boltzman contant*/

typedef struct actions {
        double jh;
        double jh2;
        double jv;
        double jv2;
        double jdu;
        double jdd;
        double h;
        double ku;
        double kd;
        char type;
} ACT;

typedef struct coordinate {
        float temp;
        double p;
        double q;
        double r;
        double tripu;
        double tripd;
        struct coordinate *next;
} COORD;

//Prototypes of our functions
int setup_chain(char *);
double setup_gen(char *, char *);
void initialise_chain(int [length][height]);
void *safe_malloc(size_t, char *);
void *safe_realloc (int *, size_t, char *); /* Does the same for realloc */
void metropolis(int [length][height], float, ACT, double [3], int [3]);
double energy(int [length][height], ACT);
double energyb(int [length][height], ACT, int, int, int);
void write_2d_array (char *, ACT);
void set_temp(float , float);
void free_cond (void);          /* Frees the memory taken by linked list */

extern COORD *hol1; 
\end{lstlisting}
\newpage
\subsection{isling2dt.c}
\begin{lstlisting}
%auto-ignore
/* We link our files with a standard header */
#include "isling2dt.h"

COORD *hol1= NULL;

int main() {
        // intialising variables
	int **chain; // for our lattice
        int i, point, count, cycle, ignore, hours, minutes, seconds, *lattice;
        int timeleft;
        time_t timer;
        char input, save;
        COORD *coord;
	double j, k;
        float min, max, minmax, percent;
        ACT vars;
        double *split;
        char filename[100];
        min= 0;
        max= min;
        // some intial values incase of square lattices
        vars.ku= 0;
        vars.kd= 0;
        vars.jv= 0;
        vars.jh= 0;
        vars.jdu= 0;
        vars.jdd= 0;
        // Finding out the type of lattice required
        while(vars.type != 'S' && vars.type != 'T' && vars.type != 'U' && vars.type != 'N' && vars.type != 'H' && vars.type != 'D') {
                printf("What type of lattice would you like to study? \n ([S]quare, [T]riangular, [U]nion Jack, [N]ext Nearest, [H]exagonal, [D]ouble Hexagonal) ");
                fscanf(stdin, "%1s", &vars.type);
                vars.type= toupper(vars.type);
        }
        while(input != 'A' && input != 'I') {
                printf("Is the lattice [I]somorpic of [A]ntimorphic? (I or A) ");
                fscanf(stdin, "%1s", &input);
                input= toupper(input);
        }
        // Finding out the attraction forces between particles
        if(input == 'I') {
                // for the isotopic lattice
                j= setup_gen("the attraction \n between two particle", "real");
                vars.jh= j;  // if isotopic all j values are the same
                vars.jv= j;
                vars.jh2= j;
                vars.jv2= j;
                if(vars.type != 'S' && vars.type != 'H' && vars.type != 'D') {
                        // specfic qualities that are different from square lattice
                        vars.jdu= j;
                        k= setup_gen("the attraction between three particles", "real");
                        vars.ku= k;    // as all values of k are the same too
                        if(vars.type == 'U' || vars.type == 'N') {
                                // specfic qualities for Union Jack and NN lattices
                                vars.jdd= j;
                                vars.kd= k;
                        }
                } else if(vars.type == 'D') {
                        k= setup_gen("the attraction between two particles second hexagon", "real");
                        vars.ku= k;
                        vars.kd= k;
                }
        } else if(input == 'A') {
                // for the antitopic lattice
                vars.jh= setup_gen("the attraction \n between two particle, horizontal", "real");
                vars.jv= setup_gen("the attraction \n between two particle, vertical", "real");
                vars.jh2= vars.jh;
                vars.jv2= vars.jv;
                if(vars.type != 'S' && vars.type != 'H' && vars.type != 'D') {
                        // specfic qualties that are different from square lattices
                        vars.jdu= setup_gen("the attraction \n between two particle, diagonal up", "real");
                        vars.ku= setup_gen("the attraction \n between three particles up", "real");
                        if(vars.type == 'U' || vars.type == 'N') {
                                // specfic qualities for the Union Jack lattice
                                if(vars.type == 'U') {
                                        printf("Does J1 = J3 AND J2 = J4?: ");
                                        fscanf(stdin,"%1s",&save);
                                        printf("\n");
                                        save= toupper(save);
                                        if(save == 'N') {
                                                vars.jh2= setup_gen("the attraction \n between two particle, horizontal (J3)", "real");
                                                vars.jv2= setup_gen("the attraction \n between two particle, vertical (J4)", "real");
                                        }
                                }
                                vars.jdd= setup_gen("the attraction \n between two particle, diagonal down", "real");
                                vars.kd= setup_gen("the attraction \n between three particles down", "real");
                        }
                } else if(vars.type == 'D') {
                        vars.ku= setup_gen("the attraction between two particles, horizonatal second hexagon","real");
                        vars.kd= setup_gen("the attraction between two particles, vertical second hexagon", "real");
                }
        }
        // Finding the overal field
        vars.h= setup_gen("the overall field", "real");
        // Setting the number of intial values to ignore
        ignore= (int)setup_gen("how many tens of inital values to ignore", "integer");
        while(ignore >= MODELS || ignore < 0) {
                // in case they got it wrong
                printf("Error: too many values ignored. Please enter a positive number less than %d\n", MODELS);
                ignore= (int)setup_gen("How many inital values to ignore", "integer");
        }
        while(min == max) {
                //setting the minimum and maximum temperatures
                min= (float)setup_gen("minimum temperature", "real above zero");
                max= (float)setup_gen("maximum temperature", "real above zero and min");
                if(min < 0) min= 0; //As we are working in Kelvin it makes little sense to have negative numbers
                if(max < 0) max= 0;
        }
        //to save retyping if the temperatures are inputted the wrong way round
        if(min > max) {
                minmax= min;
                min= max;
                max= minmax;
        }
        printf("Enter output filename: ");
        if(fscanf(stdin, "%s", &filename) != NULL) {
                printf("\n");
                fflush(stdin);
        }
        timer= time(NULL); // starting the timer
        // seeding the random numbers with the current time
        srand((unsigned)time(NULL));
        // creating our lattice
        chain= safe_malloc(length * sizeof(int [height]),"Allocating Memory to Chain");
        // The simulation starts here
        initialise_chain(chain);
        coord= hol1;
        if(vars.type == 'U') {
                lattice= safe_malloc(5 * sizeof(int),"Holding array for spin");
                split= safe_malloc(5 * sizeof(double),"Holding array for average spin");
        } else {
                lattice= safe_malloc(3 * sizeof(int),"Holding array for spin");
                split= safe_malloc(3 * sizeof(double),"Holding array for average spin");
        }
        for(point= 0; point < array_len; point++) {
                // getting a temperature
                set_temp(/*coord,*/ min, max);
                // starting our average anew
                coord= hol1;
                lattice[0]= length*height;
                if(vars.type == 'T') {
                        lattice[1]= 2*(((length-1)*(height-1))+((length-1)+(height-1)));
                } else if (vars.type == 'U') {
                        lattice[1]= length*height*0.5;
                        lattice[2]= length*height*0.5;
                        lattice[3]= (((length-1)*(height-1))+((length-1)+(height-1)));
                        lattice[4]= (((length-1)*(height-1))+((length-1)+(height-1)));

                }
                // setting the chain to "UP"
	        if(point != 0) initialise_chain(chain);
                for(cycle= 0; cycle < MODELS; cycle++) {
                        for(count=0; count < ITERATIONS; count++) {
                                //running the metropolis algorithm
                                metropolis(chain, coord->temp, vars, split, lattice); // Performing the algorithm
                                if(cycle == ignore && count == 0) {
                                        split[0]= (double)lattice[0]/(length*height);
                                        if(vars.type == 'T') {
                                                split[1]= (double)lattice[1]/(2*(((length-1)*(height-1))+((length-1)+(height-1))));
                                        } else if (vars.type == 'U') {
                                                split[1]= (double)lattice[1]/(length*height*0.5);
                                                split[2]= (double)lattice[2]/(length*height*0.5);
                                                split[3]= (double)lattice[3]/(2*(((length-1)*(height-1))+((length-1)+(height-1))));
                                                split[4]= (double)lattice[4]/(2*(((length-1)*(height-1))+((length-1)+(height-1))));
                                        }
                                }
                        }
                        percent= ((float)(cycle+1+(MODELS*point))/(float)(MODELS*array_len));
                        seconds= (time(NULL)-timer);
                        timeleft= (long int)(((time(NULL)-timer)/percent)-seconds);
                        seconds= timeleft%60;
                        minutes= abs(timeleft/60)%60;
                        hours= abs(timeleft/(60*60));
                        printf("%.2f %% done, time left: %.2d hours, %.2d minutes, %.2d seconds\r",percent*100, hours, minutes,seconds); // To show the program is still working
                }
	        coord->p= (double)split[0]/(double)((MODELS-ignore)*ITERATIONS); // storing the average spin
                if(vars.type == 'T' || vars.type == 'U') {
                        coord->q= (double)split[1]/(double)((MODELS-ignore)*ITERATIONS);
                }
                if(vars.type == 'U') {
                        coord->r= (double)split[2]/(double)((MODELS-ignore)*ITERATIONS);
                        coord->tripu= (double)split[3]/(double)((MODELS-ignore)*ITERATIONS);
                        coord->tripd= (double)split[4]/(double)((MODELS-ignore)*ITERATIONS);
                }
	        printf("\nThe average is %.2f\n",coord->p); // a check to see if it is resonable
                printf("The temperature is %.2f\n",coord->temp);
                printf("Point: %d\n", point); // where we are up to
        }
        write_2d_array (filename, vars); // outputing to our data file
        free_cond();
	printf ("Now run maple to plot graph\n"); // handy hint to the user
        free(chain);
        free(lattice);
        free(split);
        seconds= (time(NULL)-timer)%60;
        minutes= (abs((time(NULL)-timer-seconds)/60))%60;
        hours= abs((time(NULL)-timer-seconds-(minutes*60))/(60*60));
        printf("Runtime: %ld hours, %ld minutes, %ld seconds\n", hours, minutes, seconds); // how long it too
	return 0; // finish
}
\end{lstlisting}
\newpage
\subsection{setup2d.c}
\begin{lstlisting}
%auto-ignore
/* We link our files with a standard header */
#include "isling2dt.h"

void *safe_malloc (size_t size, char *location) {
	/* This function performs a malloc operation but checks to
	see if there is enough free memory for this to be written to. */
	void *ptr;
	ptr= malloc(size);
	if (ptr == NULL) {
		printf("Out of memory at function: %s\n",location);
		exit(-1);
	}
	return ptr;
}

void *safe_realloc(int *array, size_t size, char *location) {
	/* This is like the safe_malloc but for realloc. */
	void *ptr;
	ptr= realloc(array, size);
	if (ptr == NULL) {
		printf("Out of memory at function: %s\n",location);
		exit(-1);
	}
	return ptr;
}

int setup_chain(char *quant) {
        /* This function is a left over from a previous program for
        the one dimensional Ising model. A two dimensional version
        is being looked into.*/
	int num;
	printf("Enter the %s of chain required: (0,1,...) ",quant);
	fscanf(stdin, "%d", &num);
	printf("\n");
	return num;
}

void initialise_chain(int chain[length][height]) {
        /* This function take the chain and set all site in the "UP"
        position */
	int i,j;
        for(j=0; j < height; j++){
        	for(i=0; i < length; i++){
                        if(chain[i][j] != UP) chain[i][j]= UP;
	        }
        }
}

double setup_gen(char *item, char *units) {
        /* This is the basic input function, as so many of our variables
        have similar questions I decided to write a function for them*/
	double i;
	printf("Enter the value of the %s: (%s) ",item,units);
	if (fscanf(stdin, "%lf", &i)) {
        printf("\n");
        } else {
        printf("\n ERROR: No input found. \n");
        clearerr(stdin);
        fflush(stdin);
        setup_gen(item,units);
        fflush(stdin);
        }
	return i;
}

void write_2d_array (char output[], ACT action)
/* Take the arrays in x and y both of length array_len and write a file in
maple output format to the file named filename */
{
	int i;
	int m= array_len;
        COORD *coord;
	FILE *fptr;
        coord= hol1;
	fptr= fopen (output, "w");
        fprintf (fptr, "p:=[\n");
        while(coord != NULL) {
              fprintf (fptr, "[%.3f,%.15f]",coord->temp,coord->p);
	      if (/*m > 1*/coord->next != NULL) {
	       	 fprintf (fptr, ",");
              }
	      fprintf (fptr, "\n");
	      m=array_len-1-i;
              coord= coord->next;
        }
        fprintf (fptr, "];\n");
        m= array_len;
        if(action.type == 'T' || action.type == 'U') {
                if(action.type == 'T') {
                        fprintf (fptr, "trip:= [\n");
                } else {
                        fprintf (fptr, "q:=[\n");
                }
                coord= hol1;
                while(coord != NULL) {
                        fprintf (fptr, "[%.3f,%.12f]",coord->temp,coord->q);
	                if (/*m > 1*/coord->next != NULL) {
	       	                fprintf (fptr, ",");
                        }
	                fprintf (fptr, "\n");
	                m=array_len-1-i;
                        coord= coord->next;
                }
                fprintf (fptr, "];\n");
                m= array_len;
                if(action.type == 'U') {
                        fprintf (fptr, "r:=[\n");
                        coord= hol1;
                        while(coord != NULL) {
                                fprintf (fptr, "[%.3f,%.12f]",coord->temp,coord->r);
                                if (/*m > 1*/ coord->next != NULL) {
	       	                        fprintf (fptr, ",");
                                }
	                        fprintf (fptr, "\n");
	                        m=array_len-1-i;
                                coord= coord->next;
                        }
                        fprintf (fptr, "];\n");
                        m= array_len;
                        fprintf (fptr, "tripu:=[\n");
                        coord= hol1;
                        while(coord != NULL) {
                                fprintf (fptr, "[%.3f,%.12f]",coord->temp,coord->tripu);
                                if (/*m > 1*/coord->next != NULL) {
	       	                        fprintf (fptr, ",");
                                }
	                        fprintf (fptr, "\n");
	                        m=array_len-1-i;
                                coord= coord->next;
                        }
                        fprintf (fptr, "];\n");
                        m= array_len;
                        fprintf (fptr, "tripd:=[\n");
                        coord= hol1;
                        while(coord != NULL) {
                                fprintf (fptr, "[%.3f,%.12f]",coord->temp,coord->tripd);
                                if (/*m > 1*/coord->next != NULL) {
	       	                        fprintf (fptr, ",");
                                }
	                        fprintf (fptr, "\n");
	                        m=array_len-1-i;
                                coord= coord->next;
                        }
                        fprintf (fptr, "];\n");
                }
        }
        fprintf (fptr, "J:=[\n");
        fprintf (fptr, "%.12f,%.12f,%.12f,%.12f",action.jh, action.jv, action.jh2, action.jv2);
        fprintf (fptr, "];\n");
        fprintf (fptr, "Jay:= %.12f;", action.jdu);
        fprintf (fptr, "Jdash:= %.12f;", action.jdd);
        fprintf (fptr, "TripJay:= %.12f;", action.ku);
        fprintf (fptr, "TripJdash:= %.12f;", action.kd);
        fclose (fptr);
}

void set_temp(/*float temp[4][array_len], int point,*/ float min, float max) {
        /* A little function, to calculate the temperature points over the
        range between the maximum and minimum temperature levels. The random
        number is between 0 and 1000, and then divided by 1000 and multiplied
        by the range so that we can get 200 unique points */
        int i;
        float range, test;
        COORD *coord;
        range= max-min; //working out the range of the temperatures
        test= 0; //so  our loop runs at least once.
        //To avoid divide by zero errors
        while (test == 0) {
                test= (((float)(rand()%1000)/1000) * range)+ min;
        }
        coord= hol1;
        //To guarentee unique temperatures we check the current value against all previous values
        while(coord != NULL) {
                while (test == 0 || test == coord->temp) {
                        test= (((float)(rand()%1000)/1000) * range)+ min;
                        coord= hol1;
                }
                coord= coord->next;
        }
        coord= safe_malloc(sizeof(COORD),"Setting a temperature point");
        coord->temp= test;
        coord->next= hol1;
        hol1= coord;
}

void free_cond (void) {
	/* This frees up the memory taken by the linked list of 
	conditions by deleting each list item one after the other
	from the begining to the end. */
	COORD *del_ptr;
	while (hol1 != NULL) {
		del_ptr= hol1;
		hol1= hol1->next;
		free(del_ptr);
	}
}
\end{lstlisting}
\newpage
\subsection{mach2dt.c}
\begin{lstlisting}
%auto-ignore
/* We link our files with a standard header */
#include "isling2dt.h"

void metropolis(int chain[length][height], float temp, ACT var,double split[3],
 int lattice[3]) {
        /* This is the same as the previous function but now for the Union Jack lattice */
	int i= rand()%length;
        int j= rand()%height;
	int h, k;
        double changef= 0;
        double changeo= 0;
        double tripo;
        double tripf;
        double tripod;
        double tripfd;
	double prob;
	double condition;


        changeo= energyb(chain, var, chain[i][j], i, j);
        changef= energyb(chain, var, -chain[i][j], i, j);
        if(var.type == 'T') {
                tripo= chain[i][j] * ((chain[(i+1)%length][(j+1)%height]*(chain[i][(j+1)%height]
                + chain[(i+1)%length][j])) + (chain[(i+length-1)%length][(j+height-1)%height]*
                (chain[i][(j+height-1)%height]+chain[(i+length-1)%length][j]))+ (chain[(i+1)%length][j]
                * chain[i][(j+height-1)%height])+(chain[i][(j+1)%height]*chain[(i+length-1)%length][j]));
                tripf= (-chain[i][j]) * ((chain[(i+1)%length][(j+1)%height]*(chain[i][(j+1)%height]
                + chain[(i+1)%length][j])) + (chain[(i+length-1)%length][(j+height-1)%height]*
                (chain[i][(j+height-1)%height]+chain[(i+length-1)%length][j]))+ (chain[(i+1)%length][j]
                * chain[i][(j+height-1)%height])+(chain[i][(j+1)%height]*chain[(i+length-1)%length][j]));
        }else if(var.type == 'U' && (i+j)%2 == 0) {
                tripo= chain[i][j]*((chain[(i+1)%length][(j+1)%height]*(chain[(i+1)%length][j]
                + chain[i][(j+1)%height]))+ (chain[(i+length-1)%length][(j+height-1)%height]
                *(chain[i][(j+height-1)%height]+chain[(i+length-1)%length][j])));
                tripod= chain[i][j]*((chain[(i+length-1)%length][(j+1)%height]
                *(chain[(i+length-1)%length][j]+chain[i][(j+1)%height]))
                + (chain[(i+1)%length][(j+height-1)%height]*(chain[i][(j+height-1)%height]+chain[(i+1)%length][j])));
                tripf= (-chain[i][j])*((chain[(i+1)%length][(j+1)%height]*(chain[(i+1)%length][j]
                + chain[i][(j+1)%height]))+ (chain[(i+length-1)%length][(j+height-1)%height]
                *(chain[i][(j+height-1)%height]+chain[(i+length-1)%length][j])));
                tripfd= (-chain[i][j])*((chain[(i+length-1)%length][(j+1)%height]
                *(chain[(i+length-1)%length][j]+chain[i][(j+1)%height]))
                + (chain[(i+1)%length][(j+height-1)%height]*(chain[i][(j+height-1)%height]+chain[(i+1)%length][j])));
        } else if (var.type == 'U' && (i+j)%2 != 0) {
                tripo= chain[i][j]*((chain[i][(j+1)%height]*chain[(i+length-1)%length][j])+(chain[(i+1)%length][j] * chain[i][(j+height-1)%height]));
                tripod= chain[i][j]* ((chain[i][(j+1)%height]*chain[(i+1)%length][j])+(chain[(i+length-1)%length][j]*chain[i][(j+height-1)%height]));
                tripf= (-chain[i][j])*((chain[i][(j+1)%height]*chain[(i+length-1)%length][j])+(chain[(i+1)%length][j] * chain[i][(j+height-1)%height]));
                tripfd= (-chain[i][j])* ((chain[i][(j+1)%height]*chain[(i+1)%length][j])+(chain[(i+length-1)%length][j]*chain[i][(j+height-1)%height]));

        }
        if(0 >= - changeo + changef) {
                lattice[0]= lattice[0] - 2*chain[i][j];
                if((i+j)%2 == 0  && var.type == 'U') {
                        lattice[1]= lattice[1] - 2*chain[i][j];
                } else if ((i+j)%2 !=0 && var.type == 'U') {
                        lattice[2]= lattice[2] - 2*chain[i][j];
                }
                if(var.type == 'T') {
                        lattice[1]= lattice[1] - tripo + tripf;
                } else if (var.type == 'U') {
                        lattice[3]= lattice[3] - tripo + tripf;
                        lattice[4]= lattice[4] - tripod + tripfd;
                }
		chain[i][j]= -chain[i][j];
	} else {
		prob= exp(-(double)(1/(BOLT * temp))*(-changeo + changef));
		condition= COND*prob;
		if(rand()%COND < condition) {
                        lattice[0]= lattice[0] - 2*chain[i][j];
                        if((i+j)%2 == 0 && var.type == 'U') {
                                lattice[1]= lattice[1] - 2*chain[i][j];
                        } else if ((i+j)%2 !=0 && var.type == 'U') {
                                lattice[2]= lattice[2] - 2*chain[i][j];
                        }
                        if(var.type == 'T') {
                                lattice[1]= lattice[1] - tripo + tripf;
                        } else if(var.type == 'U') {
                                lattice[3]= lattice[3] - tripo + tripf;
                                lattice[4]= lattice[4] - tripod + tripfd;
                        }
			chain[i][j]= -chain[i][j];
		}
	}
        split[0]= split[0] + (double)lattice[0]/(length*height);
        if(var.type == 'T') {
                split[1]= split[1] + (double)lattice[1]/(2*(((length-1)*(height-1))+((length-1)+(height-1))));
        } else if(var.type == 'U') {
                split[1]= split[1] + (double)lattice[1]/(0.5*length*height);
                split[2]= split[2] + (double)lattice[2]/(0.5*length*height);
                split[3]= split[3] + (double)lattice[3]/(((length-1)*(height-1))+((length-1)+(height-1)));
                split[4]= split[4] + (double)lattice[4]/(((length-1)*(height-1))+((length-1)+(height-1)));
        }
}

double energy(int chain[length][height], ACT cond) {
        /* This is the energy calculation for the triangular and square lattices
        and returns the resultant value back */
        int i, j, k, m;
        double total;
        double holdu= cond.ku;
        double holdd= cond.kd;
        double hu= cond.jdu;
        double hd= cond.jdd;
        double hv= cond.jv;
        double hh= cond.jh;
        double sum1= 0;
        double sum2= 0;
        double sum3= 0;
        k= 0;
        for (i=0; i < length; i++) {
                if(k == 1) m= 3;
                if(k == 0) m= 0;
                for (j=0; j < height; j++) {
                        if(cond.type == 'U') {
                                switch(k) {
                                        case 0:
                                                holdd= cond.kd;
                                                holdu= 0;
                                                hd= cond.jdd;
                                                hu= 0;
                                                hh= cond.jh2;
                                                hv= cond.jv2;
                                                k= 1;
                                                break;
                                        case 1:
                                                holdd= 0;
                                                holdu= cond.ku;
                                                hd= 0;
                                                hu= cond.jdu;
                                                hh= cond.jh;
                                                hv= cond.jv;
                                                k= 0;
                                                break;
                                        default: // well you never know it could happen!
                                                printf("An error has occured in the calculation\n");
                                                //return(-1);
                                }
                        }
                        if(cond.type == 'H') {
                                switch(k) {
                                        case 0:
                                                hv= cond.jv;
                                                hu= 0;
                                                hd= 0;
                                                hh= 0;
                                                k= 1;
                                                break;
                                        case 1:
                                                hv= cond.jv;
                                                hu= 0;
                                                hd= 0;
                                                hh= cond.jh;
                                                k= 0;
                                                break;
                                        default: // well you never know it could happen!
                                                printf("An error has occured in the calculation\n");
                                                return(-1);
                                }
                        }
                        if(cond.type == 'D') {
                                holdu= 0;
                                holdd= 0;
                                switch(m) {
                                        case 0:
                                                hv= cond.kd;
                                                hu= cond.jv;
                                                hd= 0;
                                                hh= cond.ku;
                                                m= 1;
                                                break;
                                        case 1:
                                                hv= 0;
                                                hu= cond.kd;
                                                hd= 0;
                                                hh= 0;
                                                m= 0;
                                                break;
                                        case 2:
                                                hv= cond.jv;
                                                hu= 0;
                                                hd= 0;
                                                hh= cond.jh;
                                                m= 0;
                                                break;
                                        case 3:
                                                hv= cond.kd;
                                                hu= 0;
                                                hd= cond.jv;
                                                hh= 0;
                                                m= 4;
                                                break;
                                        case 4:
                                                hv= 0;
                                                hu= 0;
                                                hd= cond.kd;
                                                hh= cond.jh;
                                                m= 5;
                                                break;
                                        case 5:
                                                hv= cond.jv;
                                                hu= 0;
                                                hd= 0;
                                                hh= cond.ku;
                                                m= 3;
                                                break;
                                        default: // well you never know it could happen!
                                                printf("An error has occured in the calculation\n");
                                                return(-1);
                                }
                        }
                        // Calculation for nearest neighbours
                        sum1= sum1 + (chain[i][j]*((hh*chain[(i+1)%length][j])
                        + (hv*chain[i][(j+1)%height]) + (hu*chain[(i+1)%length][(j+1)%height])))
                        + (hd*chain[i][(j+1)%height]*chain[(i+1)%length][j]);
                        // Calculation for triplet neighbours
                        sum3= sum3 + (holdu*chain[i][j]*(chain[(i+1)%length][(j+1)%height]
                        * (chain[(i+1)%length][j]+chain[i][(j+1)%height])))
                        + (holdd*chain[i][(j+1)%height]*(chain[(i+1)%length][j]
                        * (chain[(i+1)%length][(j+1)%height]+chain[i][j])));;
                        // Calculation for sites
                        sum2= sum2 + chain[i][j];
                }
                if(i%2 == 0) { // to allow the intial diagonal of the row to alternate
                        k= 1;
                } else {
                        k= 0;
                }
        }
        // adding the three sums together and applying required factors
        total= (-sum3)- sum1 - (cond.h*sum2);
        // returning the result
        return total;
}


double energyb(int chain[length][height], ACT cond, int change, int i, int j) {
        /* This is the energy calculation for the triangular and square lattices
        and returns the resultant value back */
        int k, m;
        double total;
        double holdu= cond.ku;
        double holdd= cond.kd;
        double holdua= holdu;
        double holdda= holdd;
        double hu= cond.jdu;
        double hd= cond.jdd;
        double hv= cond.jv;
        double hh= cond.jh;
        double hua= hu;
        double hva= cond.jv2;
        double hda= hd;
        double hha= cond.jh2;
        double sum1= 0;
        double sum2= 0;
        double sum3= 0;
        if((i+j)%2 == 0) {
                k= 1;
        } else {
                k= 0;
        }
        if(i%2 == 0 && cond.type == 'D') {
                switch(j%3) {
                        case 0:
                                m= 0;
                                break;
                        case 1:
                                m= 1;
                                break;
                        case 2:
                                m= 2;
                                break;
                        default:
                                printf("An error has occured");
                                exit(-1);
                }
        } else if(cond.type == 'D') {
                switch(j%3) {
                        case 0:
                                m=3;
                                break;
                        case 1:
                                m= 4;
                                break;
                        case 2:
                                m= 5;
                                break;
                        default:
                                printf("An error has occured");
                                exit(-1);
                }
        }

        if(cond.type == 'U') {
                switch(k) {
                        case 0:
                                holdd= 0;
                                holdda= cond.kd;
                                holdu= 0;
                                holdua= cond.ku;
                                hd= 0;
                                hda= 0;
                                hu= 0;
                                hua= 0;
                                hh= cond.jh2;
                                hv= cond.jv2;
                                hha= cond.jh;
                                hva= cond.jv;
                                k= 1;
                                break;
                        case 1:
                                holdd= cond.kd;
                                holdda= 0;
                                holdu= cond.ku;
                                holdua= 0;
                                hd= cond.jdd;
                                hda= cond.jdd;
                                hu= cond.jdu;
                                hua= cond.jdu;
                                hh= cond.jh;
                                hv= cond.jv;
                                hha= cond.jh2;
                                hva= cond.jv2;
                                k= 0;
                                break;
                        default: // well you never know it could happen!
                                printf("An error has occured in the calculation\n");
                                //return(-1);
                }
        }
        if(cond.type == 'H') {
                holdd= 0;
                holdda= 0;
                holdu= 0;
                holdua= 0;
                switch(k) {
                        case 0:
                                hu= 0;
                                hua= 0;
                                hd= 0;
                                hda= 0;
                                hh= 0;
                                hha= cond.jh;
                                k= 1;
                                break;
                        case 1:
                                hu= 0;
                                hua= 0;
                                hd= 0;
                                hda= 0;
                                hh= cond.jh;
                                hha= 0;
                                k= 0;
                                break;
                        default: // well you never know it could happen!
                                printf("An error has occured in the calculation\n");
                                return(-1);
                }
        }
        if(cond.type == 'D') {
                holdu= 0;
                holdua= 0;
                holdd= 0;
                holdda= 0;
                switch(m) {
                        case 0:
                                hv= cond.kd;
                                hva= cond.kd;
                                hu= cond.jv;
                                hua= cond.jv;
                                hd= 0;
                                hh= cond.ku;
                                hha= cond.jh;
                                m= 1;
                                break;
                        case 1:
                                hv= 0;
                                hu= cond.kd;
                                hua= cond.kd;
                                hd= 0;
                                hh= 0;
                                hha= cond.ku;
                                m= 0;
                                break;
                        case 2:
                                hv= cond.jv;
                                hva= cond.jv;
                                hu= 0;
                                hd= 0;
                                hh= cond.jh;
                                m= 0;
                                break;
                        case 3:
                                hv= cond.kd;
                                hva= cond.kd;
                                hu= 0;
                                hd= 0;
                                hh= 0;
                                hha= cond.ku;
                                m= 4;
                                break;
                        case 4:
                                hv= 0;
                                hu= 0;
                                hd= cond.kd;
                                hda= cond.kd;
                                hh= cond.jh;
                                m= 5;
                                break;
                        case 5:
                                hv= cond.jv;
                                hva= cond.jv;
                                hu= 0;
                                hd= cond.kd;
                                hda= cond.kd;
                                hh= cond.ku;
                                hha= cond.jh;
                                m= 3;
                                break;
                        default: // well you never know it could happen!
                                printf("An error has occured in the calculation\n");
                                return(-1);
                }
        }
        // Calculation for nearest neighbours
        sum1= sum1 + (change*((hh*(chain[(i+1)%length][j])
        + (hv*chain[i][(j+1)%height]) + (hu*chain[(i+1)%length][(j+1)%height])
        + (hd*chain[(i+1)%length][(j+(height-1))%height])
        + (hha*chain[(i+(length-1))%length][j]) + (hva*chain[i][(j+(height-1))%height])
        + (hua*chain[(i+(length-1))%length][(j+height-1)%height])
        + (hda*chain[(i+(length-1))%length][(j+1)%height]))));
        // Calculation for triplet neighbours
        sum3= sum3 + (change*((holdu*chain[(i+1)%length][(j+1)%height]
        * (chain[(i+1)%length][j]+chain[i][(j+1)%height]))
        + (holdu*chain[(i+length-1)%length][(j+height-1)%height]
        * (chain[(i+length-1)%length][j]+chain[i][(j+height-1)%height]))
        + (holdd*chain[(i+1)%length][(j+height-1)%height]
        * (chain[i][(j+height-1)%height]+chain[(i+1)%length][j]))
        + (holdd*chain[(i+length-1)%length][(j+1)%height]
        * (chain[i][(j+1)%height]+chain[(i+length-1)%length][j]))
        + (holdda*chain[(i+1)%length][j]*chain[i][(j+1)%height])
        + (holdua*chain[(i+1)%length][j]*chain[i][(j+height-1)%height])
        + (holdda*chain[(i+length-1)%length][j]*chain[i][(j+height-1)%height])
        + (holdua*chain[(i+length-1)%length][j]*chain[i][(j+1)%height])));
        // Calculation for sites
        sum2= sum2 + change;
        // adding the three sums together and applying required factors
        total= (-sum3)- sum1 - (cond.h*sum2);
        // returning the result
        return total;
}
\end{lstlisting}
\newpage
\chapter{Programming II \\ Mean Field Approximation}
\label{ch:ProgII}
\section{Programmers guide}
The broad purpose of this code is to perform mean field simulations of the Ising model on various lattices. As with the last program, it does this by asking the user initially for the conditions for the system that is to be studied. These include the shape of the lattice, the temperature range to be studied over and the interaction strengths of the lattice. It then performs simulations of systems at temperatures at regular intervals along the temperature range. The program then outputs the results to a file which can be read into a mathematical program.

\subsection{Flow}
Initially the user is asked for the initial conditions of the system they wish to study, and then the temperature range they would like to study. After these have been stored, the user is then asked how many data points they require. After this the program calculates each data point by iteratively using the formulas from Chapter \ref{ch:MFT} until either the value of the sublattice magnetisation is constant, or 10,000 iterations have taken place, whichever occurs first. The value of magnetisation is then recorded for both the partially uncoupled and coupled equations along with the temperature. The program then moves on to the next data point, using the previous magnetisation value as the initial value. The temperature values are calculated at regular intervals in the temperature range. After the required number of data points have been calculated, the results are output to a data file which can then be read into a mathematical program. Again the simulation time is given on the screen after the program has finished.

\newpage
\section{Code listings}
\subsection{mean1.h}
\begin{lstlisting}
%auto-ignore
#include <stdio.h>
#include <stdlib.h>
#include <math.h>
#include <time.h>

#define	BOLT 8.617343*pow(10,-5)   /*The boltzman contant*/

typedef struct coordinate {
        long double sigma;
        long double mu;
        long double unsigma;
        long double unmu;
        float temp;
        struct coordinate *next;
} COORD;

double setup_gen(char *, char *);
void *safe_malloc(size_t, char *);
void *safe_realloc (int *, size_t, char *); /* Does the same for realloc */
void write_2d_array (char *, double, double, double, char);
void free_cond (void);          /* Frees the memory taken by linked list */
long double ltanh(long double);

extern COORD *hol1;

\end{lstlisting}
\newpage
\subsection{mean1.c}
\begin{lstlisting}
%auto-ignore
#include "mean1.h"

COORD *hol1= NULL;

int main() {
        int i, timeleft, points, seconds, minutes, hours, c, d;
        time_t timer;
        COORD *coord;
        float min, max, minmax, percent, interval, T, q;
        char type;
        double J, K, B;
        long double x, y, u, v, beta, m, n, a, b;
        char filename[100];

        x= 0;
        y= 0;
        while(type != 'S' && type != 'T' && type != 'U' && type != 'N' && type != 'H' && type != 'D') {
                printf("What type of lattice would you like to study? \n ([S]quare, [T]riangular, [U]nion Jack, [N]ext Nearest, [H]exagonal, [D]ouble Hexagonal) ");
                fscanf(stdin, "%1s", &type);
                type= toupper(type);
        }
        switch(type) {
                case 'S':
                        q= 2;
                        break;
                case 'T':
                        q= 3;
                        break;
                case 'U':
                        q= 2;
                        break;
                case 'H':
                        q=1.5;
                        break;
                default :
                        printf("There has been an error");
                        exit(-1);
        }
        if(type == 'U') {
                J= setup_gen("interaction of the square lattice", "real");
                K= setup_gen("interaction of the diagonal lattice", "real");
        } else {
                J= setup_gen("interaction of nearest neighbours", "real");
        }
        B= setup_gen("strength of overal field", "real");
        x= (long double)setup_gen("intial value for sigma", "real, less than or equal to +/- 1");
        if(type == 'U') {
                y= (long double)setup_gen("intial value for mu", "real, less than or equal to +/- 1");
        }
        min= 0;
        max= min;
        while(min == max) {
                min= (float)setup_gen("minimum temperature", "real above zero");
                max= (float)setup_gen("maximum temperature", "real above zero");
                if(min < 0) min= 0; //As we are working in Kelvin it makes little sense to have negative numbers
                if(max < 0) max= 0;
        }
        //to save retyping if the temperatures are inputted the wrong way round
        if(min > max) {
                minmax= min;
                min= max;
                max= minmax;
        }
        points= (int)setup_gen("data points", "integer above zero");
        printf("Enter output filename: ");
        if(fscanf(stdin, "%s", &filename) != NULL) {
                printf("\n");
                fflush(stdin);
        }
        timer= time(NULL); // starting the timer
        // creating our temperature array
        interval= (max-min)/points;
        c= 0;
        d= 0;
        coord= hol1;
        coord= safe_malloc(sizeof(COORD), "New data point");
        coord->temp= min;
        coord->sigma= x;
        coord->mu= y;
        coord->unsigma= x;
        coord->unmu= y;
        coord->next= hol1;
        hol1= coord;
        u= x;
        v= y;
        m= 10;
        n= 10;
        a= 10;
        b= 10;
        for(i= 1; i < points; i++) {
               T= min+(i*interval);
               beta= (long double)(1/(BOLT * T));
               while(x != m || y != n) {
                         m= x;
                         n= y;
                         if(type == 'U') {
                                x= beta*((2*K*m)+(2*J*n)+B);
                                y= beta*((2*J*m)+B);
                         } else {
                                x= beta*((q*J*m)+B);
                         }
                         if(c == 0) {
                              x= ltanh(x);
                              y= ltanh(y);
                         } else {
                           x= (ltanh(x)+ m)/2;
                           y= (ltanh(y)+ n)/2;
                         }
                         c++;
                         if(c == pow(10,8)) break;
               }
               c= 0;
               while(u != a || v != b && type == 'U') {
                         a= u;
                         b= v;
                         if(d== 0) {
                                u= ltanh(beta*((2*K*a)+B));
                                v= ltanh(beta*(4*J*a)+B);
                         } else {
                           u= (ltanh(beta*((2*K*a)+B))+a)/2;
                           v= (ltanh(beta*((4*J*a)+B))+b)/2;
                         }
                         d++;
                         if(d == pow(10,8)) break;
               }
               d= 0;
               coord= safe_malloc(sizeof(COORD), "New data point");
               coord->temp= T;
               coord->sigma= x;
               coord->mu= y;
               coord->unsigma= u;
               coord->unmu= v;
               coord->next= hol1;
               hol1= coord;
               m= 10;
               n= 10;
               a= 10;
               b= 10;
               percent= (float) (i+1)/points;
               seconds= (time(NULL)-timer);
               timeleft= (long int)(((time(NULL)-timer)/percent)-seconds);
               seconds= timeleft%60;
               minutes= abs(timeleft/60)%60;
               hours= abs(timeleft/(60*60));
               printf("%.2f %% done, time left: %.2d hours, %.2d minutes, %.2d seconds\r",percent*100, hours, minutes,seconds); // To show the program is still working
        }
        write_2d_array(filename, J, K, B, type);
        free_cond();
        printf ("\n Now run maple to plot graph\n"); // handy hint to the user
        seconds= (time(NULL)-timer)%60;
        minutes= (abs((time(NULL)-timer-seconds)/60))%60;
        hours= abs((time(NULL)-timer-seconds-(minutes*60))/(60*60));
        printf("Runtime: %ld hours, %ld minutes, %ld seconds\n", hours, minutes, seconds); // how long it too
        return 0;
}

long double ltanh(long double x) {
     if (fabsl(x) < powl(2.0,(-55.0))) {
        x= x;
     } else {
        x= tanhl(x);
     }
     return x;
}
\end{lstlisting}
\newpage
\subsection{setup.c}
\lstinputlisting{setup2d.c}

\end{document}